%% file: main.tex
\documentclass[aps,prl,twocolumn,amsmath,amssymb,nofootinbib,superscriptaddress,floatfix,reprint]{revtex4-1}

\usepackage{amsmath,amsfonts,amssymb,amsthm,epsfig,array}
\usepackage{dsfont}
\usepackage{ulem}
\usepackage{graphics}
\usepackage{float}
\usepackage{bm}% bold math
\usepackage{verbatim}%\begin{comment} ... \end{comment}
\usepackage[dvipsnames]{xcolor}
\usepackage[hidelinks,colorlinks,linkcolor=blue,
citecolor=blue,urlcolor=blue]{hyperref}
\usepackage[titletoc,title]{appendix}
\usepackage{physics}
\usepackage{latexsym}
\usepackage{tikz-feynman}
\usepackage{tikz,tikz-feynhand}
\usepackage{extpfeil}
\usepackage{subfigure}
\usepackage{titlesec}
\usepackage{siunitx}
\allowdisplaybreaks
\titleformat{\section}{\normalfont\normalsize\bfseries\centering}{\thesection.}{1em}{}
\titleformat{\subsection}{\normalfont\normalsize\itshape}{\thesubsection.}{1em}{}
\titleformat{\subsubsection}{\normalfont\normalsize\itshape}{\thesubsubsection.}{1em}{}

\newcommand{\cre}{{\dag}}
\newcommand{\ann}{{\vphantom{\dag}}}

\graphicspath{./figures/}

\newcommand{\1}{\hspace*{-1pt}}
\newcommand{\2}{\hspace*{-2pt}}

\usepackage{float}

\def\ket#1{\left|#1\right\rangle}

\newcommand\boldvector[1]{%
	\ifcat\noexpand#1\relax % check if the argument calling a command
	\boldsymbol{#1}
	\else
	\mathbf{#1}% Else it is probably a letter
	\fi
}

\def\frac#1#2{{\textstyle{#1 \over #2}}}

\newcommand{\rom}[1]{\uppercase\expandafter{\romannumeral #1\relax}}

\begin{document}

\title{Loop Current Order on the Kagome Lattice}
\author{Jun Zhan}
\thanks{These authors contributed equally to this work.}
\affiliation{Beijing National Laboratory for Condensed Matter Physics and Institute of Physics, Chinese Academy of Sciences, Beijing 100190, China}
\affiliation{School of Physical Sciences, University of Chinese Academy of Sciences, Beijing 100049, China}

\author{Hendrik Hohmann}
\thanks{These authors contributed equally to this work.}
\affiliation{Institut f\"{u}r Theoretische Physik und Astrophysik, Universit\"{a}t W\"{u}rzburg, Am Hubland Campus S\"{u}d, W\"{u}rzburg 97074, Germany}

\author{Matteo D\"{u}rrnagel}
\affiliation{Institut f\"{u}r Theoretische Physik und Astrophysik, Universit\"{a}t W\"{u}rzburg, Am Hubland Campus S\"{u}d, W\"{u}rzburg 97074, Germany}
\affiliation{Institute for Theoretical Physics, ETH Zürich, 8093 Zürich, Switzerland}

\author{Ruiqing Fu}
\affiliation{Institute of Theoretical Physics, Chinese Academy of Sciences, Beijing 100190, China}
\affiliation{School of Physical Sciences, University of Chinese Academy of Sciences, Beijing 100049, China}

\author{Sen Zhou}
\affiliation{Institute of Theoretical Physics, Chinese Academy of Sciences, Beijing 100190, China}

\author{Ziqiang Wang}
\affiliation{Department of Physics, Boston College, Chestnut Hill, Massachusetts 02467, USA}

\author{Ronny Thomale}
\affiliation{Institut f\"{u}r Theoretische Physik und Astrophysik, Universit\"{a}t W\"{u}rzburg, Am Hubland Campus S\"{u}d, W\"{u}rzburg 97074, Germany}

\author{Xianxin Wu}
\email{xxwu@itp.ac.cn}
\affiliation{Institute of Theoretical Physics, Chinese Academy of Sciences, Beijing 100190, China}

\author{Jiangping Hu}
\email{jphu@iphy.ac.cn}
\affiliation{Beijing National Laboratory for Condensed Matter Physics and Institute of Physics, Chinese Academy of Sciences, Beijing 100190, China}
\affiliation{School of Physical Sciences, University of Chinese Academy of Sciences, Beijing 100049, China}
% \affiliation{Kavli Institute for Theoretical Sciences, University of Chinese Academy of Sciences, Beijing 100190, China}
\affiliation{ New Cornerstone Science Laboratory, Institute of Physics, Chinese Academy of Sciences, Beijing 100190, China}

\begin{abstract}
Recent discoveries in kagome materials have unveiled their capacity to harbor exotic quantum
states, including intriguing charge density wave (CDW) and superconductivity. Notably,
accumulating experimental evidence suggests time-reversal symmetry breaking within the
CDW, hinting at the long-pursued loop current order (LCO). Despite extensive research efforts, achieving its model realization and understanding the mechanism through unbiased many-body simulations have remained both elusive and challenging. 
In this Letter, we develop a microscopic model for LCO on the spinless kagome lattice with nonlocal interactions, utilizing unbiased functional renormalization group calculations to explore ordering tendencies across all two-particle scattering channels. At the Van Hove filling, we identify sublattice interference to suppress onsite CDW order, leaving LCO, charge bond order, and nematic CDW state as the main competitors. Remarkably, a $2\times2$ LCO emerges as the many-body ground state over a significant parameter space with strong second nearest-neighbor repulsion, stemming from the unique interplay between sublattice characters and lattice geometry. The resulting electronic model with LCO bears similarities to the Haldane model and culminates in a quantum anomalous Hall state. We also discuss potential experimental implications for kagome metals.

\end{abstract}
\maketitle

Magnetic transition-metal based kagome materials, consisting of corner-sharing triangles, present an exciting platform to explore intriguing correlated and topological phenomena, including quantum spin liquid~\cite{RevModPhys.88.041002,AMielke_1991,PhysRevLett.100.136404} and Dirac and Weyl semimetals~\cite{ye2018massive,yin2020quantum,kang2020dirac}. They arise 
from the inherent features of the kagome lattice, including substantial geometric spin frustration, flat bands and Dirac cones. However, despite theoretical studies that have spanned over a decade, the exploration of Van Hove singularities (VHS), another critical feature of kagome lattices, has been hampered experimentally by the scarcity of suitable materials. Consequently, the recent discovery of kagome metals that feature VHS in the vicinity of the Fermi level, leading to the emergence of diverse quantum states such as superconductivity, charge density wave (CDW) order and nematicity, have garnered significant attention. In particular, the CDW in AV$_3$Sb$_5$ (A = K,Rb,Cs)~\cite{AV3Sb5_Ortiz_first_paper,AV3Sb5_nature_review,jiangping_hu_review} and FeGe~\cite{PhysRevLett.129.166401,TengXK2022,TengXK2023,han2024orbital} has been found to exhibit intriguing properties. In AV$_3$Sb$_5$, a CDW transition occurs below $T_\mathrm{CDW}=78$--$103$~K, characterized by a breaking of translational symmetry, resulting in either a $2\times2\times2$ or $2\times2\times4$ configuration. Intriguingly, various measurements, such as muon spin resonance ($\mu$SR)~\cite{Mielke2022,YuL2021} and magneto-optical Kerr effect experiments~\cite{2021arXiv211011306W,2022arXiv220410116X,GuoCY2022,xing2024optical}, have provided evidence of time-reversal symmetry (TRS) breaking within the CDW in the absence of magnetic ordering. Superconductivity subsequently emerges within the CDW phase, reaching a maximum $T_\mathrm{c}$ of 2.5 K at ambient pressure, with double superconducting domes observed under pressure and doping~\cite{PhysRevLett.126.247001,PhysRevB.103.224513,Chen_2021,PhysRevLett.127.237001,PhysRevMaterials.6.L041801,2021arXiv211012651L}. In the antiferromagnet FeGe with ferromagnetic kagome layers, a CDW order occurs below the magnetic transition temperature~\cite{PhysRevLett.129.166401}. The CDW exhibits topological characters with a $2\times2\times2$ reconstruction and an enhancement in the magnetic moment and anomalous Hall effect are observed upon the CDW transition~\cite{TengXK2022,han2024orbital}. This experimental evidence indicates the occurrence of long-pursued loop current order (LCO) within the kagome lattice~\cite{feng2021chiral,PhysRevLett.127.217601,PhysRevB.104.045122,PhysRevB.104.035142}, reminiscent of flux states suggested in cuprates~\cite{PhysRevB.55.14554,PhysRevB.63.094503,PhysRevLett.83.3538} and Haldane model~\cite{haldane1988model}.

The concept of a current-carrying phase, a rare state of matter, has been proposed in the pseudogap phase of cuprate and honeycomb systems through phenomenological or mean-field analyses~\cite{PhysRevB.55.14554,PhysRevB.63.094503,PhysRevLett.83.3538,haldane1988model,PhysRevLett.100.156401}. Whether
this loop current phase could serve as a precursor or parent state to exotic quantum phenomena like high-$T_\mathrm{c}$ superconductivity and quantum anomalous Hall effect has long been debated. However, the experimental verification of LCO in cuprate and its theoretical stability beyond mean-field level in square and honeycomb lattice remains controversial. 
 Unbiased many-body calculations have indicated that the true ground state is not LCO but onsite charge or spin order under conventional interaction settings~\cite{PhysRevB.70.113105,PhysRevB.71.075103,PhysRevLett.102.017005,PhysRevLett.99.027005,PhysRevLett.112.117001,PhysRevB.88.245123,PhysRevB.89.035103,PhysRevB.92.085147,PhysRevB.92.085146,PhysRevB.92.155137}. This has relegated LCO to a phenomenological hypothesis rather than a microscopic reality.
However, various TRS breaking evidence within CDW of kagome metals offers strong motivation to explore the possibility of LCO in the kagome lattice.
 The VHS of the kagome lattice exhibit unique sublattice texture on the Fermi surface (FS)~\cite{Kiesel2012,Wu2021}, which significantly impacts the electronic instabilities. Importantly, the implied sublattice interference effect suppresses local interaction scales, thereby enhancing the role of the long-range tail of the screened Coulomb repulsion~\cite{Kiesel2013,SYu2012,Wang2013}. Nevertheless, prior studies have shown the absence of LCO in simple interacting kagome models, rendering its realization challenging~\cite{Kiesel2013,SYu2012,Wang2013,FerrariPRB2022}. Recent theoretical analyses imply that long-range interactions and frozen spin degrees of freedom may promote LCO~\cite{Dong2023,RPA_paper}. 
Therefore, whether the simple kagome lattice can realize LCO as the many-body ground state and what the microscopic mechanism entails are outstanding questions. This necessitates an unbiased many-body investigation that adequately accounts for competing fluctuations in all channels.

Motivated by these fundamental questions, we concentrate on the inherent CDW phenomena in the kagome lattice at Van Hove filling. We adopt the spinless kagome model with
nonlocal interactions, which can avoid complex spin fluctuations and have direct experimental relevance to materials, like magnet FeGe, hosting spin-polarized low-energy electronic structures. To comprehensively address fluctuations across channels, we employ the unbiased functional renormalization group (FRG) approach to analyze FS instabilities. Our calculations reveal that the second
nearest-neighbor repulsion can promote fluctuations of imaginary bond charge order, driven by the sublattice
interference and frustrated kagome geometry. This results in a $2\times2$ LCO ground state, which prevails over a substantial parameter space when the second nearest-neighbor repulsion is notably strong. Remarkably, this TRS breaking $2\times2$ LCO is validated for the first time, to our knowledge, as the many-body ground state through rigorous many-body calculations, confirming its existence in the kagome lattice. We also observe the nematic CDW, charge bond order, and $f$-wave superconductivity (SC) for different parameters, which are potentially relevant for kagome metals.

{\it Interacting kagome model and formalism of FRG.--}
To explore the intrinsic charge orders driven by electronic interactions within the kagome lattice, we study the model under a spinless scenario that involves density-density interactions. The full Hamiltonian including kinetic
energy and nonlocal interactions reads
\begin{equation}
	\begin{split}
		H=&-t\sum_{\langle i, j\rangle }  (c_{i}^{\dagger} c_{j} + \text{h.c.})  -\mu\sum_{i}n_{i} \\  
		+&V_1\sum_{\langle i, j\rangle } n_i n_j + V_2 \sum_{\langle\langle i, j\rangle \rangle } n_i n_j,
	\end{split}
	\label{eqn:hamiltonian}
\end{equation}
where $c^{\dagger}_{i}$ creates an electron on lattice site $i$, $t$ is the nearest-neighbor hopping amplitude, $n_i=c_{i}^{\dagger} c_{i}$ is the electron density operator which couples to the chemical potential $\mu$, and $V_n$ is the $n$th nearest-neighbor ($n$nn) Coulomb repulsion. Distinct from the spinful case, the spin degrees of freedom are frozen here, and the onsite interaction is absent in Eq.~\eqref{eqn:hamiltonian} as fermionic antisymmetry precludes double occupation of a single site~\cite{RPA_paper,PhysRevB.82.075125}.
The single-particle dispersion features a flat band, Dirac cones and two different
kinds of VHS, which are characterized by distinct sublattice texture on the FS~\cite{Kiesel2013,Wu2021}, distinct from
triangular and honeycomb lattices. In this Letter we focus on the p-type VHS with $\mu=0$, where each inequivalent M point
consists of states solely attributed to one sublattice, and the nesting vectors $\bm{Q}_{A, B, C}$ connect
saddle points with distinct sublattice characters [see Supplemental
Material (SM)~\cite{SM}].

To properly address the intricate competition between many-body correlated states within the interacting kagome model, we employ FRG calculations, which take into account
all intertwined particle-hole and particle-particle fluctuations on equal
footing~\cite{Metzner2012,Platt2013,Beyer2022}. The
leading instability is extracted by the eigenvector
associated with the most divergent eigenvalue of the effective vertex function at $\Lambda_c$, where FRG flow exhibits a divergence. To monitor the
propensity of the system toward various symmetry broken states
$\ket{\Phi_\theta}$ within the FRG flow, we determine the expectation~value~$\Xi \,=\, \expval{V_\Lambda}{\Phi_{\theta}}$ of the effective FRG vertex $V$
at each flow step. 
%For a detailed description of the FRG and the employed
%implementation, we refer to Ref.~\cite{Beyer2022} and SM~\cite{SM}.
The detailed descriptions of the FRG formalism and the employed
implementation are provided in SM~\cite{SM}.

\begin{figure}[t]
	\centering
	\includegraphics[width=1.\columnwidth]{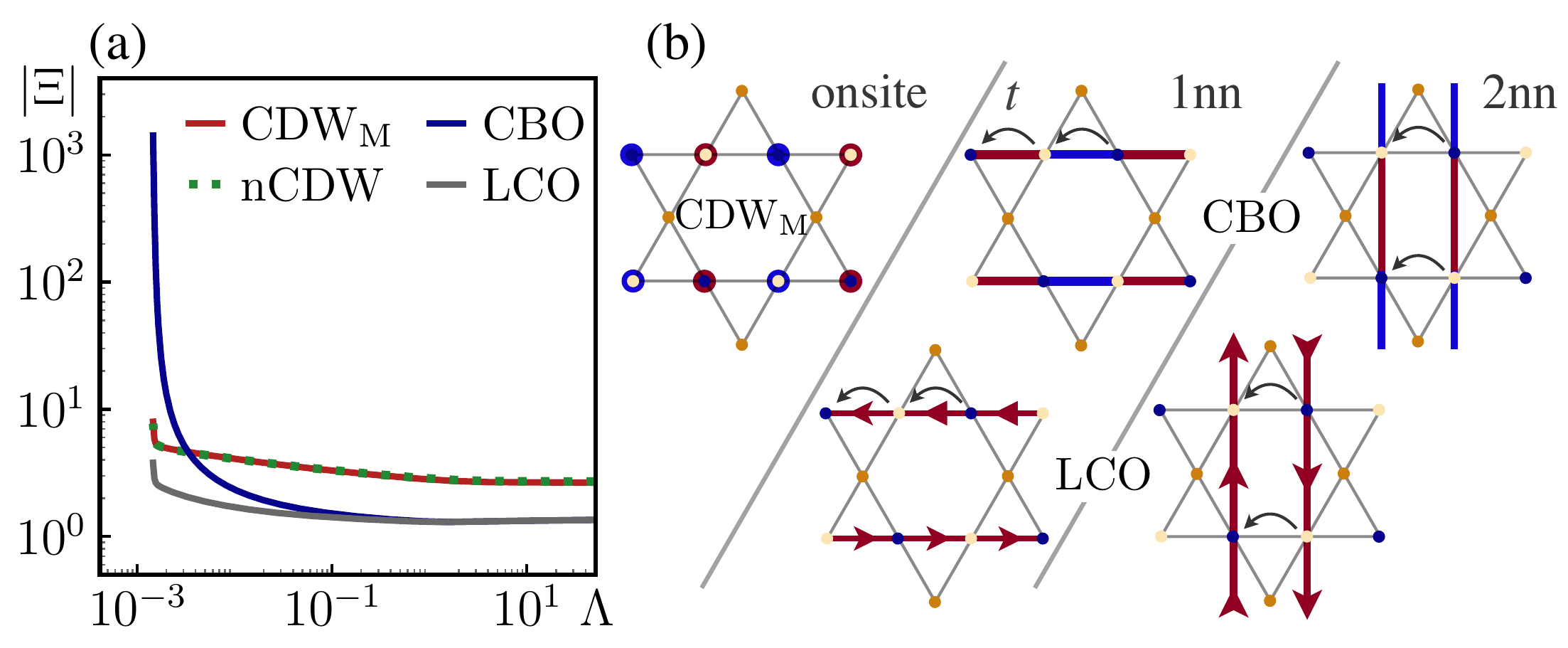}
	\caption{ Representative FRG flow with 1nn repulsion and real-space CDW patterns. (a) FRG flow of the expectation values of CBO, LCO, nCDW, and $\mathrm{CDW}_\text{M}$ and for $V_1=t$ and $V_2=0$.~(b)~Real-space configurations of CBO and LCO on 1nn and 2nn bond as well as the onsite modulated CDW with a wave vector of $\bm{Q}_C$. The real-space pattern at two other nesting vectors can be obtained by the application of sixfold rotational operation. Red (blue) denotes positive (negative) onsite or bond order parameters, while arrows denote directions of current flows. Only the bond orders within the antisymmetric channel are depicted. }
	\label{fig:CBO}
\end{figure}

{\it Intrinsic bond order fluctuations.--}
In the triangular and honeycomb lattices with nonlocal Coulomb interactions, the
dominant instability associated with VHS is the onsite CDW order at the nesting vectors~\cite{PhysRevB.92.155137,O2021,Gneist2022}, owing to its significant reduction of the potential energy of nonlocal
repulsion.
However, in the
kagome lattice, onsite CDW fluctuations at the nesting vector are significantly
suppressed due to the sublattice texture at Van Hove filling and the associated sublattice interference (SI) effect
gives rise to a pronounced inclination toward bond charge
orders~\cite{Kiesel2013,Wang2013,RPA_paper}.
The SI effect on charge states can be clearly demonstrated in the FRG flow with a typical
nearest-neighbor repulsion $V_1=t$ shown in Fig.~\ref{fig:CBO}(a). Starting from the bare interacting system at a high energy scale, onsite CDW fluctuations at $\bm{Q}=0$ (nCDW) and
$\bm{Q}=\text{M}$ (CDW$_{\text{M}}$, see Fig.~\ref{fig:CBO}(b)) are initially dominant, in accordance with other hexagonal lattice systems~\cite{O2021, Gneist2022,Kiesel2012}.
By successively integrating out the high-energy modes, the eigenvalue of
charge bond orders (CBO, Fig.~\ref{fig:CBO}(b)) increases dramatically and eventually exceeds those of onsite CDW orders at
an intermediate scale and finally diverges, indicating a leading instability of
real charge bond order. Strikingly, this is accompanied by the
emergence of a competing imaginary charge bond order. Figure~\ref{fig:CBO}(b) shows the real-space pattern of relevant onsite CDW, real and imaginary bond orders with the wave vector of $\bm{Q}_C$, where onsite charge and bond modulations are represented by different colors and the arrows denote the direction of currents.

\begin{figure}[h]
	\centering
	\includegraphics[width=0.9\columnwidth]{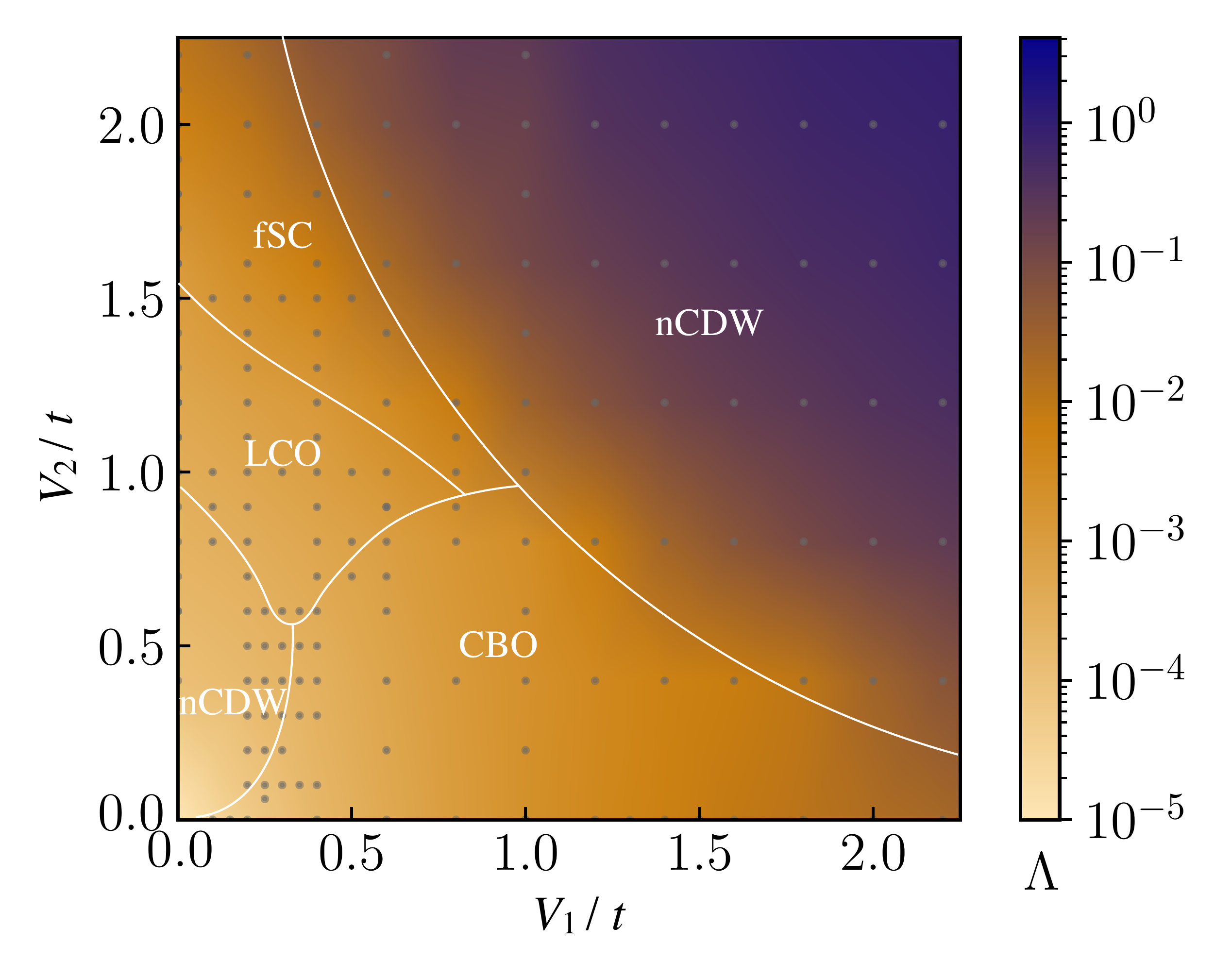}
	\caption{ Phase diagram of the spinless interacting kagome model at the p-type Van Hove filling. The critical scales $\Lambda_{\mathrm{c}}$ proportional
	to the expected transition temperature $T_{\mathrm{c}}$ are indicated by color. Gray dots indicate the calculated points in interaction parameter space.
	}
	\label{fig:phase_diagram}
\end{figure}

{\it FRG phase diagram.--}
We conducted thorough FRG calculations for the model described in
Eq.\ref{eqn:hamiltonian}. The correlated phases identified within the
interaction parameter space are depicted in
Fig.~\ref{fig:phase_diagram}, where the color represents the critical cutoff.
In the regime of strong interactions, we recover a twofold nematic charge
density wave~(nCDW) as the classical solution of the interacting part of the
Hamiltonian, that can acquire significant electrostatic energy gain by introducing
unequal occupation between the three sublattices~\cite{RPA_paper}.
This energy gain is linear with respect to the nonlocal repulsion, leading to the
nCDW becoming the dominant order with high critical scales at large repulsion.
This twofold nCDW mainly involves an intraunit cell charge density wave.
At lower scales, electronic screening effects dictate the realized phases:
In the $V_1$ dominated regime, the system develops an instability toward
CBO as known from prior studies of the spinful case~\cite{Wang2013,
	Kiesel2013, Profe2024}.
Remarkably, we observe the emergence  of a TRS breaking LCO state in the
intermediate $V_2$ regime. This state is situated between regions of unconventional
$f$-wave superconducting order in the strong $V_2$ regime and twofold nCDW in
the weak $V_2$ regime.
In the following, we will focus on the loop current order, elucidating both
its underlying mechanism and physical properties of the symmetry
broken state. In particular, we will identify the competition and cooperation between
the different phases present in the $V_2$ dominated regime and address the
apparent asymmetry to the CBO phase in the $V_1$ regime.

\begin{figure}[t]
	\centering
	\includegraphics[width=0.9\columnwidth]{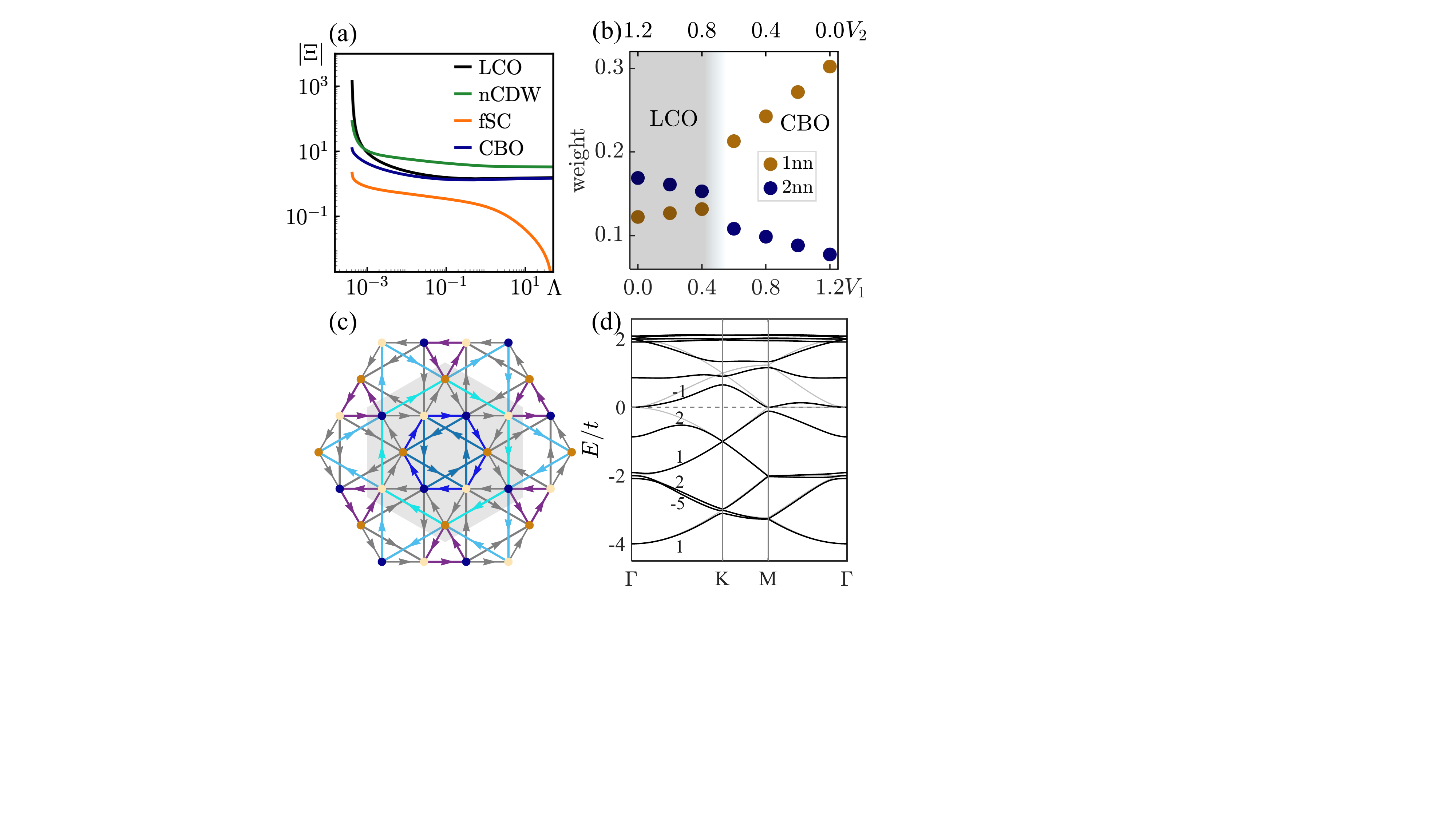}
	\caption{  %RG flows, real-space pattern and electronic structure of the Loop current order.
		(a) Representative FRG flow of the expectation values of nCDW,  2nn CBO, 2nn LCO and fSC
		for $V_1=0$ and $V_2=1.1t$.
		(b) Variation of 1nn and 2nn contributions to the LCO and CBO phases as a
		function of changing $V_1$ and $V_2$ at fixed $V_1+V_2=1.2t$.
		(c) Real-space pattern of the representative $3\bm{Q}$ LCO on both 1nn and 2nn bonds.
		The enlarged $2 \times 2$ supercell is indicated by gray shading.
		(d) Emergent Chern bands of the LCO phase  in the folded BZ with order parameters
		$\Delta_{\mathrm{1nn}}^{\mathrm{LCO}}=0.1t$ and $\Delta_{\mathrm{2nn}}^{\mathrm{LCO}}=0.15t$, exhibiting a
		full gap around the Fermi level. The filled bands feature a total Chern
		number of $C=1$. Gray curves indicate the backfolded dispersion.}
	\label{fig:LCO}
\end{figure}

{\it Loop current order.--}
In the intermediate $V_2$ regime, the representative flows with LCO as leading instability are illustrated in Fig.~\ref{fig:LCO}(a) with $V_2=1.1t$ and $V_1=0$. The nCDW occurs as close competitor, distinct from the $V_1$ dominant case in Fig.~\ref{fig:CBO}(a). The observed LCO mainly involves a mixture between the 1nn and 2nn bonds from diagonalizing renormalized susceptibility, in contrast to dominant 1nn bond of loop current pattern reported in previous studies~\cite{Dong2023,Tazai2023}. 
The inclusion of 1nn repulsion $V_1$ modifies real-space structure of LCO, which is evidenced by the changing weights of 1nn and 2nn bond orders along the parameter line $V_1+V_2=1.2t$, as shown in Fig.~\ref{fig:LCO}(b). An increase in $V_1$ enhances the 1nn component in the LCO, which in turn significantly enhances the CBO due to pronounced 1nn real bond fluctuations and triggers a transition from LCO to CBO, where the 1nn and 2nn bond components exhibit a jump and drop, respectively. %The LCO is dominant in a relatively large regime and is stabilized by stronger imaginary bond fluctuation on the 2nn bond compared to that of real bond order at the bare level, a direct consequence of the SI effect and unique kagome lattice geometry~\cite{RPA_paper}. 
The LCO is dominant in a relatively large regime and is stabilized by stronger imaginary bond fluctuation on the 2nn bond compared to that in real bond channel, which is a direct consequence of the SI effect and unique kagome lattice geometry~\cite{RPA_paper}.
Intuitively, the intimate relation between the bond length and the prevalence of real or imaginary bond orders can be directly inferred from the different transformation behavior of LCO and CBO under short range hopping fluctuations of particle-hole pairs, that result in an energy splitting already at second order in Ginzburg-Landau theory (details in Sec.~III of SM~\cite{SM}).

In accordance with the nesting vectors and sublattice makeup of the FS, the LCO
establishes at the three inequivalent nesting vectors $\bm{Q}_{A,B,C}$, involving particle-hole pairs
between three different bonds. To analyze the ground state, we minimize the
Landau free energy containing three degenerate order parameters $\Delta_{1,2,3}$, each associated with one of the nesting vectors. In the Ginzburg-Landau expansion, their couplings at quartic order govern the ground state energy and are given by,
\begin{equation}
	\begin{aligned}
		f^{(4)}
		&= \frac{1}{2} Z_1 |\mathbf{\Delta}|^4 + (Z_2-Z_1) (\Delta_1^2\Delta_2^2 +\Delta_2^2\Delta_3^2 +\Delta_3^2\Delta_1^2)
	\end{aligned}
\end{equation}
with $Z_1 - Z_2 > 0$ as explicated in the SM~\cite{SM}. Different from the CBO case, the
trilinear term vanishes owing to the TRS breaking. These quartic terms favor an equal contribution of the 3$\bm{Q}$ modulations, leading to an enlarged $2\times2$
pattern. The characteristic pattern of the LCO maintaining the sixfold
rotational symmetry is shown in Fig.~\ref{fig:LCO}(c), where currents of equal color form closed loops around the rotational center. This TRS breaking
3$\bm{Q}$ LCO fully gaps out the FS and generates a Chern insulator, analogous to the Haldane model.
This state features topologically nontrivial bands close to the Fermi level, as shown in Fig.~\ref{fig:LCO}(d), and Chern Fermi pockets can be introduced upon electron or hole doping.

Surprisingly, lowering the interaction scale triggers a transition from the LCO to a reentrant twofold CDW order at $\bm{q}=0$, corresponding to the two-dimensional $E_{2g}$ representation of $D_{6h}$.
The minimization of the free energy within the associated two-dimensional eigenspace leads to a nematic order that breaks the sixfold rotational symmetry. There are three possible configurations with an unequal charge occupation of three sublattices, as shown in SM~\cite{SM}. According to our analysis, the projection of particle-particle fluctuations onto the particle-hole channels promotes nCDW but simultaneously suppresses LCO (see Sec.~VI of SM for more details~\cite{SM}), rendering the nCDW slightly dominant.

\begin{figure}[t]
	\centering
	\includegraphics[width=1\columnwidth]{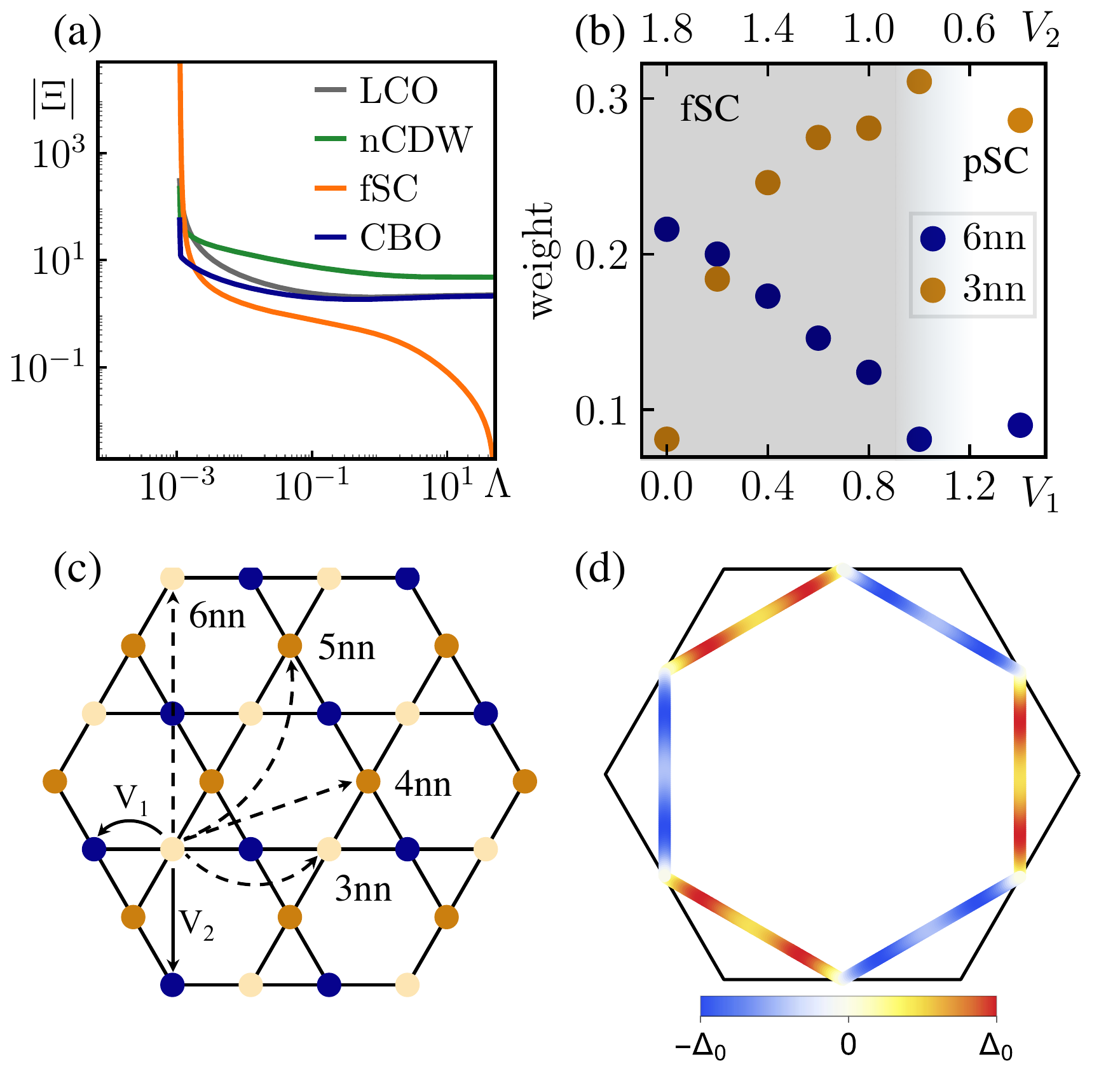}
	\caption{%{\bf RG flow, gap functions in real and momentum space of the $f$-wave superconducting order.} 
	(a) FRG flow of leading instabilities for $V_1=0$ and $V_2=1.6\,t$. (b) Variation of 3nn and 6nn components
	in the $f$- and $p$-wave superconductivity as a function of changing $V_1$ and $V_2$. (c) Schematics for the real-space pairing configuration, mediated by a combination of high-order virtual $V_1$ and $V_2$ processes. (d) $f$-wave gap function on the FS transforming under the $B_{2u}$ representation in the $V_2$ dominant regime.}
	\label{fig:fSC}
\end{figure}

{\it f-wave superconductivity.--}
For larger values of $V_2$ the LCO is surpassed by a superconducting
instability mediated by the particle-hole fluctuations within LCO and nCDW channels (cf. Fig.~\ref{fig:fSC}(a)).
Caused by the spinless restriction, the pairing order parameter is odd in real
space due to fermionic anti-commutation, which implies a vanishing gap at time
reversal invariant momenta.
Since the singular DOS at the M points cannot be gapped out by the
superconducting order, SC phases are less prominent in the phase diagram of
Fig.~\ref{fig:phase_diagram} compared to prior studies on the spinful kagome lattice~\cite{Wang2013, Schwemmer2023, Profe2024}.

The remaining FS is fully gapped by an exotic $f$-wave SC order corresponding to
the $B_{2u}$ irreducible representation (cf. Fig.~\ref{fig:fSC}(c)).
In real space, this corresponds to Cooper pairing between 5th and 6th nn sites.
In the $V_1=0$ case, these represent the closest sites available for pair
formation mediated by the second order $V_2$ processes as indicated in
Fig.~\ref{fig:fSC}(c).
The sublattice texture on the Fermi surface favors Cooper pair formation on the 6nn compared to the 5nn bonds, since the states at $\mathbf k$ and $-\mathbf k$ near the VHS are mainly constituted by the same sublattice.
Consequently, the intrasublattice pairing, like the 6nn bond, can gap out Fermi segments with high DOS, resulting in a significant condensation energy gain.

A sizable $V_1$ induces pairing between the 3nn sites due to mixed $V_1-V_2$ processes as depicted in Fig.~\ref{fig:fSC}(b).
The decreased distance between Cooper pairs reduces the gap modulation along the FS and increases the condensation energy gain by the SC order.
A substantial $V_1$ leads to a transition from the $f$-wave to $p$-wave pairing symmetry in the particle-particle channel (Fig.~\ref{fig:fSC}(b)), which is subleading to the CBO in the $V_1$-dominant regime.
The SC order can thereby exploit $V_1$ to gain energy opposed to the LCO, which exclusively relies on $V_2$ for its emergence (compare Fig.~\ref{fig:fSC}(a) and Fig.~\ref{fig:LCO}(b)).
Hence, the phase boundary to the SC state is shifted to lower $V_2$ values with increasing $V_1$, thereby bounding the LCO domain from above.

{\it Discussion and conclusion.--}
Our minimal model isolates the kagome-intrinsic ingredients that stabilize LCO---the interplay between VHS-related sublattice texture and lattice geometry---a mechanism that is generic to frustrated lattices (e.g., checkerboard, pyrochlore). The resulting sublattice interference effect not only promotes bond fluctuations at the nesting vectors but also impacts the superconducting pairing. In the spinful case, either superconductivity or nCDW is leading on the $V_1$ axis stemming from enhanced onsite CDW fluctuation at $\bm{q}=0$, in contrast to the sole CBO in the spinless case~\cite{PhysRevLett.127.217601,Wang2013}. The inclusion of $V_2$ in the spinful case yields a similar effect, making LCO subleading, as confirmed by our FRG calculations. Compensating for this overscreening by introducing the onsite repulsion $U$ will suppress the CDW but a large $U$ triggers magnetic orders as previously reported for $U\gg V_1$~\cite{Wang2013,Profe2024}.

We discuss experimental implications of our model calculations and identified mechanism for charge orders in kagome materials. The intriguing CDW in kagome metals FeGe and AV$_3$Sb$_5$ can be linked to instabilities associated with VHS, which have been identified in the vicinity of the Fermi level in ARPES measurements~\cite{Hu2022,Kang2022,TengXK2022,TengXK2023}. 
In these materials, strong $d$-$p$ hybridization leads to delocalized $d$-orbital Wannier functions, resulting in comparable 1nn and 2nn Coulomb repulsion. Indeed, cRPA calculations find a ratio about $V_1/V_2=1.1$~\cite{PhysRevB.111.125163}, placing them near the boundary where LCO and CBO can coexist. Moreover, electron-phonon coupling~\cite{Tan2021,Gutierrez-Amigo2024} can generate an effective 1nn attraction which counteracts the repulsion $V_1$, potentially realizing an effective $V_2>V_1$ regime.
The multiorbital nature and multiple VHS in realistic materials may further enhance loop-current fluctuations and thus facilitate LCO formation.
In FeGe, the large ferromagnetic splitting in each layer results in a spin-polarized $p$-type VHS in proximity to the Fermi level, analogous to our spinless model. Even when employing the spinful model with large ferromagnetic splitting to simulate FeGe, our FRG calculations reveal that LCO occurs with an intermediate 2nn repulsion, showing almost quantitative consistency with the spinless model (see SM~\cite{SM}). According to our calculations, the $2\times2$ LCO in FeGe can generate an orbital magnetic moment as large as $0.03~\mu_B$ per site, which may account for the observed magnetic moment change upon the CDW transition (details in SM~\cite{SM}). 
For the multiorbital AV$_3$Sb$_5$, both $p$- and $m$-type VHS from the V-kagome net occur near the Fermi level~\cite{TengXK2022,TengXK2023}. Signatures of TRS breaking inside the CDW have been reported in different experimental measurements. 
The LCO, while subdominant in the single-orbital spinful model, may become stable in multiorbital AV$_3$Sb$_5$ through the combined effect of loop-current fluctuations amplification from distinct types of VHS~\cite{PhysRevLett.132.146501,Tazai2023} and reduced nn repulsion due to electron-phonon coupling~\cite{Tan2021,Gutierrez-Amigo2024}.
The exotic interplay between LCO and superconductivity can generate intriguing pair density wave states~\cite{Zhou2022}.

Our Letter offers critical insights into intrinsic quantum fluctuations and their interplay with lattice geometry within the kagome lattice. By identifying $2\times2$ LCO through unbiased many-body calculations, we transform it from a speculative concept into a microscopically validated quantum state, establishing a solid theoretical foundation for the experimental exploration of LCO within kagome metals. This highlights the kagome metals as an ideal platform to study exotic correlated phenomena.

{\it Acknowledgments.--} J.~Z. and J.~H. are supported by the Ministry of Science and Technology (Grant No.~2022YFA1403901), the National Natural Science Foundation of China (Grant No.~12494594) and the New Cornerstone Investigator Program.
R.~F., S.~Z., and X.~W. are supported by the National Key R\&D Program of China (Grants No.~2023YFA1407300 and No.~2022YFA1403800) and the National Natural Science Foundation of China (Grants No.~12574151, No.~12447103, No.~12447101, No.~12374153, and No.~12047503).
Z.~W. is supported by U.S. Department of Energy, Basic Energy Sciences Grant No.~DE-FG02-99ER45747 and the Cottrell SEED Award No.~27856 from Research Corporation for Science Advancement. M.~D. is grateful for support from a Ph.D. scholarship of the Studienstiftung des deutschen Volkes. The work is funded by the Deutsche Forschungsgemeinschaft (DFG, German Research Foundation) through Project No.~258499086--SFB 1170, and through the research unit QUAST, FOR 5249, Project No.~449872909, and through the W\"urzburg-Dresden Cluster of Excellence on Complexity and Topology in Quantum Matter--\textit{ct.qmat} Project No.~390858490--EXC 2147.
We acknowledge HPC resources provided by the Erlangen National High Performance Computing Center (NHR@FAU) of the Friedrich-Alexander-Universit\"at Erlangen-N\"urnberg (FAU). NHR funding is provided by federal and Bavarian state authorities. NHR@FAU hardware is partially funded by DFG Grant No.~440719683.

\input{main_refs.bbl}
\clearpage
\onecolumngrid
\setcounter{page}{1}
\providecommand{\theHpage}{\arabic{page}}
\renewcommand{\theHpage}{SM.\arabic{page}}
\pagestyle{myheadings}

\setcounter{section}{0}
\setcounter{subsection}{0}
\setcounter{subsubsection}{0}
\setcounter{secnumdepth}{3}
\setcounter{equation}{0}
\setcounter{figure}{0}
\setcounter{table}{0}
\renewcommand{\thesection}{Appendix \Alph{section}}
\renewcommand{\thesubsection}{\Alph{section}.\arabic{subsection}}
\renewcommand{\thesubsubsection}{\Alph{section}.\arabic{subsection}.\arabic{subsubsection}}
\renewcommand{\theequation}{S\arabic{equation}}
\renewcommand{\thefigure}{S\arabic{figure}}
\renewcommand{\thetable}{S\arabic{table}}
\renewcommand{\theHsection}{supp.\arabic{section}}
\renewcommand{\theHsubsection}{supp.\arabic{section}.\arabic{subsection}}
\renewcommand{\theHsubsubsection}{supp.\arabic{section}.\arabic{subsection}.\arabic{subsubsection}}
\renewcommand{\theHequation}{supp.\arabic{equation}}
\renewcommand{\theHfigure}{supp.\arabic{figure}}
\renewcommand{\theHtable}{supp.\arabic{table}}
\titleformat{\section}[hang]{\normalfont\normalsize\bfseries\centering}{\thesection.}{0.6em}{}
\titleformat{\subsection}[hang]{\normalfont\normalsize\itshape}{\thesubsection.}{0.6em}{}
\newcommand{\smtocsec}[2]{\noindent\textcolor{BrickRed}{\hyperref[#1]{#2}}\hfill\pageref{#1}\par}
\newcommand{\smtocsubsec}[2]{\noindent\hspace*{1em}\textcolor{BrickRed}{\hyperref[#1]{#2}}\hfill\pageref{#1}\par}

\begin{center}
{\large\bfseries Supplementary Material for ``Loop Current Order on the Kagome Lattice''\par}
\vspace{0.8em}
{\normalsize Jun Zhan$^{1,2,*}$, Hendrik Hohmann$^{3,*}$, Matteo D\"{u}rrnagel$^{3,4}$, Ruiqing Fu$^{5,2}$, Sen Zhou$^{5}$, Ziqiang Wang$^{6}$, Ronny Thomale$^{3}$, Xianxin Wu$^{5,\dagger}$, and Jiangping Hu$^{1,2,7,\ddagger}$\par}
\vspace{0.6em}
{\small
$^{1}$Beijing National Laboratory for Condensed Matter Physics and Institute of Physics, Chinese Academy of Sciences, Beijing 100190, China\par
$^{2}$School of Physical Sciences, University of Chinese Academy of Sciences, Beijing 100049, China\par
$^{3}$Institut f\"{u}r Theoretische Physik und Astrophysik, Universit\"{a}t W\"{u}rzburg, Am Hubland Campus S\"{u}d, W\"{u}rzburg 97074, Germany\par
$^{4}$Institute for Theoretical Physics, ETH Z\"{u}rich, 8093 Z\"{u}rich, Switzerland\par
$^{5}$Institute of Theoretical Physics, Chinese Academy of Sciences, Beijing 100190, China\par
$^{6}$Department of Physics, Boston College, Chestnut Hill, Massachusetts 02467, USA\par
$^{7}$New Cornerstone Science Laboratory, Institute of Physics, Chinese Academy of Sciences, Beijing 100190, China\par
}
\end{center}

\vspace{0.8em}
\begin{center}
{\normalsize\bfseries CONTENTS}
\end{center}
\begingroup
\hypersetup{linkcolor=BrickRed}
\small
\smtocsec{supp:sec:vhs}{Appendix A. Sublattice texture at van Hove filling}
\smtocsec{supp:sec:frg}{Appendix B. FRG formalism and numerical details}
\smtocsubsec{supp:subsec:frg-flow}{B.1. FRG flow equations}
\smtocsubsec{supp:subsec:frg-tu}{B.2. Truncated unity FRG}
\smtocsubsec{supp:subsec:frg-reg}{B.3. Regulators}
\smtocsubsec{supp:subsec:frg-init}{B.4. Initial conditions}
\smtocsubsec{supp:subsec:frg-inst}{B.5. Instability analysis}
\smtocsubsec{supp:subsec:frg-identify}{B.6. Identifying CBO and LCO parameters from vertex eigenstates}
\smtocsec{supp:sec:bondorders}{Appendix C. An intuitive picture of the competing real and imaginary bond orders on the kagome lattice}
\smtocsubsec{supp:subsec:bond-mean}{C.1. Mean field energy gap}
\smtocsubsec{supp:subsec:bond-nn}{C.2. Nearest neighbour bond orders}
\smtocsubsec{supp:subsec:bond-gl}{C.3. Ginzburg-Landau analysis of real and imaginary bond orders}
\smtocsubsec{supp:subsec:bond-cbo}{C.4. CBO in the $V_1$ dominated regime}
\smtocsec{supp:sec:lco}{Appendix D. Physical properties of the LCO state}
\smtocsubsec{supp:subsec:lco-gl}{D.1. Ginzburg-Landau analysis and real space pattern}
\smtocsubsec{supp:subsec:lco-mag}{D.2. Orbital magnetization of LCO}
\smtocsec{supp:sec:ncdw-gl}{Appendix E. Ginzburg-Landau analysis for nCDW}
\smtocsec{supp:sec:ncdw-enhance}{Appendix F. Enhancement of nCDW through SC channel fluctuations for weak $V_2$}
\smtocsec{supp:sec:spinful}{Appendix G. LCO in spinful model with Zeeman splitting}
\smtocsec{supp:sec:refs}{References}
\endgroup

\vspace{0.8em}
In this Supplementary Material, we present the technical details of the functional renormalization group (FRG) methods employed in this work, together with an in-depth analysis of the emergent orders. In particular, we provide a Ginzburg--Landau (GL) analysis to determine the real-space structures of the triple-$\mathbf{Q}$ bond order and the nematic charge density wave (CDW) phases.
In addition, we evaluate the doping dependence of the orbital magnetization in the LCO phase and make direct comparisons with the experimental systems FeGe and AV$_3$Sb$_5$. The realization of LCO in a spinful model with Zeeman splitting is discussed as well.

	\section{Sublattice texture at van Hove filling}\label{supp:anchor:vhs}\label{supp:sec:vhs}
\begin{figure}[t]
		\centering
		\includegraphics[width=0.6\columnwidth]{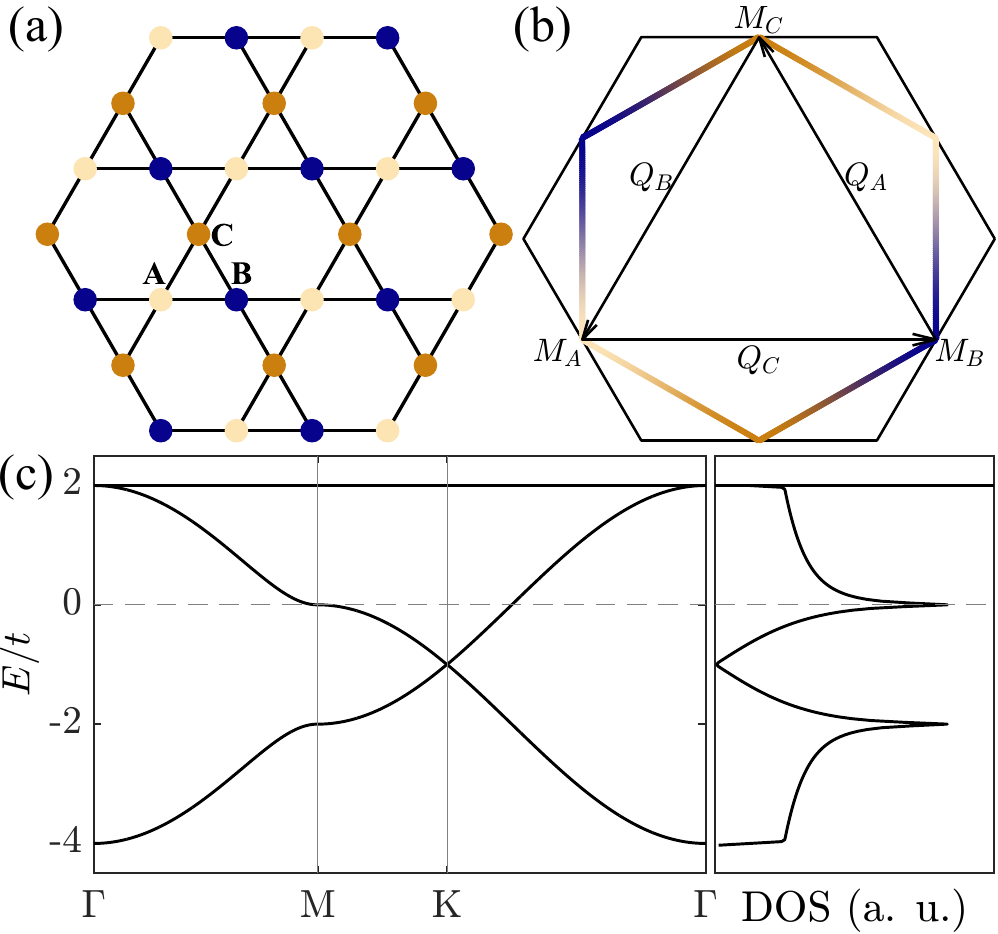}
		\caption{Kagome lattice, electronic structure and sublattice-resolved Fermi surface. (a)~Real space structure of the kagome lattice with three sublattices A, B and C indicated by different colors. (b)~Fermi surface at the upper van-Hove filling and corresponding sublattice makeup of the FS states. (c)~Energy dispersion of the tight binding model along the high symmetry path and density of state (DOS) peaking featured by van-Hove singularities.
			% \textit{TODO: Color of Kagome site: light="#FDE5B3" blue="#07038D" orange='#CB7F0F'? similar in FS plot and CBO, LCO, fSC real space plots. Also use this real space mesh for all real space plots}
		}
		\label{fig:model}
	\end{figure}

	On the kagome lattice (Fig.~\ref{fig:model}(a)), a simple tight binding Hamiltonian with nearest
	neighbour hybridisation features a bandstructure consisting of a flat band and
	two dispersive bands crossing at the Dirac point while hosting two different
	kinds of van-Hove singularities (vHs) displayed in Fig.~\ref{fig:model}(c).
	The two vHs are characterized by distinct sublattice texture on the Fermi
	surface, setting them apart from those observed in
	triangular and honeycomb lattices.
	Within this work we focus on the p-type vHs, where each inequivalent M-point
	consists of states solely attributed to one sublattice (Fig.~\ref{fig:model}(b)).
	Correspondingly, the nesting vectors $\mathbf{Q}_{A, B, C}$ connect
	saddle points with distinct sublattice characters.

	\section{FRG formalism and numerical details}\label{supp:anchor:frg}\label{supp:sec:frg}
	\subsection{FRG flow equations }\label{supp:subsec:frg-flow}
	The functional renormalization group (FRG) is an unbiased method to determine Fermi liquid instabilty of interacting fermionic systems \cite{SM_Metzner2012,SM_Platt2013,SM_Beyer2022}. 
	The basic idea of FRG is to introduce a scale dependence of the effective action or generating functional of the one particle irreducible (1PI) vertex functions, $\Gamma\to\Gamma^{\Lambda}$. Differentiating $\Gamma^{\Lambda}$ with respect to $\Lambda$ leads to exact functional flow equations from which one can interpolate between a microscopic bare action at $\Lambda\to \infty$ and a low energy effective action at small $\Lambda$. Expanding in
	powers of the fields one obtains an exact hierarchy of flow equations for 1PI vertex functions. We adopt the standard approximations which truncate the hierarchy after two-particle vertex function and discard self-energy feedback, 
	%and neglect the frequency dependence of vertex functions.
	which is appropriate in the weak to intermediate coupling regime. 
	%The two particle part of effective action can be represented by effective interaction $V^{\Lambda}$ and Grassmann filed $\psi$ and $\bar{\psi}$ as
	For multi-orbital spinless fermionic systems, the two particle part of the effective action can be expressed by effective interaction $V^{\Lambda}$ and Grassmann fields $\psi$ and $\bar{\psi}$ as
	%\begin{equation}
	%	\begin{aligned}
		%		\Gamma^{(4) \Lambda}&[\bar{\psi},\psi] =\frac{1}{2!} \int \prod_{i=1}^{4} \dd \xi_{i}
		%		V^{\Lambda}_{o_1o_2o_3o_4}(k_1,k_2;k_3,k_4)  \\
		%		&\times\delta(k_1+k_2-k_3-k_4)\bar{\psi}(\xi_1)\bar{\psi}(\xi_2)\psi(\xi_4)\psi(\xi_3)
		%	\end{aligned}
	%\end{equation}
	\begin{equation}
		\begin{aligned}
			\Gamma^{(4) \Lambda}&[\bar{\psi},\psi] =\frac{1}{(2!)^2} \int \prod_{i=1}^{4} \dd \xi_{i}
			V^{\Lambda}_{o_1o_2o_3o_4}(k_1,k_2;k_3,k_4)  
			%		\times
			\delta(k_1+k_2-k_3-k_4)\bar{\psi}(\xi_1)\bar{\psi}(\xi_2)\psi(\xi_4)\psi(\xi_3)
			\, ,
		\end{aligned}
	\end{equation}
	where $k_i=(\omega_{i},\mathbf{k}_{i})$ and $\xi_i = (\omega_i, \mathbf{k}_i, o_i )$ are multi-indices
	gathering a Matsubara frequency $\omega_i$, wave vector $\mathbf{k}_i$,
	and orbital/sublattice index $o_i$ and
	$\dd \xi_i$ stands for $\int \frac{d \mathbf{k}_i}{S_{B Z}} \frac{1}{\beta} \sum \omega_i \sum o_i$ with the Brillouin zone area $S_{BZ}$ and inverse
	temperature $\beta$.
	%Then one can obtain flow equations for effective two particle interaction 
	With this, one can obtain flow equations for the effective two particle interactions
	%\begin{equation}\label{spinless}
	%	\begin{aligned}
		%		\frac{\dd}{\dd \Lambda}&V^{\Lambda}_{\left\lbrace  o_i \right\rbrace  }(k_1,k_2;k_3,k_4) = \mathcal{T}^{\mathrm{pp}}_{\left\lbrace  o_i \right\rbrace  }\left( k_1,k_2;k_3,k_4\right)  \\
		%		&+\mathcal{T}^{\mathrm{cph}}_{\left\lbrace  o_i \right\rbrace  }\left( k_1,k_2;k_3,k_4\right)  +\mathcal{T}^{\mathrm{dph}}_{\left\lbrace  o_i \right\rbrace  } \left(k_1,k_2;k_3,k_4\right)
		%	\end{aligned}
	%\end{equation}
	\begin{equation}\label{spinless}
		\begin{aligned}
			\frac{\dd}{\dd \Lambda}&V^{\Lambda}_{\left\lbrace  o_i \right\rbrace  }(k_1,k_2;k_3,k_4) = \mathcal{T}^{\mathrm{pp}}_{\left\lbrace  o_i \right\rbrace  }\left( k_1,k_2;k_3,k_4\right)  
			&+\mathcal{T}^{\mathrm{cph}}_{\left\lbrace  o_i \right\rbrace  }\left( k_1,k_2;k_3,k_4\right)  +\mathcal{T}^{\mathrm{dph}}_{\left\lbrace  o_i \right\rbrace  } \left(k_1,k_2;k_3,k_4\right) \, .
		\end{aligned}
	\end{equation}
	The three terms on the right hand side of the flow equation are
		\begin{equation}
			\begin{aligned}
				\mathcal{T}_{\left\lbrace o_i \right\rbrace}^{\mathrm{pp}}\left(k_{1}, k_{2}; k_{3},k_{4}\right)&=-\frac{1}{2}\int_{q} V^{\Lambda}_{o_1o_2oo'}\left(k_1,k_2;k_1+k_2+q,-q\right) \frac{\mathrm{d}}{\mathrm{d}\Lambda}\left[ G^{\Lambda}_{o\tilde{o}}(k_1+k_2-q) G^{\Lambda}_{o'\tilde{o}'}(-q) \right] V^{\Lambda}_{\tilde{o}\tilde{o}'o_3o_4}\left( k_1+k_2+q,-q ; k_3,k_4 \right) ,\\
				\mathcal{T}_{\left\lbrace o_i \right\rbrace}^{\mathrm{cph}}\left(k_{1}, k_{2}; k_{3},k_{4}\right)&=-\int_{q}V^{\Lambda}_{o_1o'oo_4}\left(k_1,q;k_1-k_4+q,k_4\right) \frac{\mathrm{d}}{\mathrm{d}\Lambda} \left[ G^{\Lambda}_{o\tilde{o}}(k_1-k_4+q) G^{\Lambda}_{\tilde{o}'o'}(q) \right] V^{\Lambda}_{\tilde{o}o_2o_3\tilde{o}'}\left( k_1-k_4+q,k_2; k_3,q \right) ,\\
				\mathcal{T}_{\left\lbrace o_i \right\rbrace}^{\mathrm{dph}}\left(k_{1}, k_{2}; k_{3},k_{4}\right)&=+\int_{q}V^{\Lambda}_{o_1o'o_3o}\left(k_1,q;k_3,k_1-k_3+q\right) \frac{\mathrm{d}}{\mathrm{d}\Lambda} \left[ G^{\Lambda}_{o\tilde{o}}(k_1-k_3+q) G^{\Lambda}_{\tilde{o}'o'}(q) \right]  V^{\Lambda}_{\tilde{o}o_2\tilde{o}'o_4}\left( k_1-k_3+q,k_2; q,k_4 \right) \ , \\
			\end{aligned}
		\end{equation}
	%\begin{widetext}
	%	\begin{figure}[tpb]
		%		\includegraphics[width=1\textwidth]{Fig1.pdf}
		%		\caption{(a) Structure of the kagome lattice. (b) Fermi surface of the positive-helicity band hosting higher-order VHSs and its spin texture. (c) 2D dispersion around the higher-order saddle point with $E(\mathbf{k})=A(-k_x^4+k_y^2)$}.
		%		\label{fig:Fig1}
		%	\end{figure}
	%\end{widetext}
	which represent contributions to the interactions in particle-particle, cross particle-hole, and  direct particle-hole channels, respectively, with the diagrammatic representations shown in Fig. \ref{fig:fRG}. 
	Here  $G^{\Lambda}_{oo'}$ is the bare propagator in orbital basis at scale $\Lambda$ and $\int_q=\sum_{oo' \tilde{o}\tilde{o}'}\frac{1}{\beta}\sum_{\omega}\int \frac{\dd \mathbf{q}}{S_{BZ}}$. We will further ignore the frequency dependence of effective interaction since interactions at finite frequencies are irrelevant under the renormalization group transformations and we focus on static ground-state properties. 
	
	\begin{figure}[tpb]
		\centering
		\subfigure[]{
			\begin{tikzpicture}[> = Stealth,scale=2]
				\color{black}
				\begin{feynhand}
					%				\node at (-0.8,0) {$-$};
					\vertex [dot] (i) at (0,0) {};
					\vertex (k1) at (0,1) {$\xi_2$};
					\vertex (k2) at (0,-1) {$\xi_1$};
					\vertex (i1) at (0.5,0.5) {};
					\vertex (i2) at (0.5,-0.5) {};
					\propag [fer] (k1) to  (i);
					\propag [fer] (k2) to  (i);
					
					\vertex [dot] (ip) at (1.,0) {};
					\vertex (k3) at (1,1) {$\xi_4$};
					\vertex (k4) at (1,-1) {$\xi_3$};
					\propag [antfer] (k3) to  (ip);
					\propag [antfer] (k4) to  (ip);
					%			\vertex [dot] (b) at (0,0.5) {};
					
					%				\propag[fer] (i) to [edge label = $-q$]  [out=45, in=135,looseness =1] (ip);
					\propag[fer] (i) to  [out=45, in=135,looseness =1] (ip);
					\propag[fer] (i) to [out=315, in=225,looseness =1] (ip);
					%				\node at (0.5,-0.37) {$k_1+k_2+q$};
					
					%				\vertex [] (s1) at (0.58,0.105) {};
					%				\vertex [] (s2) [above = 0.2  of s1] {};
					%				\draw (s1) to [] (s2);
					
				\end{feynhand}
			\end{tikzpicture}
		}
		\subfigure[]{
			\begin{tikzpicture}[> = Stealth,scale=2]
				\color{black}
				\begin{feynhand}
					%				\node at (-0.8,0.5) {$-$};
					
					\vertex [dot] (i) at (0,0) {};
					\vertex (k1) at (-0.5,-0.5) {$\xi_1$};
					\vertex (k2) at (0.5,-0.5) {$\xi_4$};
					\propag [fer] (k1) to  (i);
					\propag [antfer] (k2) to  (i);
					
					\vertex [dot] (ip) at (0,1) {};
					\vertex (k4) at (-0.5,1.5) {$\xi_2$};
					\vertex (k2) at (0.5,1.5) {$\xi_3$};
					\propag [fer] (k4) to  (ip);
					\propag [antfer] (k2) to  (ip);
					
					%				\propag[antfer] (i) to [edge label = $q$]  [out=135, in=225,looseness =1] (ip);
					\propag[antfer] (i) to   [out=135, in=225,looseness =1] (ip);
					\propag[fer] (i) to  [out=45, in=315,looseness =1] (ip);
					%				\node at (0.7,0.5) {$k_1-k_2'+q$};
					
					%				\vertex [] (s1) at (-0.11,0.43) {};
					%				\vertex [] (s2) [left = 0.2  of s1] {};
					%				\draw (s1) to [] (s2);
					
				\end{feynhand}
			\end{tikzpicture}
		}
		\subfigure[]{
			\begin{tikzpicture}[> = Stealth,scale=2]
				\color{black}
				\begin{feynhand}
					%				\node at (-0.8,0.5) {$-\zeta$};
					
					\vertex [dot] (i) at (0,0) {};
					\vertex (k1) at (-0.5,-0.5) {$\xi_1$};
					\vertex (k2) at (0.5,-0.5) {$\xi_3$};
					\propag [fer] (k1) to  (i);
					\propag [antfer] (k2) to  (i);
					
					\vertex [dot] (ip) at (0,1) {};
					\vertex (k4) at (-0.5,1.5) {$\xi_2$};
					\vertex (k2) at (0.5,1.5) {$\xi_4$};
					\propag [fer] (k4) to  (ip);
					\propag [antfer] (k2) to  (ip);
					
					%				\propag[antfer] (i) to [edge label = $q$]  [out=135, in=225,looseness =1] (ip);
					\propag[antfer] (i) to  [out=135, in=225,looseness =1] (ip);
					\propag[fer] (i) to  [out=45, in=315,looseness =1] (ip);
					%				\node at (0.7,0.5) {$k_1-k_1'+q$};
					
					%				\vertex [] (s1) at (-0.11,0.43) {};
					%				\vertex [] (s2) [left = 0.2  of s1] {};
					%				\draw (s1) to [] (s2);
					
				\end{feynhand}
			\end{tikzpicture}
		}
		\caption{One-loop contributions to the renormalized interaction: (a)~The particle-particle term. (b)~The crossed particle-hole term. (c)~The direct particle-hole term.}
		\label{fig:fRG}
	\end{figure} 
	
%	\subsection{TUfRG}
%	In the truncated unity functional renormalizational group (TUfRG) formalism, the static four point vertex function is decomposed into a scale independent initial bare interaction $V^{(0)}$ plus three coupling functions $\Phi^{X}, X\in \left\lbrace P,C,D \right\rbrace $,
	\subsection{Truncated unity FRG}\label{supp:subsec:frg-tu}
	In the singular mode or truncated unity functional renormalizational group (TUFRG) formalism~\cite{SM_QHWangGraphene,SM_Lichtenstein2017}, the static four point vertex function is decomposed into a scale independent initial bare interaction $V^{(0)}$ plus three coupling functions $\Phi^{X}, X\in \left\lbrace P,C,D \right\rbrace $,
	%\begin{equation}
	%	\begin{aligned}
		%		V^{\Lambda}_{\left\lbrace  o_i \right\rbrace  }(\mathbf{k}_1,\mathbf{k}_2;\mathbf{k}_3,\mathbf{k}_4) &= V^{(0)}_{\left\lbrace  o_i \right\rbrace  }(\mathbf{k}_1,\mathbf{k}_2;\mathbf{k}_3,\mathbf{k}_4) \\
		%		&+\Phi^{\mathrm{P}}_{\left\lbrace  o_i \right\rbrace  }\left( \mathbf{k}_1 +\mathbf{k}_2, -\mathbf{k}_2 ,-\mathbf{k}_4 \right) \\
		%		&+\Phi^{\mathrm{C}}_{\left\lbrace  o_i \right\rbrace  }\left( \mathbf{k}_1 -\mathbf{k}_4, \mathbf{k}_4 ,-\mathbf{k}_2 \right)  \\
		%		&+\Phi^{\mathrm{D}}_{\left\lbrace  o_i \right\rbrace  }\left( \mathbf{k}_1 -\mathbf{k}_3,  \mathbf{k}_3 ,-\mathbf{k}_2 \right)
		%	\end{aligned}
	%\end{equation}
	\begin{equation} \label{TUdecomposition}
		\begin{aligned}
			V^{\Lambda}_{\left\lbrace  o_i \right\rbrace  }(\mathbf{k}_1,\mathbf{k}_2;\mathbf{k}_3,\mathbf{k}_4) &= V^{(0)}_{\left\lbrace  o_i \right\rbrace  }(\mathbf{k}_1,\mathbf{k}_2;\mathbf{k}_3,\mathbf{k}_4) 
			+\Phi^{\mathrm{P}}_{\left\lbrace  o_i \right\rbrace  }\left( \mathbf{k}_1 +\mathbf{k}_2, -\mathbf{k}_2 ,-\mathbf{k}_4 \right) \\
			&+\Phi^{\mathrm{C}}_{\left\lbrace  o_i \right\rbrace  }\left( \mathbf{k}_1 -\mathbf{k}_4, \mathbf{k}_4 ,-\mathbf{k}_2 \right)  
			+\Phi^{\mathrm{D}}_{\left\lbrace  o_i \right\rbrace  }\left( \mathbf{k}_1 -\mathbf{k}_3,  \mathbf{k}_3 ,-\mathbf{k}_2 \right) \, ,
		\end{aligned}
	\end{equation}
	%\begin{equation}
	%	\begin{aligned}
		%		V^{\Lambda}_{o_1o_2o_3o_4 }(\mathbf{k}_1,\mathbf{k}_2;\mathbf{k}_3,\mathbf{k}_4) &= V^{(0)}_{o_1o_2o_3o_4  }(\mathbf{k}_1,\mathbf{k}_2;\mathbf{k}_3,\mathbf{k}_4) 
		%		+\Phi^{\mathrm{P}}_{o_1o_2o_3o_4  }\left( \mathbf{k}_1 +\mathbf{k}_2, -\mathbf{k}_2 ,-\mathbf{k}_4 \right) \\
		%		&+\Phi^{\mathrm{C}}_{o_1o_4o_3o_2  }\left( \mathbf{k}_1 -\mathbf{k}_4, \mathbf{k}_4 ,-\mathbf{k}_2 \right)  
		%		+\Phi^{\mathrm{D}}_{o_1o_3o_4o_2  }\left( \mathbf{k}_1 -\mathbf{k}_3,  \mathbf{k}_3 ,-\mathbf{k}_2 \right)
		%	\end{aligned}
	%\end{equation}
	which are generated during the flows according to
	\begin{equation}
		\begin{aligned}
			&\frac{\dd}{\dd \Lambda}\Phi^{\mathrm{P}}_{\left\lbrace  o_i \right\rbrace  }\left( \mathbf{k}_1 +\mathbf{k}_2, -\mathbf{k}_2 ,-\mathbf{k}_4 \right) = \mathcal{T}_{\left\lbrace o_i \right\rbrace}^{\mathrm{pp}}\left(\mathbf{k}_1,\mathbf{k}_2;\mathbf{k}_3,\mathbf{k}_4\right) \\
			&\frac{\dd}{\dd \Lambda}\Phi^{\mathrm{C}}_{\left\lbrace  o_i \right\rbrace  }\left( \mathbf{k}_1 -\mathbf{k}_4, \mathbf{k}_4 ,-\mathbf{k}_2 \right) =\mathcal{T}_{\left\lbrace o_i \right\rbrace}^{\mathrm{cph}}\left(\mathbf{k}_1,\mathbf{k}_2;\mathbf{k}_3,\mathbf{k}_4\right) \\
			&\frac{\dd}{\dd \Lambda}\Phi^{\mathrm{D}}_{\left\lbrace  o_i \right\rbrace  }\left( \mathbf{k}_1 -\mathbf{k}_3,  \mathbf{k}_3 ,-\mathbf{k}_2 \right) =\mathcal{T}_{\left\lbrace o_i \right\rbrace}^{\mathrm{dph}}\left(\mathbf{k}_1,\mathbf{k}_2;\mathbf{k}_3,\mathbf{k}_4\right) \, .
		\end{aligned}
	\end{equation}
	%\begin{equation}
	%	\begin{aligned}
		%		&\frac{\dd}{\dd \Lambda}\Phi^{\mathrm{P}}_{o_1o_2,o_3o_4 }\left( \mathbf{k}_1 +\mathbf{k}_2, -\mathbf{k}_2 ,-\mathbf{k}_4 \right) = \mathcal{T}_{o_1o_2o_3o_4}^{\mathrm{pp}}\left(\mathbf{k}_1,\mathbf{k}_2;\mathbf{k}_3,\mathbf{k}_4\right) \\
		%		&\frac{\dd}{\dd \Lambda}\Phi^{\mathrm{C}}_{o_1o_4o_3o_2  }\left( \mathbf{k}_1 -\mathbf{k}_4, \mathbf{k}_4 ,-\mathbf{k}_2 \right) =\mathcal{T}_{o_1o_2o_3o_4}^{\mathrm{cph}}\left(\mathbf{k}_1,\mathbf{k}_2;\mathbf{k}_3,\mathbf{k}_4\right) \\
		%		&\frac{\dd}{\dd \Lambda}\Phi^{\mathrm{D}}_{o_1o_3o_4o_2  }\left( \mathbf{k}_1 -\mathbf{k}_3,  \mathbf{k}_3 ,-\mathbf{k}_2 \right) =\mathcal{T}_{o_1o_2o_3o_4}^{\mathrm{dph}}\left(\mathbf{k}_1,\mathbf{k}_2;\mathbf{k}_3,\mathbf{k}_4\right)
		%	\end{aligned}
	%\end{equation}

	The first argument of $\Phi^{X}$ is transfer momentum and the last two are non-transfer momenta. We then expand the less important non-transfer momenta into a formfactor basis $\left\lbrace f_m(\mathbf{k}) \right\rbrace  $,
	\begin{equation} \label{FFexpansion}
		\Phi^{\mathrm{X}}_{\left\lbrace  o_i \right\rbrace  }\left( \mathbf{q}, \mathbf{k} ,\mathbf{k}' \right) = 
		\sum_{m,n}f_m^{*}(\mathbf{k}) X_{o_1o_2o_3o_4,mn}(\mathbf{q})f_n(\mathbf{k}')
	\end{equation}
	%\begin{equation}
	%	\Phi^{\mathrm{X}}_{o_1o_2o_3o_4  }\left( \mathbf{q}, \mathbf{k} ,\mathbf{k}' \right) = 
	%	\sum_{m,n}f_m^{*}(\mathbf{k}) X_{mo_1o_2,no_3o_4}(\mathbf{q})f_n(\mathbf{k}')
	%\end{equation}
	%\begin{equation}
	%	\Phi^{\mathrm{P}}_{\left\lbrace  o_i \right\rbrace  }\left( \mathbf{q}, \mathbf{k} ,\mathbf{k}' \right) = 
	%	\sum_{m,n}f_m^{*}(\mathbf{k}) P_{mo_1o_2,no_3o_4}(\mathbf{q})f_n(\mathbf{k}')
	%\end{equation}
	which is exact as long as the formfactors are orthogonal and form a representation of unity,
	\begin{equation}
		\begin{aligned}
			&S_{\text{BZ}}^{-1}\sum_{m}f_m(\mathbf{k}) f_m^*(\mathbf{k'})=\delta(\mathbf{k}-\mathbf{k}')\\
			&S_{\text{BZ}}^{-1}\int \dd \mathbf{k} f_m(\mathbf{k}) f_n^*(\mathbf{k})=\delta_{mn} \, .
		\end{aligned}
	\end{equation}
	In practice the infinite basis of form factors is truncated. Then the flow equations for vertex functions turn into flow equations for exchange propagators $P,C $ and $D$,

	%\subsection{Truncated-Unity FRG}
	%The TUfRG flow equations reads
	
	%~\\\\
	%\vspace{20pt}
		\begin{equation} \label{TUfloweq1}
			\begin{aligned}
				\frac{d}{d \Lambda} P_{o_1o_2o_3o_4,mn}(\mathbf{q})&= \frac{1}{2} \sum_{oo'\tilde{o}\tilde{o}'m'n'} V^{\text{P}}_{o_1o_2oo',m m'}(\mathbf{q}) \dot{\chi}_{oo'\tilde{o}\tilde{o}',m'n'}^{\text{pp}}(\mathbf{q}) V^{\text{P}}_{\tilde{o}\tilde{o}'o_3o_4,n'n}(\mathbf{q}) ,\\
				\frac{d}{d \Lambda} C^\ann_{o_1o_2o_3o_4,mn}(\mathbf{q})&=   \sum_{oo'\tilde{o}\tilde{o}'m'n'} V^{\text{C}}_{o_1o'oo_4,mm'}(\mathbf{q})  \dot{\chi}_{oo'\tilde{o}\tilde{o}',m'n'}^{\text{ph}}(\mathbf{q}) V^{\text{C}}_{\tilde{o}o_2o_3\tilde{o}',n'n}(\mathbf{q}) \\
				\frac{d}{d \Lambda} D_{o_1o_2o_3o_4,mn}(\mathbf{q})&= - \sum_{oo'\tilde{o}\tilde{o}'m'n'}
				V^{\text{D}}(\mathbf{q})_{o_1o'o_3o,mm'} \dot{\chi}(\mathbf{q})_{oo'\tilde{o}\tilde{o}',m'n'}^{\text{ph}} V^{\text{D}}(\mathbf{q})_{\tilde{o}o_2\tilde{o}'o_4,n'n} \, ,
			\end{aligned}
		\end{equation}
		where $\dot{\chi}^{\text{pp/ph}}$ are particle-particle and particle-hole bubble integrals in form-factor basis
		\begin{equation}
			\begin{aligned}
				& \dot{\chi}_{oo'\tilde{o}\tilde{o}',mn}^{\text{pp}}(\mathbf{q})=-\int_p  f_m(\mathbf{p}) \frac{d}{d \Lambda}\left[G^{\Lambda}_{o\tilde{o}}(i \omega, \mathbf{q}+\mathbf{p})\right. \left. G^{\Lambda}_{o'\tilde{o}'}( - i \omega, -\mathbf{p})\right]  f_{n}^*(\mathbf{p})\\
				& \dot{\chi}_{oo'\tilde{o}\tilde{o}',mn}^{\text{ph}}(\mathbf{q})=-\int_p  f_m(\mathbf{p}) \frac{d}{d \Lambda}\left[G^{\Lambda}_{o\tilde{o}}(i \omega, \mathbf{q}+\mathbf{p})\right. \left. G^{\Lambda}_{\tilde{o}'o'}(  i \omega,  \mathbf{p})\right]  f_{n}^*(\mathbf{p}) 
			\end{aligned}
		\end{equation}
		% and $V^{P,C,D} $ are projections of the effective interactions onto the form of the three channels
        and $V^{P,C,D} $ are projections of the effective interactions in three channels
		\begin{equation}\label{InitialV}
			\begin{aligned}
				& V^{\text{P}}_{o_1o_2o_3o_4,mn}(\mathbf{q})=S_{\text{BZ}}^{-2} \int \dd \mathbf{k} \dd \mathbf{k}' f_m(\mathbf{k})V^{\Lambda}_{o_1o_2o_3o_4  }(\mathbf{k}+\mathbf{q},-\mathbf{k};\mathbf{k}'+\mathbf{q},-\mathbf{k}') f_n^*(\mathbf{k}') \\
				& V^{\text{C}}_{o_1o_2o_3o_4,mn}(\mathbf{q}) =S_{\text{BZ}}^{-2} \int \dd \mathbf{k} \dd \mathbf{k}' f_m(\mathbf{k})V^{\Lambda}_{o_1o_2o_3o_4  }(\mathbf{k}+\mathbf{q},\mathbf{k}';\mathbf{k}'+\mathbf{q},\mathbf{k}) f_n^*(\mathbf{k}') \\
				&V^{\text{D}}_{o_1o_2o_3o_4,mn}(\mathbf{q})= S_{\text{BZ}}^{-2} \int \dd \mathbf{k} \dd \mathbf{k}' f_m(\mathbf{k})V^{\Lambda}_{o_1o_2o_3o_4  }(\mathbf{k}+\mathbf{q},\mathbf{k}';\mathbf{k},\mathbf{k}'+\mathbf{q})f_n^*(\mathbf{k}') \, .
			\end{aligned}
		\end{equation}
	Using Eqs. \eqref{TUdecomposition} and \eqref{FFexpansion}, the projections can be expressed by $P, C $ and $D$ such that Eq.~\eqref{TUfloweq1} becomes a closed set of differential equations. In plane wave formfactor basis, $f_m(\mathbf{k})=e^{i \mathbf{k} \cdot \mathbf{R}_m }$ where $\mathbf{R}_m$ is Bravais lattice vector and the projections can be simplified to 
	\begin{equation}
		\begin{aligned}
			V^{\mathrm{P}}(\mathbf{q})&=V^{\mathrm{P},0}(\mathbf{q})+P(\mathbf{q})+V^{\mathrm{P} \leftarrow \mathrm{C}}(\mathbf{q})+V^{\mathrm{P} \leftarrow \mathrm{D}}(\mathbf{q}) \\
			V_{o_1o_2o_3o_4,mn}^{\mathrm{P} \leftarrow \mathrm{C}}(\mathbf{q})&=\sum_l \tilde{C}_{o_1o_2o_3o_4, \mathbf{R}_l  ,\mathbf{R}_m+\mathbf{R}_n-\mathbf{R}_l}\left(\mathbf{R}_n-\mathbf{R}_l\right) e^{i\left(\mathbf{R}_n-\mathbf{R}_l\right) \cdot \mathbf{q}} \\
			V_{o_1o_2o_3o_4,mn}^{\mathrm{P} \leftarrow \mathrm{D}}(\mathbf{q})&=\sum_l \tilde{D}_{o_1o_2o_3o_4, \mathbf{R}_l, \mathbf{R}_m-\mathbf{R}_n-\mathbf{R}_l}\left(-\mathbf{R}_n-\mathbf{R}_l\right) e^{-i \mathbf{R}_l \cdot \mathbf{q}} \\
			V^{\mathrm{C}}(\mathbf{q})&=V^{\mathrm{C},0}(\mathbf{q})+C(\mathbf{q})+V^{\mathrm{C} \leftarrow \mathrm{P}}(\mathbf{q})+V^{\mathrm{C} \leftarrow \mathrm{D}}(\mathbf{q}) \\
			V_{o_1o_2o_3o_4,mn}^{\mathrm{C} \leftarrow \mathrm{P}}(\mathbf{q})&=\sum_l \tilde{P}_{o_1o_2o_3o_4,\mathbf{R}_l, \mathbf{R}_m+\mathbf{R}_n-\mathbf{R}_l}\left(\mathbf{R}_n-\mathbf{R}_l\right) e^{i\left(\mathbf{R}_n-\mathbf{R}_l\right) \cdot \mathbf{q}} \\
			V_{o_1o_2o_3o_4,mn}^{\mathrm{C} \leftarrow \mathrm{D}}(\mathbf{q})&=\sum_l \tilde{D}_{o_1o_2o_3o_4, \mathbf{R}_l , \mathbf{R}_n+\mathbf{R}_l-\mathbf{R}_m}\left(-\mathbf{R}_m\right) e^{-i \mathbf{R}_l \cdot \mathbf{q}} \\
			V^{\mathrm{D}}(\mathbf{q})&=V^{\mathrm{D},0}(\mathbf{q})+D(\mathbf{q})+V^{\mathrm{P} \leftarrow \mathrm{D}}(\mathbf{q})+V^{\mathrm{D} \leftarrow \mathrm{C}}(\mathbf{q}) \\
			V_{o_1o_2o_3o_4,mn}^{\mathrm{D} \leftarrow \mathrm{P}}(\mathbf{q})&=\sum_l \tilde{P}_{o_1o_2o_3o_4, \mathbf{R}_l , \mathbf{R}_l-\mathbf{R}_m-\mathbf{R}_n}\left(-\mathbf{R}_m\right) e^{i\left(\mathbf{R}_n-\mathbf{R}_l\right) \cdot \mathbf{q}} \\
			V_{o_1o_2o_3o_4,mn}^{\mathrm{D} \leftarrow \mathrm{C}}(\mathbf{q})&=\sum_l \tilde{C}_{o_1o_2o_3o_4, \mathbf{R}_l , \mathbf{R}_n+\mathbf{R}_l-\mathbf{R}_m}\left(-\mathbf{R}_m\right) e^{-i \mathbf{R}_l \cdot \mathbf{q}}, \\
		\end{aligned}
	\end{equation}
	where $\tilde{X}_{o_1o_2o_3o_4,mn}(\mathbf{R}_m)$ are are the Fourier transformed channels
	\begin{equation}
		\tilde{X}_{o_1o_2o_3o_4,mn}(\mathbf{R}_m)=S_{\mathrm{BZ}}^{-1} \int d \mathbf{q} \tilde{X}_{o_1o_2o_3o_4,mn}(\mathbf{q}) e^{-i \mathbf{q} \mathbf{R}_m}
	\end{equation}
	and $V^{\mathrm{X},0}$ are projections of $V^{0}$ into the $X$ channel.
	By regrouping the orbital and formfactor indices of $X, V^{X} $ and $\chi$,  
	\begin{equation}
		\begin{aligned}
			&P_{o_1o_2o_3o_4,mn}=P_{o_1o_2m,o_3o_4n},V^P_{o_1o_2o_3o_4,mn}=V^P_{o_1o_2m,o_3o_4n}\\
			&C^\ann_{o_1o_2o_3o_4,mn}=C^\ann_{o_1o_4m,o_3o_2n},V^C_{o_1o_2o_3o_4,mn}=V^C_{o_1o_4m,o_3o_2n}\\
			&D_{o_1o_2o_3o_4,mn}=D_{o_1o_3m,o_4o_2n},V^D_{o_1o_2o_3o_4,mn}=V^D_{o_1o_3m,o_4o_2n}\\
			&\dot{\chi}_{oo'\tilde{o}\tilde{o}',mn}^{\text{pp}}=\dot{\chi}_{o\tilde{o}m,o'\tilde{o}'n}^{\mathrm{pp}}, \dot{\chi}_{oo'\tilde{o}\tilde{o}',mn}^{\text{pp}}=\dot{\chi}_{o\tilde{o}m,o'\tilde{o}'n}^{\mathrm{ph}},
		\end{aligned}
	\end{equation}
	the TUFRG flow equation can be written in a concise matrix form
	%\begin{equation}
	%	\begin{aligned}
		%		& \frac{d P(\mathbf{q})}{d \Lambda}=\frac{1}{2}V^{\mathrm{P}}(\mathbf{q}) \dot{\chi}^{\mathrm{pp}}(\mathbf{q}) V^{\mathrm{P}}(\mathbf{q}), \\
		%		& \frac{d C(\mathbf{q})}{d \Lambda}=V^{\mathrm{C}}(\mathbf{q}) \dot{\chi}^{\mathrm{ph}}(\mathbf{q}) V^{\mathrm{C}}(\mathbf{q}), \\
		%		& \frac{d D(\mathbf{q})}{d \Lambda}=V^{\mathrm{D}}(\mathbf{q}) \dot{\chi}^{\mathrm{ph}}(\mathbf{q}) V^{\mathrm{D}}(\mathbf{q}),
		%	\end{aligned}
	%\end{equation}
	\begin{equation}
		\begin{aligned}
			&  \dot{ P}(\mathbf{q})=\frac{1}{2}V^{\mathrm{P}}(\mathbf{q}) \dot{\chi}^{\mathrm{pp}}(\mathbf{q}) V^{\mathrm{P}}(\mathbf{q}) \\
			&\dot{ C}(\mathbf{q})=V^{\mathrm{C}}(\mathbf{q}) \dot{\chi}^{\mathrm{ph}}(\mathbf{q}) V^{\mathrm{C}}(\mathbf{q}) \\
			& \dot{ D}(\mathbf{q})=-V^{\mathrm{D}}(\mathbf{q}) \dot{\chi}^{\mathrm{ph}}(\mathbf{q}) V^{\mathrm{D}}(\mathbf{q}).
		\end{aligned}
	\end{equation}
	
	\subsection{Regulators}\label{supp:subsec:frg-reg}
	The scale dependence of the bare propagator $G^{}_{oo'}$ is facilitated by a cutoff function $\Theta^{\Lambda}$ (regulator) according to $G^{\Lambda}_{oo'} =  \Theta^{\Lambda}\, G^{}_{oo'}\,$ \cite{SM_PhysRevB.79.195125, SM_Beyer2022}. The choice of regulators is limited by the behavior in its limits of $\Lambda$ i.e. $\Phi^{\Lambda \rightarrow \infty} = 0$ and $\Phi^{\Lambda \rightarrow 0} = 1$.
	The approximations applied to numerically evaluate the flow equations introduce a regulator dependence, even though the general scheme of FRG is independent on the choice of regulators. Within this work, we observed qualitative agreement of $\Omega$-cutoff, sharp-frequency-cutoff \cite{SM_PhysRevB.79.195125, SM_Beyer2022} and temperature cutoff \cite{SM_Honerkamp2001}. The results displayed in the main part and supplemental material are obtained by FRG calculations utilizing the $\Omega$-cutoff.
	An exception from this is Fig.~\ref{fig:FigCBO2}, here the temperature cutoff \cite{SM_Honerkamp2001} is applied.
	In the temperature flow scheme, the fermionic fields are rescaled
	such that only the quadratic part of the action is temperature dependent. This leads to a temperature-scaled free propagator and the temperature is treated as 
	flow parameters, i.e.
	\begin{equation}
		G_0(i \omega, \mathbf{k}) \rightarrow G_0^T(i \omega, \mathbf{k})=\frac{T^{1 / 2}}{i \omega-\xi(\mathbf{k})}
	\end{equation}
	and the particle-particle(-) and particle-hole(+) bubbles are
	\begin{equation}
		\begin{aligned}
			\dot{\chi}_{\pm}^{\Lambda}(\vb{k_1},\vb{k_2})&=\frac{\dd}{\dd{T}}\sum_{i\omega_n}G^{T}(\vb{k_1},i\omega_n)G^{T}(\vb{k_2},\pm i\omega_n)
			=\frac{\dd}{\dd{T}}\sum_{i\omega_n} T \frac{1}{i\omega_n-\xi_{\vb{k}_1}}\frac{1}{\pm i\omega_n-\xi_{\vb{k}_2}} \\
			&=\frac{\dd}{\dd{T}}\frac{n_F(\pm\xi_{\vb{k}_1})-n_F(\xi_{\vb{k}_2})}{\xi_{\vb{k}_1}\mp\xi_{\vb{k}_2}} 
			=\frac{n_F^{\prime}(\pm\xi_{\vb{k}_1})-n_F^{\prime}(\xi_{\vb{k}_2})}{\xi_{\vb{k}_1}\mp\xi_{\vb{k}_2}} \, ,
		\end{aligned}
	\end{equation}
	where $n_F^{'}$ stands for temperature derivative Fermi function.
	
	\subsection{Initial conditions}\label{supp:subsec:frg-init}
	Due to the antisymmetry of Grassman fields, the spinless two particle vertex is antisymmetric with respect to exchange of two incoming or outgoing particles
	\begin{equation}
		\begin{aligned}
			V_{o_1o_2o_3o_4}(\mathbf{k}_1,\mathbf{k}_2;\mathbf{k}_3,\mathbf{k}_4)
			=-V_{o_2o_1o_3o_4}(\mathbf{k}_2,\mathbf{k}_1;\mathbf{k}_3,\mathbf{k}_4),
			V_{o_1o_2o_3o_4}(\mathbf{k}_1,\mathbf{k}_2;\mathbf{k}_3,\mathbf{k}_4)
			=-V_{o_1o_2o_4o_3}(\mathbf{k}_1,\mathbf{k}_2;\mathbf{k}_4,\mathbf{k}_3) \, .
		\end{aligned}
	\end{equation}
	To obtain initial conditions for the FRG flow, the bare interaction terms need to be antisymmetrized properly,
	\begin{equation}
		\begin{aligned}
			&V_n\sum_{\mathbf{R}_i,\mathbf{R}_n} n_o(\mathbf{R}_i) n_{o'}(\mathbf{R}_i+\mathbf{R}_n)
			=V_n\sum_{\mathbf{R}_i,\mathbf{R}_n} c^\cre_{o}(\mathbf{R}_i)c^\cre_{o'}(\mathbf{R}_i+\mathbf{R}_n) c^\ann_{o'}(\mathbf{R}_i+\mathbf{R}_n) c^\ann_{o}(\mathbf{R}_i) \\
			&=\frac{1}{4}V_n\sum_{\mathbf{k}_1,\mathbf{k}_2,\mathbf{k}_3} 
			\left(e^{i(\mathbf{k}_1-\mathbf{k}_3)\cdot \mathbf{R}_n}c^\cre_{o}(\mathbf{k}_1)c^\cre_{o'}(\mathbf{k}_2) c^\ann_{o'}(\mathbf{k}_4) c^\ann_{o}(\mathbf{k}_3)
			-e^{i(\mathbf{k}_2-\mathbf{k}_3)\cdot \mathbf{R}_n}c^\cre_{o'}(\mathbf{k}_1)c^\cre_{o}(\mathbf{k}_2) c^\ann_{o'}(\mathbf{k}_4) c^\ann_{o}(\mathbf{k}_3) \right. \\
			& \left. \quad -e^{i(\mathbf{k}_1-\mathbf{k}_4)\cdot \mathbf{R}_n}c^\cre_{o}(\mathbf{k}_1)c^\cre_{o'}(\mathbf{k}_2) c^\ann_{o}(\mathbf{k}_4) c^\ann_{o'}(\mathbf{k}_3)
			+e^{i(\mathbf{k}_2-\mathbf{k}_4)\cdot \mathbf{R}_n})c^\cre_{o'}(\mathbf{k}_1)c^\cre_{o}(\mathbf{k}_2) c^\ann_{o}(\mathbf{k}_4) c^\ann_{o'}(\mathbf{k}_3 \right)
			\, ,
		\end{aligned}
	\end{equation}
	where Bravais lattice vectors $\mathbf{R}_n$ are shown Fig. $\ref{fig:Fig6NNfSC}$(b).
	Then vertex functions are obtained from the above equation and inserted into Eq. \eqref{InitialV} to determine valid initial conditions.
	For an inter-sublattice density-density interaction with $o \neq o'$, the initial interactions in $P$, $C$ and $D$ channel are given by 
	\begin{equation}
		\begin{aligned}
			&V^{\mathrm{P},0}_{oo'oo',-\mathbf{R}_n,-\mathbf{R}_n}(\mathbf{q})=V_n , \
			V^{\mathrm{P},0}_{o'ooo',\mathbf{R}_n,-\mathbf{R}_n}(\mathbf{q})=-V_ne^{-i\mathbf{q}\cdot \mathbf{R}_n}, \
			V^{\mathrm{P},0}_{oo'o'o,-\mathbf{R}_n,\mathbf{R}_n}(\mathbf{q})=-V_ne^{i\mathbf{q}\cdot \mathbf{R}_n}, \
			V^{\mathrm{P},0}_{o'oo'o,\mathbf{R}_n,\mathbf{R}_n}(\mathbf{q})=V_n, \\
			&V^{\mathrm{C},0}_{oo'oo',-\mathbf{R}_n,-\mathbf{R}_n}(\mathbf{q})=V_n, \
			V^{\mathrm{C},0}_{o'ooo',\mathbf{R}_1,\mathbf{R}_1}(\mathbf{q})=-V_n\sum_{\mathbf{R}_n}e^{-i\mathbf{q}\cdot \mathbf{R}_n}, \
			V^{\mathrm{C},0}_{oo'o'o,\mathbf{R}_1,\mathbf{R}_1}(\mathbf{q})=-V_n\sum_{\mathbf{R}_n}e^{i\mathbf{q}\cdot \mathbf{R}_n}, \
			V^{\mathrm{C},0}_{o'oo'o,\mathbf{R}_n,\mathbf{R}_n}(\mathbf{q})=V_n, \\
			&V^{\mathrm{D},0}_{oo'oo',\mathbf{R}_1,\mathbf{R}_1}(\mathbf{q})=V_n\sum_{\mathbf{R}_n}e^{i\mathbf{q}\cdot \mathbf{R}_n}, \
			V^{\mathrm{D},0}_{o'ooo',\mathbf{R}_n,\mathbf{R}_n}(\mathbf{q})=-V_n, \
			V^{\mathrm{D},0}_{oo'o'o,-\mathbf{R}_n,-\mathbf{R}_n}(\mathbf{q})=-V_n, \
			V^{\mathrm{D},0}_{o'oo'o,\mathbf{R}_1,\mathbf{R}_1}(\mathbf{q})=V_n\sum_{\mathbf{R}_n}e^{-i\mathbf{q}\cdot \mathbf{R}_n} \, .
		\end{aligned} 
	\end{equation}

	\subsection{Instability analysis}\label{supp:subsec:frg-inst}
	% We then integrate
	% the flow equations and get the renormalized interaction by reducing the RG scale $\Lambda$.
    The FRG flow equations can be integrated to get the renormalized effective interaction by reducing the FRG scale $\Lambda$.
    The flow will be  terminated at scale $\Lambda_c$ upon encountering a divergence in a channel, which indicates the normal metallic phase may become unstable towards a symmetry-broken state. The critical scale $\Lambda_c$ provides an estimate for the critical Temperature $T_c$. The divergent vertex component
	%which diverge signals the type of symmetry-broken phase the system might enter. Projecting effective interactions into superconducting and charge channels, the effective interaction for pair density wave(PDW) and charge density wave(CDW) orders are 
	signals the type of symmetry-broken phase the system might enter. Projecting effective interactions of the employed spinless model into superconducting and charge channels, the effective interaction for superconductivity (SC) and CDW orders are 
	\begin{equation} \label{lable112}
		\begin{aligned} 
			\hat{ V}^{\mathrm{SC}} &= \sum_{\mathbf{q}, \mathbf{k} ,\mathbf{k}' ,\left\lbrace o_i \right\rbrace }V^{\mathrm{P}}_{o_1o_2o_3o_4}\left( \mathbf{q}, \mathbf{k} ,\mathbf{k}' \right) 
			c^\cre_{\mathbf{k}+\mathbf{q},o_1}c^\cre_{-\mathbf{k},o_2} c^\ann_{-\mathbf{k}',o_4}c^\ann_{\mathbf{k}'+\mathbf{q},o_3}
			\\
			\hat{ V}^{\mathrm{CDW}} &= \sum_{\mathbf{q}, \mathbf{k} ,\mathbf{k}' ,\left\lbrace o_i \right\rbrace }V^{\mathrm{D}}_{o_1o_2o_3o_4}\left( \mathbf{q}, \mathbf{k} ,\mathbf{k}' \right)
			c^\cre_{\mathbf{k}+\mathbf{q},o_1}c^\ann_{\mathbf{k},o_3} c^\cre_{\mathbf{k}',o_2}c^\ann_{\mathbf{k}'+\mathbf{q},o_4}
		\end{aligned}
	\end{equation}
	with 
	\begin{equation} \label{lable111}
		\begin{aligned}
			\hat{ V}^{\mathrm{X}}_{o_1o_2o_3o_4}\left( \mathbf{q}, \mathbf{k} ,\mathbf{k}' \right) =   \sum_{m,n}f_m^{*}(\mathbf{k}) V^{\mathrm{X}}_{o_1o_2o_3o_4,mn} (\mathbf{q})f_n(\mathbf{k}') \, .
		\end{aligned}
	\end{equation}
	%Then one can get the mead field gap equation by decoupling effective interaction in most divergent channel, 
	With plane wave formfactors, plugging Eq.~\eqref{lable112} into Eq.~\eqref{lable111} leads to effective interactions in real space
	\begin{equation} 
		\begin{aligned} 
			&\hat{V}^{\mathrm{SC}} = \sum_{\mathbf{q}, \mathbf{R}_i, \mathbf{R}_j  ,m,n,\left\lbrace o_i \right\rbrace }V^{\mathrm{P}}_{o_1o_2m,o_3o_4n}\left( \mathbf{q} \right) 
			e^{i\mathbf{q}\cdot (\mathbf{R}_i+\mathbf{R}_m)}
			c^\cre_{o_1}(\mathbf{R}_i+\mathbf{R}_m)c^\cre_{o_2}(\mathbf{R}_i) 
			c^\ann_{o_4}(\mathbf{R}_j)c^\ann_{o_3}(\mathbf{R}_j+\mathbf{R}_n) e^{-i\mathbf{q}\cdot (\mathbf{R}_j+\mathbf{R}_n)}
			\\
			& \hat{ V}^{\mathrm{CDW}} = \sum_{\mathbf{q}, \mathbf{R}_i, \mathbf{R}_j ,m,n\left\lbrace o_i \right\rbrace }V^{\mathrm{D}}_{o_1o_3m,o_4o_2n}\left( \mathbf{q} \right)
			e^{i\mathbf{q}\cdot (\mathbf{R}_i+\mathbf{R}_m)}
			c^\cre_{o_1}(\mathbf{R}_i+\mathbf{R}_m)c^\ann_{o_3}(\mathbf{R}_i) c^\cre_{o_2}(\mathbf{R}_j)c^\ann_{o_4}(\mathbf{R}_j+\mathbf{R}_n) e^{-i\mathbf{q}\cdot (\mathbf{R}_j+\mathbf{R}_n)}.
		\end{aligned}
	\end{equation}
	%Defining order parameter in real space 
	%\begin{equation} 
	%	\begin{aligned} 
		%		&\Delta^{\mathrm{SC},\mathbf{q}}_{o_1o_2,m} = 
		%%		\frac{1}{N}
		%		-\sum_{j,n,o_4,o_4}
		%		V^{\mathrm{P}}_{o_1o_2m,o_3o_4n}\left( \mathbf{q} \right) 
		%		\left\langle   c^\ann_{o_4}(\mathbf{R}_j)c^\ann_{o_3}(\mathbf{R}_j+\mathbf{R}_n) \right\rangle 
		%		e^{-i\mathbf{q}\cdot (\mathbf{R}_j+\mathbf{R}_n)} ,	\\
		%		&\Delta^{\mathrm{CDW},\mathbf{q}}_{o_1o_2,m} = 
		%		%		\frac{1}{N}
		%		-\sum_{j,n,o_4,o_4}
		%		V^{\mathrm{C}}_{o_1o_2m,o_3o_4n}\left( \mathbf{q} \right) 
		%		\left\langle   c^\cre_{o_4}(\mathbf{R}_j)c^\ann_{o_3}(\mathbf{R}_j+\mathbf{R}_n)\right\rangle 
		%		e^{-i\mathbf{q}\cdot (\mathbf{R}_j+\mathbf{R}_n)},
		%	\end{aligned}
	%\end{equation}
	%the above effective interaction can be mean-field decoupled to(up to constant terms)
	%\begin{equation} 
	%	\begin{aligned} 
		%		 V^{\mathrm{SC}} &= -\sum_{\mathbf{q}, \mathbf{R}_i,m, o_1 ,o_2} 
		%		\Delta^{\mathrm{SC},\mathbf{q}}_{o_1o_2,m} 
		%		e^{i\mathbf{q}\cdot (\mathbf{R}_i+\mathbf{R}_m)}
		%		c^\cre_{o_1}(\mathbf{R}_i+\mathbf{R}_m)c^\cre_{o_2}(\mathbf{R}_i) 
		%		+ h.c. ,
		%		\\
		%		 V^{\mathrm{CDW}} &= -\sum_{\mathbf{q}, \mathbf{R}_i,m, o_1 ,o_2 }\Delta^{\mathrm{CDW},\mathbf{q}}_{o_1o_2,m} 
		%		e^{i\mathbf{q}\cdot (\mathbf{R}_i+\mathbf{R}_m)}
		%		c^\cre_{o_1}(\mathbf{R}_i+\mathbf{R}_m)c^\ann_{o_2}(\mathbf{R}_i) + h.c. .
		%	\end{aligned}
	%\end{equation}
	Defining order parameter in momentum space 
	\begin{equation} 
		\begin{aligned} 
			&\Delta^{\mathrm{SC},\mathbf{q}}_{o_1o_2}(\mathbf{k}) = 
			%		\frac{1}{N}
			-\sum_{\mathbf{k}',o_4,o_4}
			V^{\mathrm{P}}_{o_1o_2o_3o_4}\left( \mathbf{q}, \mathbf{k} ,\mathbf{k}' \right)
			\left\langle   c^\ann_{-\mathbf{k}',o_4}c^\ann_{\mathbf{k}'+\mathbf{q},o_3} \right\rangle 
			,	\\
			&\Delta^{\mathrm{CDW},\mathbf{q}}_{o_1o_2}(\mathbf{k}) = 
			%		\frac{1}{N}
			-\sum_{\mathbf{k}',o_4,o_4}
			V^{\mathrm{D}}_{o_1o_2o_3o_4}\left( \mathbf{q}, \mathbf{k} ,\mathbf{k}' \right)
			\left\langle  c^\cre_{\mathbf{k}',o_4}c^\ann_{\mathbf{k}'+\mathbf{q},o_3}  \right\rangle 
			,
		\end{aligned}
	\end{equation}
	the above effective interaction can be mean field decoupled, which leads to linearized gap equations
	\begin{equation}
		\begin{aligned}
			&-\sum_{\mathbf{k}'o_1'o_2',o_3'o_4'} V^{\text{P}}_{o_1o_2,o_3o_4}(\mathbf{q},\mathbf{k},\mathbf{k}') L^{\text{pp}}_{o_3o_4, o_1'o_2'}(\mathbf{q},\mathbf{k}') \Delta^{\mathrm{SC},\mathbf{q}}_{o_1'o_2'}(\mathbf{k}') =\lambda \Delta^{\mathrm{SC},\mathbf{q}}_{o_1o_2}(\mathbf{k}). \\
			&-\sum_{\mathbf{k}'o_1'o_2',o_3'o_4'} V^{\text{D}}_{o_1o_2,o_3o_4}(\mathbf{q},\mathbf{k},\mathbf{k}') L^{\text{ph}}_{o_3o_4, o_1'o_2'}(\mathbf{q},\mathbf{k}') \Delta^{\mathrm{CDW},\mathbf{q}}_{o_1'o_2'}(\mathbf{k}') =\lambda \Delta^{\mathrm{CDW},\mathbf{q}}_{o_1o_2}(\mathbf{k}).
		\end{aligned}
	\end{equation}
	In formfactor basis, this equation has the following matrix multiplication form
	\begin{equation} \label{LGEinRS}
		\begin{aligned}
			&-\sum_{o_1'o_2'm',o_3'o_4'n'} V^{\text{P}}_{o_1o_2m,o_3o_4 m'}(\mathbf{q}) \chi^{\text{pp}}_{o_3o_4 m', o_1'o_2' n'}(\mathbf{q}) \Delta^{\mathrm{SC},\mathbf{q}}_{o_1'o_2' n'} =\lambda \Delta^{\mathrm{SC},\mathbf{q}}_{o_1o_2m}. \\
			&-\sum_{o_1'o_2'm',o_3'o_4'n'} V^{\text{D}}_{o_1o_2m,o_3o_4 m'}(\mathbf{q}) \chi^{\text{ph}}_{o_3o_4 m', o_1'o_2' n'}(\mathbf{q}) \Delta^{\mathrm{CDW},\mathbf{q}}_{o_1'o_2' n'} =\lambda \Delta^{\mathrm{CDW},\mathbf{q}}_{o_1o_2m}.
		\end{aligned}
	\end{equation}
	Here we define the real space order parameter
	\begin{equation}
		\begin{aligned}
			%		\Delta^{\mathrm{SC/CDW},\mathbf{q}}_{o_1o_2 m} = \int \dd \mathbf{k} f_m(\mathbf{k})\Delta^{\mathrm{SC/CDW},\mathbf{q}}_{o_1o_2}(\mathbf{k}).
			\Delta^{\mathrm{SC},\mathbf{q}}_{o_1o_2 m} = \int \dd \mathbf{k} f_m(\mathbf{k})\Delta^{\mathrm{SC},\mathbf{q}}_{o_1o_2}(\mathbf{k})
			=-\sum_{j,n,o_4,o_4}
			V^{\mathrm{P}}_{o_1o_2m,o_3o_4n}\left( \mathbf{q} \right) 
			\left\langle   c^\ann_{o_4}(\mathbf{R}_j)c^\ann_{o_3}(\mathbf{R}_j+\mathbf{R}_n) \right\rangle 
			e^{-i\mathbf{q}\cdot (\mathbf{R}_j+\mathbf{R}_n)} \\
			\Delta^{\mathrm{CDW},\mathbf{q}}_{o_1o_2 m} = \int \dd \mathbf{k} f_m(\mathbf{k})\Delta^{\mathrm{CDW},\mathbf{q}}_{o_1o_2}(\mathbf{k})
			=-\sum_{j,n,o_4,o_4}
			V^{\mathrm{D}}_{o_1o_2m,o_3o_4n}\left( \mathbf{q} \right) 
			\left\langle   c^\cre_{o_4}(\mathbf{R}_j)c^\ann_{o_3}(\mathbf{R}_j+\mathbf{R}_n)\right\rangle 
			e^{-i\mathbf{q}\cdot (\mathbf{R}_j+\mathbf{R}_n)}.
		\end{aligned}
	\end{equation}
	%Defining order parameter in real space 
	%\begin{equation} 
	%	\begin{aligned} 
		%		&\Delta^{\mathrm{SC},\mathbf{q}}_{o_1o_2,m} = 
		%		%		\frac{1}{N}
		%		-\sum_{j,n,o_4,o_4}
		%		V^{\mathrm{P}}_{o_1o_2m,o_3o_4n}\left( \mathbf{q} \right) 
		%		\left\langle   c^\ann_{o_4}(\mathbf{R}_j)c^\ann_{o_3}(\mathbf{R}_j+\mathbf{R}_n) \right\rangle 
		%		e^{-i\mathbf{q}\cdot (\mathbf{R}_j+\mathbf{R}_n)} ,	\\
		%		&\Delta^{\mathrm{CDW},\mathbf{q}}_{o_1o_2,m} = 
		%		%		\frac{1}{N}
		%		-\sum_{j,n,o_4,o_4}
		%		V^{\mathrm{C}}_{o_1o_2m,o_3o_4n}\left( \mathbf{q} \right) 
		%		\left\langle   c^\cre_{o_4}(\mathbf{R}_j)c^\ann_{o_3}(\mathbf{R}_j+\mathbf{R}_n)\right\rangle 
		%		e^{-i\mathbf{q}\cdot (\mathbf{R}_j+\mathbf{R}_n)},
		%	\end{aligned}
	%\end{equation}
	Then the effective interaction in real space can be mean-field decoupled to yield 
	\begin{equation} 
		\begin{aligned} 
			V^{\mathrm{SC}} &= -\sum_{\mathbf{q}, \mathbf{R}_i,m, o_1 ,o_2} 
			\Delta^{\mathrm{SC},\mathbf{q}}_{o_1o_2,m} 
			e^{i\mathbf{q}\cdot (\mathbf{R}_i+\mathbf{R}_m)}
			c^\cre_{o_1}(\mathbf{R}_i+\mathbf{R}_m)c^\cre_{o_2}(\mathbf{R}_i) 
			+ h.c.
			\\
			V^{\mathrm{CDW}} &= -\sum_{\mathbf{q}, \mathbf{R}_i,m, o_1 ,o_2 }\Delta^{\mathrm{CDW},\mathbf{q}}_{o_1o_2,m} 
			e^{i\mathbf{q}\cdot (\mathbf{R}_i+\mathbf{R}_m)}
			c^\cre_{o_1}(\mathbf{R}_i+\mathbf{R}_m)c^\ann_{o_2}(\mathbf{R}_i) + h.c.
		\end{aligned}
	\end{equation}
	omitting constant terms.
	%Hence  $e^{i\mathbf{q}\cdot (\mathbf{R}_i+\mathbf{R}_m)}\Delta^{\mathrm{SC},\mathbf{q}}_{o_1o_2,m} $ and $e^{i\mathbf{q}\cdot (\mathbf{R}_i+\mathbf{R}_m)}\Delta^{\mathrm{CDW},\mathbf{q}}_{o_1o_2,m} $ describe SC or CDW paring on  solving linearized gap equation (\ref{LGEinRS}).
	Then by solving the linearized gap equation Eq.~\eqref{LGEinRS}, we obtain the real space pairing pattern $\Delta^{\mathrm{SC},\mathbf{q}}_{o_1o_2,m} $ and $\Delta^{\mathrm{CDW},\mathbf{q}}_{o_1o_2,m}$, which describes the SC and CDW paring on bond $\mathbf{R}_i+\mathbf{R}_m$ and $\mathbf{R}_i$ modulated by a phase $e^{i\mathbf{q}\cdot (\mathbf{R}_i+\mathbf{R}_m)}$.
	
	From effective interactions, we can also analyze leading order by diagonlizing interaction corrected renormalized susceptibility in particle-hole and particle-particle channels
	\begin{equation}
		\begin{aligned}
			& \chi^{\text{PDW},\mathrm{I}}(\mathbf{q})=\chi^{\text{pp}}(\mathbf{q}) \left( -V^{\text{P}} (\mathbf{q}) \right)  \chi^{\text{pp}}(\mathbf{q}),\\
			& \chi^{\text{CDW},\mathrm{I}}(\mathbf{q})=\chi^{\text{ph}}(\mathbf{q}) \left(-V^{\text{D}}(\mathbf{q}) \right)  \chi^{\text{ph}}(\mathbf{q}),
		\end{aligned}
	\end{equation}
	from which we obtain the components of bond and cuurent order in the main text.
	\begin{figure}[t]
		\centering
		\subfigure[]{
			\begin{tikzpicture}[> = Stealth, baseline = 0 cm,scale=1]
				\color{black}
				\begin{feynhand}
					\vertex (i1) at (0,0);
					\vertex (i2) at (1,0);
					\vertex (i3) at (3,0);
					\vertex (i4) at (4,0) ;
					\vertex (j1) at (2,1);
					\vertex (j2) at (2,-1) ;
					\propag [chabos] (i1) to (i2);
					\propag [chabos] (i3) to (i4);
					\propag [fer] (i2) to [in=180,out=90] (j1);
					\propag [fer] (i2) to [in=180,out=270] (j2);
					\propag [fer] (j1) to [in=90,out=0] (i3);
					\propag [fer] (j2) to [in=270,out=360] (i3);
					\propag [bos] (j1) to (j2);
					\node at (0.5,-0.3) {$q $};
					\node at (3.5,-0.3) {$q $};
					\node at (1.4,1.2) {$k+q$};
					\node at (2.6,1.2) {$k'+q$};
					\node at (1.6,-1.2) {$-k$};
					\node at (2.4,-1.2) {$-k'$};
					\node at (4,0) {$\quad$};
					
					%					\propag[fer] (i2) to  [half left, looseness=1.7] (i3);
					%					\propag[fer] (i2) to  [half right, looseness=1.7] (i3);
					%					\node at (1.5,0.75) {$o_1,k+q,o_3 $};
					%					\node at (1.5,-0.75) {$o_2,-k,o_4 $};
					%					\node at (0.8,0.25) {$f_m $};
					%					\node at (2.2,0.25) {$f_n $};
					%					\node at (-1.5,0) {$\chi^{\text{pp}}_{o_1o_2o_3o_4,mn}(q) = $};
				\end{feynhand}
			\end{tikzpicture}
		}
		\subfigure[]{
			\begin{tikzpicture}[> = Stealth, baseline = 0 cm,scale=1]
				\color{black}
				\begin{feynhand}
					\vertex (i1) at (0,0);
					\vertex (i2) at (1,0);
					\vertex (i3) at (3,0);
					\vertex (i4) at (4,0) ;
					\vertex (j1) at (2,1);
					\vertex (j2) at (2,-1) ;
					\propag [chabos] (i1) to (i2);
					\propag [chabos] (i3) to (i4);
					\propag [fer] (i2) to [in=180,out=90] (j1);
					\propag [antfer] (i2) to [in=180,out=270] (j2);
					\propag [fer] (j1) to [in=90,out=0] (i3);
					\propag [antfer] (j2) to [in=270,out=360] (i3);
					\propag [bos] (j1) to (j2);
					\node at (0.5,-0.3) {$q $};
					\node at (3.5,-0.3) {$q $};
					\node at (1.4,1.2) {$k+q$};
					\node at (2.6,1.2) {$k'+q$};
					\node at (1.6,-1.2) {$k$};
					\node at (2.4,-1.2) {$k'$};
				\end{feynhand}
			\end{tikzpicture}
		}
		\caption{Interaction corrected susceptibility in the particle-particle (a) and particle-hole (b) channels.}
	\end{figure}

	\subsection{Identifying CBO and LCO parameters from vertex eigenstates}\label{supp:subsec:frg-identify}
	As discussed in the previous section, the outcome of an FRG flow is the low energy effective vertex in the three different fluctuation channels. The order parameter manifesting at a divergence of the FRG susceptibilities is hence obtained by solving the linearised gap equation in the respective channel. This boils down to the diagonalisation of the effective vertex at each transfer momentum $\mathbf{q}$, where the eigenstate corresponding to the largest eigenvalue gives the ordering vector of the symmetry broken phase. Since the overall phase of the eigenstates is gauge dependent, this poses the question how to distinguish between real and imaginary bond orders.
	
	On the Kagome lattice the FRG vertex diverges at the three $\mathbf{M}$ points at the edges of the BZ.
	While this usually requires a subsequent Ginzburg-Landau analysis to fix the relative phases between the degenerate states at the $\mathbf{M}$ points~\cite{SM_Denner2021}, the formfactor information encoded in the eigenstates is already sufficient to distinguish CBO from LCO states by means of a single $\mathbf{M}$ point instability.
	As discussed in the main text, the instability at each $\mathbf{M}$ point corresponds to a real space order on perpendicular bonds. As $\mathbf{M} = -\mathbf{M}$, this implies, that the order parameter on one specific bond is uniquely determined by the eigenstate of a single $\mathbf{M}$ point (compare Fig.~\ref{fig:bond_orders}).
	Furthermore, the knowledge of both $\langle c^\cre_i c^\ann_j \rangle$ and $\langle c^\cre_j c^\ann_i \rangle$ and its relative phase as provided by the formfactor indices of the FRG order parameter, allows for a direct classification as real or imaginary bond order 
	\begin{equation}
		\begin{split}
			\Delta^{\mathrm{CBO}}_{ij} = \langle c^\cre_i c^\ann_j + c^\cre_j c^\ann_i \rangle = \langle c^\cre_i c^\ann_j \rangle + \langle c^\cre_j c^\ann_i \rangle\\
			\Delta^{\mathrm{LCO}}_{ij} = \text{i} \langle c^\cre_i c^\ann_j - c^\cre_j c^\ann_i \rangle =
			\text{i} ( \langle c^\cre_i c^\ann_j \rangle - \langle c^\cre_j c^\ann_i \rangle ) \ ,
		\end{split}
	\end{equation}
	since the global gauge is fixed by the requirement for an hermitian order parameter.

	\section{An intuitive picture of the competing real and imaginary bond orders on the kagome lattice}\label{supp:anchor:bondorders}\label{supp:sec:bondorders}
	
	As discussed in the main text, the bond orders on the kagome lattice emerge from the VHS nesting at the Fermi level and can be characterised by a $\mathbf{M}$ point periodicity. On the kagome lattice, this reflects in a sign change of the order parameter between neighbouring 1D chains along the $\mathbf{M}$ point direction, while the bond order establishes along the respective chain (see Fig.~\ref{fig:bond_orders}).
	Let us study the $V_2 = 0$ limit first, \textit{i.e.} consider only nearest neighbor bond orders. In this case, it is possible to discuss the emergence of bond order instabilities by considering the 1D chains independently, since both the hopping and interaction terms only couple states within one single chain.

	% \begin{figure}[t]
		% 	\centering
		% 	\includegraphics[width=0.9\linewidth]{materials/Fig_gap_new.pdf}
		% 	\caption{Mean field energy dispersion of $A-B$ bond order at $\mathbf{Q_c}$ along fermi surface $M_B-M_C-M_A$: (a)1NN CBO; (b)1NN LCO; (b)2NN CBO; (d)2NN LCO. Around the middle point of two vHs $(M_A+M_B)/2$, 1NN and 2NN bond order open larger gap in real and imaginary channel respectively.}
		% 	\label{fig:gap}
		% \end{figure}

	\subsection{Mean field energy gap}\label{supp:subsec:bond-mean}
	Bond orders in both real and imaginary channel are three fold degenerate with ordering wave vectors $\mathbf{Q}_{A,B,C}$. We exemplarily consider a bond order at $\mathbf{Q}_A$ in the following. The effective mean field Hamiltonian in momentum space reads
	\begin{equation}
		\begin{aligned}
			H_{\text{LCO}}=\sum_{\mathbf{k}} \Delta(
			f_{\mathbf{k}} c^{\dagger}_{\mathbf{k}+\mathbf{Q}_{C},A} c_{\mathbf{k},B}
			+f_{\mathbf{k}+\mathbf{Q}_{C}}c^{\dagger}_{\mathbf{k},A} c_{\mathbf{k}+\mathbf{Q}_{C},B}
			+f_{\mathbf{k}+\mathbf{Q}_{A}}c^{\dagger}_{\mathbf{k}+\mathbf{Q}_{B},A} c_{\mathbf{k}+\mathbf{Q}_{A},B}
			+f_{\mathbf{k}+\mathbf{Q}_{B}}c^{\dagger}_{\mathbf{k}+\mathbf{Q}_{A},A} c_{\mathbf{k}+\mathbf{Q}_{B},B}) +\mathrm{H.c.} \, ,
		\end{aligned}
	\end{equation}
	where $\Delta$ is the order parameter and $f_{\mathbf{k}}$ is the corresponding formfactor. 
	% The scattering matrix between FS points $\mathbf{k}=(-\pi,k_y)$ and $\mathbf{k}+\mathbf{Q}_{C}=(\pi,k_y)$ reads
	% \begin{equation}
		% 	\begin{aligned}
			%             h=\begin{pmatrix}
				%             0&\Delta f_{\mathbf{k}} u^{*}_{1A}u_{2A} + \Delta^{*} f_{\mathbf{k}+\mathbf{Q}_{C}} u^{*}_{2A}u_{1B}
				%             \\ \Delta^{*} f_{\mathbf{k}}^{*} u^{*}_{1A}u_{2A} + \Delta f_{\mathbf{k}+\mathbf{Q}_{C}} u^{*}_{2A}u_{1B}&0
				%             \end{pmatrix}
			% 	\end{aligned}
		% \end{equation}
	Then the effective Hamiltonian describing the scattering between two Fermi points $\mathbf{k}=(-\pi,k_y)$ and $\mathbf{k}+\mathbf{Q}_{C}=(\pi,k_y)$ reads
	\begin{equation}
		\begin{aligned}
			h=(\Delta f_{\mathbf{k}} u^{*}_{A}(\mathbf{k}+\mathbf{Q}_{C})u_{B}(\mathbf{k}) + \Delta^{*} f_{\mathbf{k}+\mathbf{Q}_{C}}^{*} u^{*}_{B}(\mathbf{k}+\mathbf{Q}_{C})u_{A}(\mathbf{k}))c^{\dagger}_{\mathbf{k}+\mathbf{Q}_{C}}c_{\mathbf{k}} + \mathrm{H.c.} \, .
		\end{aligned}
	\end{equation}
	At the Van-Hove point, $k_y=-\pi/\sqrt{3}$ and $u_{A}(\mathbf{k})=u_{B}(\mathbf{k}+\mathbf{Q}_c)=1, u_{B}(\mathbf{k})=u_{A}(\mathbf{k}+\mathbf{Q}_c)=0$ and the gap opening at the Fermi level is given by $\Delta_{\text{VH}}=|\Delta f_{\mathbf{M}_A}|$. The form factor reads $f_{\mathbf{k}}=\cos(\mathbf{k}\cdot \mathbf{\delta})$ for the symmetrized bond order and $f_{\mathbf{k}}=i\sin(\mathbf{k}\cdot \mathbf{\delta})$ for anti-symmetrized bond order.
	
\begin{figure}[t]
		\centering
		\includegraphics[width=0.99\linewidth]{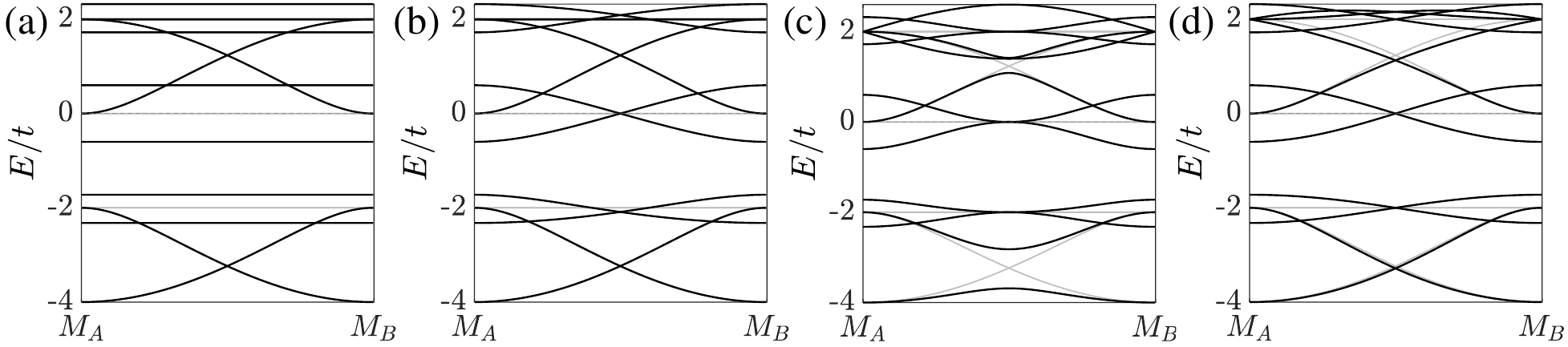}
		\caption{Mean field energy dispersion of $A-B$ bond order at $\mathbf{Q_c}$ along fermi surface $M_A-M_B$: (a)1NN CBO; (b)1NN LCO; (b)2NN CBO; (d)2NN LCO. Around the middle point of two vHs $(M_A+M_B)/2$, 1NN and 2NN bond order open larger gap in real and imaginary channel respectively.}
		\label{fig:gap}
	\end{figure}

	As $\mathbf{M}_A\cdot \mathbf{\delta}=\pi/2$ for $\delta=(1/2,0)$ or $(0,\sqrt{3}/2)$, symmetrized bond orders cannot gap out the Van-Hove points and are hence energetically unfavorable. At the center between two Van-Hove points, $k_y=0$ and $u_{A}(\mathbf{k})=u_{B}(\mathbf{k}+\mathbf{Q}_c)= u_{B}(\mathbf{k})=u_{A}(\mathbf{k}+\mathbf{Q}_c)=1/\sqrt{2}$. Considering anti-symmetric formfactors $f^{*}_{\mathbf{k}+\mathbf{Q}_{C}} = \pm f_{\mathbf{k}}$ for $\delta=(1/2,0)$ and $(0,\sqrt{3}/2)$, the gap opening of an anti-symmetrized bond order at the middle point $ \Delta_{\text{Mid}}=|\Delta \pm \Delta^*| |f_{(-\pi,0)}|/2$. Hence the bond order will open a lager gap in real channel on 1NN bond and imaginary channel on 2NN bond around the middle point of Van-Hove points as shown in Fig. \ref{fig:gap}. 
	
	\subsection{Nearest neighbour bond orders}\label{supp:subsec:bond-nn}
\begin{figure}[t]
		\centering
		\includegraphics[width=0.8\linewidth]{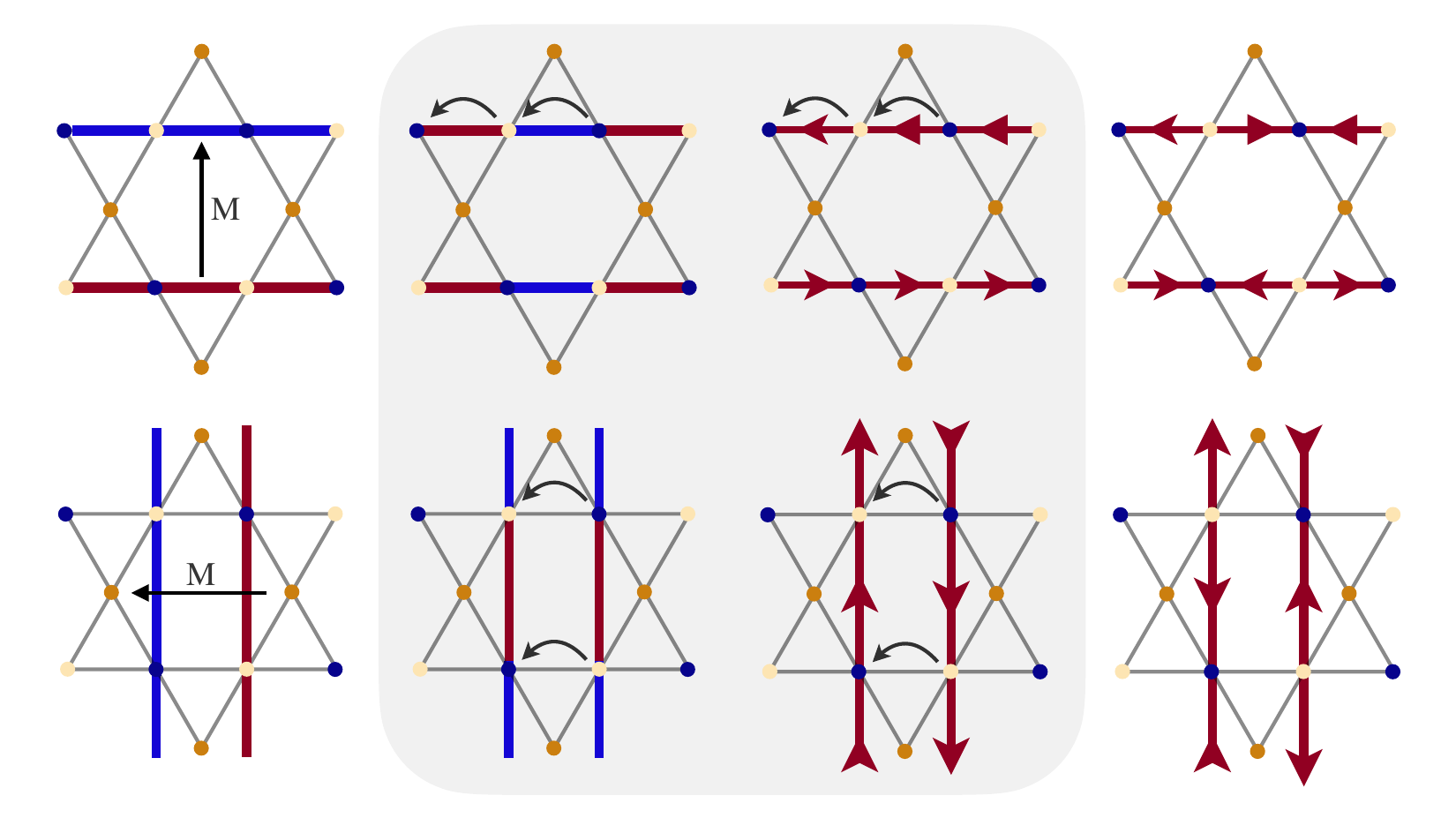}
		\caption{Bond orders on the kagome lattice. The bond order forms on the 1D chains perpendicular to the instability vector $\mathbf{M}$. For the nearest neighbor orders (first row), these chains coincide with the 1D chains making up the kagome lattice, while the second nearest neighbor bond orders (second row) establish perpendicular to kagome chains. For the symmetric (a/e) and antisymmetric (b/f) charge bond order, thick(thin) lines denote enhanced(reduced) hopping elements. For the symmetric (c/g) and antisymmetric (d/h) loop current orders, the direction of the arrows indicate the direction of the net flux along the respective bond.}
		\label{fig:bond_orders}
	\end{figure}

	The four possible nearest neighbour bond orders on the 1D chain can be cathegorized by their transformation behaviour under inversion and time reversal symmetry (TRS) into the real symmetric (even) and antisymmetric (odd) bond orders
	\begin{equation}
		\begin{split}    
			\ket{\mathrm{CBO}}_e = & \sum_i (c^\cre_{2i+1} c^\ann_{2i} + c^\cre_{2i+2} c^\ann_{2i+1} + h.c.) \ket{\Phi_0} \\
			\ket{\mathrm{CBO}}_o = & \sum_i (c^\cre_{2i+1} c^\ann_{2i} - c^\cre_{2i+2} c^\ann_{2i+1} + h.c.) \ket{\Phi_0} 
		\end{split}
		\label{eq:CBO}
	\end{equation}
	and the loop current orders
	\begin{equation}
		\begin{split}    
			\ket{\mathrm{LCO}}_e = & \sum_i (c^\cre_{2i+1} c^\ann_{2i} + c^\cre_{2i+2} c^\ann_{2i+1} - h.c.) \ket{\Phi_0} \\
			\ket{\mathrm{LCO}}_o = & \sum_i (c^\cre_{2i+1} c^\ann_{2i} - c^\cre_{2i+2} c^\ann_{2i+1} - h.c.) \ket{\Phi_0} \ ,
		\end{split}
		\label{eq:LCO}
	\end{equation}
	where $i$ labels the site position and the operators act on the ground state of the non-interacting Hamiltonian $\ket{\Phi_0} = \rho \sum_i c^\cre_i c^\ann_i \ket{0}$. The electron density per site $\rho = 5/12$ at p-type vH filling.
	Inspecting the real space structure of the orders, as depicted in Fig.~\ref{fig:bond_orders}, already two of these orders can be discarded from physical arguments:
	Interpreting the CBO as a generalised Peierls instability, \textit{i.e.} the increased hybridisation strength between two sites originates from a shift in the Wannier centers towards the bond center, the symmetric bond order is not volume preserving and hence not compatible with Born-von Karman periodic boundary conditions.
	The asymmetric LCO on the other hand, leads to charge accumulation on every second sites and can therefore not be realized in thermodynamic equilibrium.
	Indeed both of these ordering tendencies show little susceptibility in our numerical calculations~\cite{SM_RPA_paper}.
	In the main text as well as in the following, we therefore denote with CBO and LCO only the physically meaningful instabilities $\ket{\mathrm{CBO}} = \ket{\mathrm{CBO}}_o$ and $\ket{\mathrm{LCO}} = \ket{\mathrm{LCO}}_e$.
	
	\subsection{Ginzburg-Landau analysis of real and imaginary bond orders}\label{supp:subsec:bond-gl}
	To pinpoint the origin of the surprising correlation between the NNN interaction and the emergence of LCO, we expand the free energy in the symmetry broken phase up to second order in the bond order parameter $\Delta_i$ to obtain
	\begin{equation}
		f^{(2)} = \frac{1}{g} \sum_i \abs{\Delta_i}^2 + \frac{1}{2} \sum_i \operatorname{Tr}((G \Delta_i)^2)
		\label{eq:GL_FE}
	\end{equation}
	with the bare single-particle Greensfunction $G$ and some renormalized interaction strength $g$. $\Delta_i$ is the bond order modulation corresponding to the eigenstate of the vertex at $M_i$.
	While the first term is equal for real and imaginary bond orders, the second term can lead to an energy difference between the CBO and LCO states.
	
	As discussed earlier, the order parameter associated with a single $M_i$ instability uniquely determines the bond order profile along the corresponding real space direction (cf. Fig.~\ref{fig:bond_orders}) and does not couple to the other direction up to second order in the GL functional. Hence, let us for simplicity consider a single $\mathbf{M}$ order and evaluate the second term in Eq.~\eqref{eq:GL_FE} for the two competing NN orders
	\begin{equation}
		\begin{split}
			\Delta^{\mathrm{CBO}} = & \sum_i \langle (c^\cre_{2i+1} c^\ann_{2i} - c^\cre_{2i+2} c^\ann_{2i+1}) + h.c. \rangle \\
			\Delta^{\mathrm{LCO}} = & \sum_i \langle \textit{i} (c^\cre_{2i+1} c^\ann_{2i} + c^\cre_{2i+2} c^\ann_{2i+1}) + h.c. \rangle \ .
		\end{split}
	\end{equation}
	Here, $i$ labels the real space positions along the 1D chain.
	It is instructive to do this in real space to obtain
	\begin{equation}
		\begin{split}
			\operatorname{Tr}(G \Delta G \Delta) \sim \sum_i & G_{i,i} \Delta_{i,i+1} G_{i+1,i+1} \Delta_{i+1,i} + G_{i,i} \Delta_{i,i-1} G_{i-1,i-1} \Delta_{i-1,i} \\
			+ & G_{i,i+1} \Delta_{i+1,i+2} G_{i+2,i+1} G_{i+1,i} + G_{i,i-1} \Delta_{i-1,i-2} G_{i-2,i-1} \Delta_{i-1,i} \ ,
		\end{split}
	\end{equation}
	where we only take into account short range scattering terms up to first nearest neighbour.
	The energy difference between the imaginary and real bond order is then easily traced back to its distinct transformation behaviour under translation by one lattice site $\Delta_{i,i+1} \rightarrow \Delta_{i+1, i+2}$. Using the relations
	\begin{equation}
		\begin{split}
			\Delta^{\mathrm{LCO}}_{i,i+1} = (\Delta^{\mathrm{LCO}}_{i+1, i})^* \ , \
			\Delta^{\mathrm{LCO}}_{i+1,i+2} = & \hphantom{-} \ (\Delta^{\mathrm{LCO}}_{i+1, i})^* \ , \\
			\Delta^{\mathrm{CBO}}_{i,i+1} = (\Delta^{\mathrm{CBO}}_{i+1, i})^* \ , \
			\Delta^{\mathrm{CBO}}_{i+1,i+2} = & - (\Delta^{\mathrm{CBO}}_{i+1, i})^* \ ,
		\end{split}
	\end{equation}
	one can distinguish two type of terms:
	Scattering events involving translations by an even number of sites (corresponding to to equal sublattice indices of the Green functions on the kagome lattice)
	\begin{equation}
		\Pi = \abs{\Delta}^2 \sum_i G_{i,i} G_{i+1,i+1} + G_{i,i} G_{i-1,i-1}
	\end{equation}
	contribute equally to the free energy of CBO and LCO.
	Terms with translations by an odd number of sites
	\begin{equation}
		\Xi = \abs{\Delta}^2 \sum_i G_{i,i+1} G_{i+2,i+1} + G_{i,i-1} G_{i-2,i-1}
	\end{equation}
	are distinct between CBO and LCO, such that we obtain under the made approximations
	\begin{equation}
		\begin{split}
			f_{\mathrm{CBO/LCO}} = \Pi \pm \Xi \ . 
		\end{split}
	\end{equation}
	The situation is reversed for the case of second nearest neighbor order. Now the NN hopping events flip the sign of the LCO and the CBO is invariant under a translation by one lattice constant along the 1D chains (cf. Fig.~\ref{fig:bond_orders}). While again the $\Pi$ term is equivalent for both orders, the sign of $\Xi$ is reversed and the LCO has a higher energy gain. The analysis of the short ranged scattering events reveals a natural correspondence between the energy gain of the bond orders and their real space structure. The obtained result find solid ground in RPA susceptibility calculations, where the real space information is encoded in the formfactor dependence of the particle-hole bubble~\cite{SM_RPA_paper}.
	
	This sheds new light on the necessity of second nearest neighbor interactions in driving LCO in the kagome Hubbard model as already speculated on MF level in Ref.~\cite{SM_Dong2023}. The LCO can only emerge as a self-sufficient instability due to coupling of the individual 1D kagome chains by second nearest neighbour formfactors. This allows for an energy gain by screening processes mediated by a single hopping event.
	Apparently, this argument is tightly bound to the presence of nearest neighbor hybridisation only. We expect second nearest neighbor hybridisation $t'$ to strengthen the NN LCO compared to the NN CBO as this unlocks additional first order processes for energy gains within the LCO phase.
	Since the bare interaction couples predominantly to bond order fluctuations of the same length scale, the $V_2/V_1$ allows for a tuning between CBO and LCO, where the CBO (LCO) is predominantly found in the $V_1 >(<) V_2$ regime.

	\subsection{CBO in the $V_1$ dominated regime}\label{supp:subsec:bond-cbo}
\begin{figure}[t]
		\centering
		\includegraphics[width=0.99\linewidth]{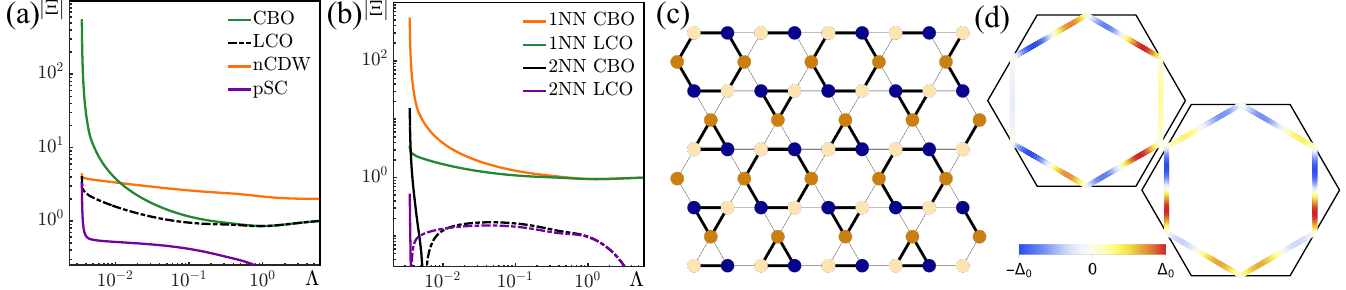}
		\caption{Charge bond order. (a) FRG flow of the expectation values of nCDW, CBO, LCO and fSC for $V_1=1t$ and $V_2=0$. (b) Nearest neighbour (1NN) and next nearest neighbour (2NN) contributions to the CBO and LCO phases throughout the flow. (c) Real-space pattern of $3\mathbf{Q}$ charge bond order forming trihexagonal bond order configuration. (d) Leading SC fluctuations on the FS displaying two-fold degenerate $E_{1}$ symmetry.}
		\label{fig:FigCBO2}
	\end{figure}

	In the $V_1$ dominated regime, we find robust charge bond order compared to the relative fragile LCO of the $V_2$ dominated regime. A typical flow in the CBO region is shown in Fig. \ref{fig:FigCBO2}(a). The obtained charge bond modulations predominantly occur on nearest neighbour (1NN) bonds as shown in Fig. \ref{fig:FigCBO2}(b). This is consistent with RPA results where $V_1$ enhances real bond susceptibilities compared to imaginary ones and favors the formation of CBO~\cite{SM_RPA_paper}.
	The CBO is three fold degenerate with ordering wave vectors $\mathbf{Q}_{A}$, $\mathbf{Q}_{B}$, $\mathbf{Q}_{C}$, which modulate bonds along $BC, CA, AB$ directions respectively.
	A trilinear term $\Delta_A \Delta_B \Delta_C$ in Ginzburg-Landau free energy will favor the formation of $3\mathbf{Q}$ other than $3\mathbf{Q}$ CBO. The effective mean field Hamiltonian is described by
	%$H_{CBO}=\sum_{\alpha\beta\gamma} \epsilon_{\alpha\beta\gamma} $
	\begin{equation} 
		\begin{aligned} 
			H_{\mathrm{CBO}}=- \sum_{\alpha\beta\gamma \mathbf{R}} \epsilon_{\alpha\beta\gamma}  \Delta_{\gamma}
			\left( 
			c^\cre_{\alpha}(\mathbf{R})c^\ann_{\beta}(\mathbf{R}) + c^\cre_{\alpha}(\mathbf{R})c^\ann_{\beta}(\mathbf{R}-\mathbf{d}^{\mathrm{NN}}_{\beta\gamma})
			\right) 
			\cos(\mathbf{Q}_{\gamma} \cdot \mathbf{R}) 
			+ h.c. \, ,
		\end{aligned}
	\end{equation}
	where $\Delta_{\gamma}$ is the real order parameter. Six fold rotational symmetry is preserved when $\Delta_{\gamma}=\Delta $ for $\gamma=A,B,C$. A real $\Delta>0 (\Delta<0)$ will correspond to trihexagonal(anti-trihexagonal) patterns. The trihexagonal pattern is shown in Fig. \ref{fig:FigCBO2}(c). 
	
	The subdominant SC fluctuations in CBO phase exhibit $p$-wave symmetry. As shown in Fig.~\ref{fig:FigCBO2}(d), the corresponding paring functions on FS exhibit nodes not only along $\Gamma-M$ but also along $\Gamma-K$ direction. Thus compared with $B_1$ $f$-wave SC, the $E_2$ $p$-wave SC has lower condensation energy and does not appear as dominant instability in $V_1$ dominated regime.
	
	\section{Physical properties of the LCO state}\label{supp:anchor:lco}\label{supp:sec:lco}
	
	\subsection{Ginzburg-Landau analysis and real space pattern}\label{supp:subsec:lco-gl}
	%For $V_1=0$ and $V_2=1.25t$, Fig. \ref{fig:lco}(a) shows the maximal absolute value of $P$, $C$ and $D$ channels with inset figure showing the flow of effective interaction strength for CDW, CBO, LCO and fSC order. The leading instability occurs in charge channel. Solving the linearied gap equation and close inspectSion  of leading eigenmode reveals the order is second nearest neighbor(2nn) antisymmetrized CBO order as shown in Fig. \ref{fig:lco}(b). Similar to CBO, the peaks of LCO along three direction $AB, BC, CA$ are attributed to three nesting momentum $\mathbf{Q}_C=(0,2\pi/\sqrt{3}), \mathbf{Q}_A=(-\pi,-\pi/\sqrt{3}), \mathbf{Q}_B=(\pi,-\pi/\sqrt{3})$ respectively which is closely related to sublattice interference in upper VHS of Kagome lattice. 
	%A Ginzburg-Landau free energy analysis will revel the formation of $3\mathbf{Q}$ LCO order.
	%A Ginzburg-Landau analysis will reveal the $3\mathbf{Q}$ LCO. 
	%In a patch model, the low energy effective interaction is dominant by 
	
	% \begin{figure}[b]
		% 	\centering
		% 	\includegraphics[width=0.6\linewidth]{old/Free_energy.pdf}
		% 	\caption{Diagrammatical representation of fourth order Ginzburg-Landau free energy coefficient in the patch model.}
		% 	\label{fig:GLcoe}
		% \end{figure}
	Within the LCO regime, the low energy effective interaction is dominated by scattering in the LCO channel. Thus the effective effective mean field Hamiltonian can be written as
	% \begin{equation} 
		% 	\begin{aligned} 
			% 		H_{\mathrm{LCO}}=- \sum_{\alpha\beta\gamma \mathbf{R}} \epsilon_{\alpha\beta\gamma} i \Delta_{\gamma}
			% 		\left( 
			% 		c^\cre_{\alpha}(\mathbf{R})c^\ann_{\beta}(\mathbf{R}+\mathbf{d}^{\mathrm{2NN}}_{\beta\gamma}) + c^\cre_{\alpha}(\mathbf{R})c^\ann_{\beta}(\mathbf{R}+\mathbf{d}^{\mathrm{2NN'}}_{\beta\gamma})
			% 		\right) 
			% 		\cos(\mathbf{Q}_{\gamma} \cdot \mathbf{R}) 
			% 		+ h.c. .
			% 	\end{aligned}
		% \end{equation}
	\begin{equation} 
		\begin{aligned} 
			H_{\mathrm{LCO}}=- \sum_{\alpha\beta\gamma \mathbf{R}} \epsilon_{\alpha\beta\gamma} i \Delta_{\gamma}
			\left( 
			c^\cre_{\alpha}(\mathbf{R}+\mathbf{d}^{\mathrm{2NN}}_{\beta\gamma})c^\ann_{\beta}(\mathbf{R}) + c^\cre_{\alpha}(\mathbf{R}+\mathbf{d}^{\mathrm{2NN'}}_{\beta\gamma})c^\ann_{\beta}(\mathbf{R})
			\right) 
			\cos(\mathbf{Q}_{\gamma} \cdot \mathbf{R}) 
			+ h.c. \, .
		\end{aligned}
	\end{equation}
	Projecting onto patches near the three Van Hove points, the effective theory in momentum space reads
	\begin{equation}
		H_{\mathrm{int}}=-\frac{g}{4} \sum_{\alpha>\beta } 
		(c^\cre_{\alpha} c^\ann_{\beta} -c^\cre_{\beta} c^\ann_{\alpha}) 
		(c^\cre_{\beta} c^\ann_{\alpha}-c^\cre_{\alpha} c^\ann_{\beta} ) \, ,
	\end{equation}
	where summations over momentum $|\mathbf{k}|< \Lambda_c$ within each patch are not written explicitly. Using a Hubbard-Stratonovich transformation, the quartic interaction can be decoupled with a real order parameter field $\Delta_{\gamma}$ and the action takes the form
	\begin{equation}
		\begin{aligned}
			S=\int_{0}^{\beta} \dd \tau \sum_{\alpha} c^\cre_{\alpha}  (\partial_{\tau} + \epsilon_{\alpha} ) c^\ann_{\alpha} +  
			\sum_{\alpha\beta\gamma } i \epsilon_{\alpha\beta\gamma}  \Delta_{\gamma} c^\cre_{\alpha} c^\ann_{\beta} 
			+ g^{-1} \sum_{\gamma} \Delta_{\gamma}^2 
		\end{aligned}
	\end{equation}
	Imposing the static condition for the order parameter field $\Delta_{\gamma}(\tau)=\Delta_{\gamma}$,
	%and integrating the fermions out in the Matsubara frequency representation $c^\ann_{\alpha} =\frac{1}{\beta} c^\ann_{n} e^{i\omega_{n}\tau} $ 
	%and integrate out the fermionic field in Matsubara frequency representation leads to  mean field free energy
	the action in Matsubara frequency representation reads
	\begin{equation}
		\begin{aligned}
			S= -T \sum_{\omega_{n}} \sum_{\alpha} c^\cre_{\alpha} (\omega_{n}) G_{\alpha} (\omega_{n})  c^\ann_{\alpha} (\omega_{n}) +  
			T \sum_{\alpha\beta\gamma }  i \epsilon_{\alpha\beta\gamma}  \Delta_{\gamma} c^\cre_{\alpha} (\omega_{n}) c^\ann_{\beta} (\omega_{n})
			+ \frac{T}{g} \sum_{\gamma} \Delta_{\gamma}^2 
		\end{aligned}
	\end{equation}
	with the bare Green's function $G_{\alpha}(\omega_{n})=1/(i\omega_n-\epsilon_{\alpha})$.
	
	\begin{figure}[t]
		\centering
		\includegraphics[width=0.6\linewidth]{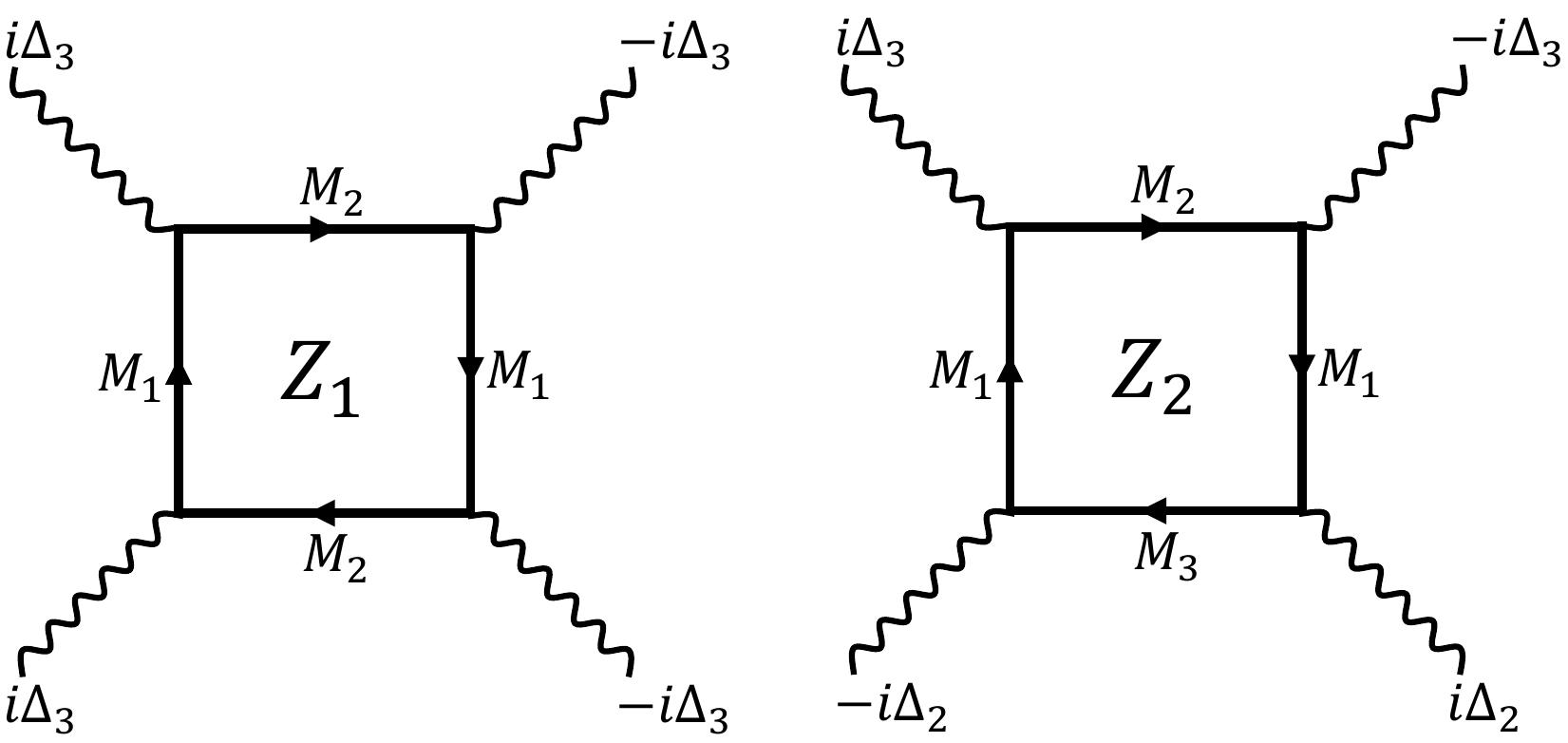}
		11	\caption{Diagrammatical representation of fourth order Ginzburg-Landau free energy coefficient in the patch model.}
		\label{fig:GLcoe}
	\end{figure}
	
	In the spinor basis $\psi=(c^\ann_1,c^\ann_2,c^\ann_3)^{\mathrm{T}}$, the action can be written as 
	\begin{equation}
		\begin{aligned}
			S= -T \sum_{\omega_{n}}  \psi^{\dagger} (\omega_{n}) \mathcal{G} ^{-1}(\omega_{n})  \psi (\omega_{n})
			+ \frac{T}{g} |\mathbf{\Delta}|^2 \, .
		\end{aligned}
	\end{equation}
	Here $\mathcal{G}^{-1}=G^{-1} - \hat{\Sigma} $ is the full single particle Green function where
	$G=\mathrm{diag}(G_1,G_2,G_3)$ and
	\begin{equation}
		\begin{aligned}
			\hat{\Sigma}=\left(\begin{array}{ccc}
				0 & i \Delta_3 & -i \Delta_2 \\
				-i \Delta_3 & 0 & i \Delta_1 \\
				i \Delta_2 & -i \Delta_1 & 0
			\end{array}\right) \, .
		\end{aligned}
	\end{equation}
	%Here $\mathcal{G}^{-1}=G^{-1} - \Sigma $ , $G_{\alpha}=1/(i\omega_n-\epsilon_{\alpha})$ is the bare Green's function and $\hat{\Sigma}$
	%where the inverse loop current propagator is defined
	%\begin{equation}
	%	\begin{aligned}
		%		\mathcal{G}^{-1}=\left(\begin{array}{ccc}
			%			G_1^{-1} & -i \Delta_3 & i \Delta_2 \\
			%			i \Delta_3 & G_2^{-1} & -i \Delta_1 \\
			%			-i \Delta_2 & i \Delta_1 & G_3^{-1}
			%		\end{array}\right)
		%	\end{aligned}
	%\end{equation}
	Integrating out the fermionic fields leads to mean field free energy
	\begin{equation}
		\begin{aligned}
			f= g^{-1} |\mathbf{\Delta}|^2 -\operatorname{Tr} \ln \left(-\mathcal{G}^{-1}\right) \, .
		\end{aligned}
	\end{equation}
	In the vicinity of phase transition, the order parameter is small and we can
	perturbatively expand the free energy in $\hat{\Sigma}$. With the constant
	part ignored, the expansion to fourth order reads
	\begin{equation}
		\begin{aligned}
			f&=  \frac{1}{g}|\mathbf{\Delta}|^2 + \frac{1}{2} \operatorname{Tr}\left(G \hat{\Sigma}\right)^2+\frac{1}{4} \operatorname{Tr}\left(G \hat{\Sigma}\right)^4 =f^{(2)}+ f^{(4)} \, .
		\end{aligned}
	\end{equation}
	The quadratic $f^{(2)}$ term reads
	\begin{equation}
		\begin{aligned}
			f^{(2)}= \frac{|\mathbf{\Delta}|^2}{g} + \operatorname{Tr}(G_1G_2) |\mathbf{\Delta}|^2 =\chi^{(2)} |\mathbf{\Delta}|^2 \, ,
		\end{aligned}
	\end{equation}
	where the coefficient 
	\begin{equation}
		\begin{aligned}
			\chi^{(2)}(T) = \frac{1}{g} +  \operatorname{Tr}(G_1G_2) =\frac{1}{2}\left( \frac{1}{g} -  \chi^{\text{ph}}(T)\right) 
		\end{aligned}
	\end{equation}
	is proportional to $T-T_c$ near $T_c$.
	The quartic term reads
	\begin{equation}
		\begin{aligned}
			f^{(4)}&=\frac{1}{2} \left( \operatorname{Tr}(G_1^2G_2^2) (\Delta_1^4 +\Delta_2^4 +\Delta_3^4) + \operatorname{Tr}(G_1G_2G_3^2 ) (\Delta_1^2\Delta_2^2 +\Delta_2^2\Delta_3^2 +\Delta_3^2\Delta_1^2) \right) \\
			&= \frac{1}{2} (Z_1 |\mathbf{\Delta}|^4 +2(Z_2-Z_1)(\Delta_1^2\Delta_2^2 +\Delta_2^2\Delta_3^2 +\Delta_3^2\Delta_1^2) )
		\end{aligned}
	\end{equation}
	with the coefficients given by $Z_1 = \operatorname{Tr}(G_1^2G_2^2)$ and $Z_2 = \operatorname{Tr}(G_1^2G_2G_3)$. Their diagrammatic representations are shown in Fig.~\ref{fig:GLcoe}. 
	Direct calculation of $Z_1$ and $Z_2$  in the patch model shows that $Z_1>Z_2>0$~\cite{SM_Nandkishore2012}. Since $Z_1-Z_2>0$, the second term in $f^{(4)}$ promotes the formation of a $3\mathbf{Q}$ LCO with $|\Delta_{\alpha}|=\Delta(\alpha=1,2,3)$.
	Additional to the $Z_4$ translational symmetry breaking, this leaves a fourfold space of degenerate ground states. The system will choose one of the four states $\Delta(1,1,1), \Delta(-1,1,1),
	\Delta(1,-1,1),\Delta(1,1,-1)$ spontaneously upon entering the symmetry broken phase.
	
	On the other hand, we can obtain the renormalized quasiparticle bands in the LCO phase as shown in Fig.~\ref{fig:gap}. Apparently the $3\mathbf{Q}$ LCO fully gaps out the Fermi surface and is hence energetically favored. Moreover, the $3\mathbf{Q}$ LCO breaks time reversal symmetry thus forming a Chern insulator with quantum anomalous hall response. Indeed, the highest valence band features a non-zero Chern number $C=1$ for 2nn $3\mathbf{Q}$ LCO. 
	
	%\begin{equation}
	%	f=Z|\Delta|^2+Z_1|\Delta|^4+2\left(Z_2-Z_1\right)\left(\Delta_1^2 \Delta_2^2+\Delta_2^2 \Delta_3^2+\Delta_3^2 \Delta_1^2\right)
	%\end{equation}
	
	% \subsection{Chern number of loop current band}
	% \subsection{Orbital magnetic moment orbital magnetization of LCO}
	
	\begin{figure}[t]
		\centering
		\includegraphics[width=0.9\linewidth]{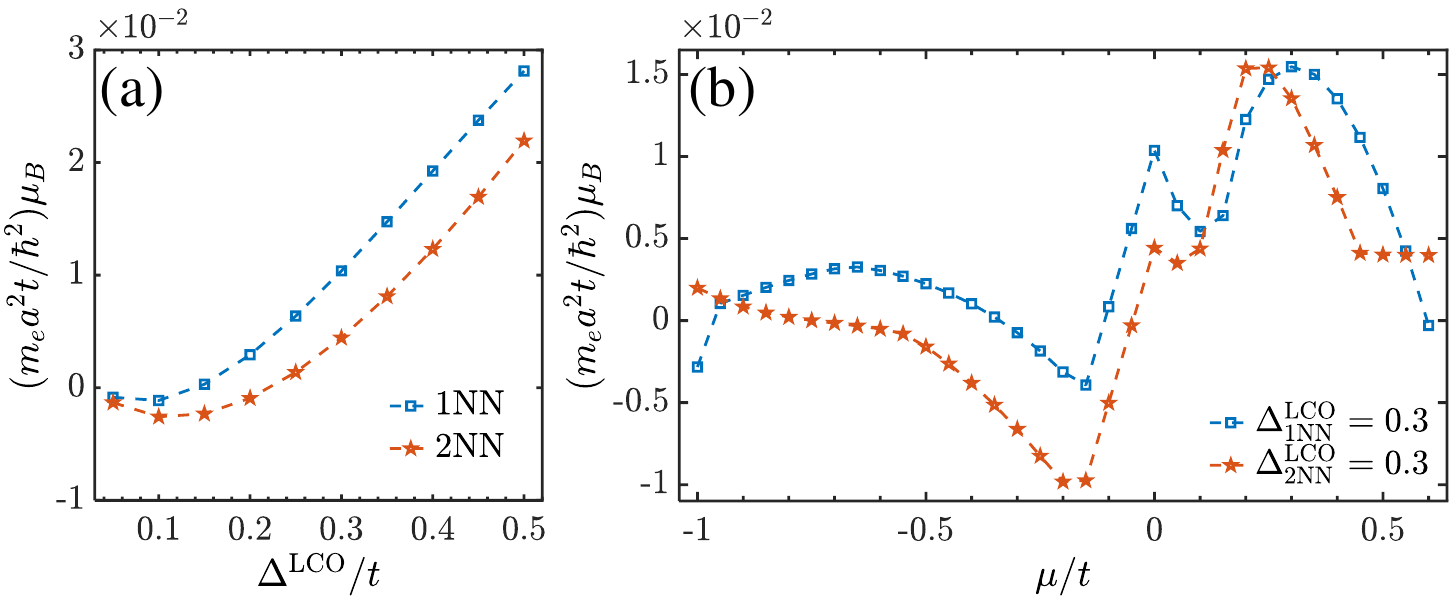}
		\caption{Orbital magnetization in unit of $\frac{m_ea^2t}{\hbar^2}\mu_B$ per Kagome site as a function of LCO order parameter on 1NN and 2NN without doping (b) chemical potential $\mu$ with doping into Chern band. For $\mathrm{AV}_3\mathrm{Sb}_5$ with the lattice constant $a=5.4\si{\angstrom}$ and $t=0.5$eV , $\frac{m_ea^2t}{\hbar^2}\mu_B \approx 1.9\mu_B$ and for $\mathrm{FeGe}$ with $a=5\si{\angstrom}$ and $t=0.3$eV, $\frac{m_ea^2t}{\hbar^2}\mu_B \approx \mu_B$. }
		\label{fig:MZ}
	\end{figure}
	
	\subsection{Orbital magnetization of LCO}\label{supp:subsec:lco-mag}
	The loop current state breaks time-reversal symmetry, which induces an orbital magnetic moment of the $n$th band at momentum $\mathbf{k}$~\cite{SM_RevModPhys.82.1959}
	\begin{equation}
		\begin{aligned}
			\mathbf{m}_n(\mathbf{k})=-i\frac e{2\hbar}\langle\nabla_{\mathbf{k}}u_{n\mathbf{k}}|\times(H_{\mathbf{k}}-\epsilon_{n\mathbf{k}})|\nabla_{\mathbf{k}}u_{n\mathbf{k}}\rangle ,
		\end{aligned}
	\end{equation}
	where $u_{n\mathbf{k}}$ is the periodic part of the Bloch wave function, $H_{\mathbf{k}}$ is the Hamiltonian of the system and $\epsilon_{n\mathbf{k}}$ is the energy dispersion of $n$th band. The orbital magnetic moment originates from a self-rotation of a wave packet around its center of mass and will couple to an external magnetic field. Besides the thermodynamic average of the orbital moment, the orbital magnetization $\mathbf{M}$ has another contribution form the center-of-mass motion of wave packet. The total contribution is given by~\cite{SM_RevModPhys.82.1959, SM_Zhou2022}
	\begin{equation}
		\mathbf{M}=\sum_n\int_{\mathrm{BZ}}\frac{d\mathbf{k}}{\left(2\pi\right)^2}\left[\mathbf{m}_n(\mathbf{k})-\frac e\hbar\epsilon_{n\mathbf{k}}\Omega_n(\mathbf{k})\right]f(\epsilon_{n\mathbf{k}}).
	\end{equation}
	where $\Omega_n(\mathbf{k})=i\langle\nabla_\mathbf{k}u_{n\mathbf{k}}|\times|\nabla_\mathbf{k}u_{n\mathbf{k}}\rangle $ is the Berry curvature of $n$th band and $f$ is the Fermi function.
	% For 2NN LCO state with $\Delta_{\mathrm{2NN}}^{\mathrm{LCO}}=0.3t$ and $t=0.5$eV, the  toatal orbital magnetization of filled bands is calculated to be 
	% % $\mathbf{M}\simeq 0.015 \frac{et}{4\hbar} = 5\times 10^{-4}\mu_B/(\si{\angstrom})^2$.
	% $\mathbf{M}\simeq 0.004 et/\hbar = 5\times 10^{-4}\mu_B/(\si{\angstrom})^2$. The orbital magnetization changes with LCO strength as shown in Fig. \ref{fig:MZ}.
	The calculated orbital magnetizations as function of gap size and doping are shown in Fig.~\ref{fig:MZ}. The orbital magnetic moment increases with increasing order parameter of LCO and varies with the evolution of chemical potential. The moment is much weaker than the local magnetic moment from the spin. By adopting experimental lattice parameters for AV$_3$Sb$_5$ and FeGe, we find that the LCO-induced orbital moment can be up to 0.03 $\mu_{\text{B}}$ per site. This calculated magnitude offers a compelling microscopic explanation for the magnetic moment enhancement observed across the CDW transition for FeGe ~\cite{SM_TengXK2022}.
	
	\section{Ginzburg-Landau analysis for nCDW}\label{supp:anchor:ncdw-gl}\label{supp:sec:ncdw-gl}
	The intra-unit cell nCDW state is two fold degenerate and corresponds to the two dimensional irreducible representation $E_2$ of $C^\ann_{6v}$. The order parameter can be written as $\Delta=\eta_1 d_1 + \eta_2 d_2$ where $\eta_{1,2}$ are two real order parameters with formfactor $d_1=(2,-1,-1)/\sqrt{6}$ and $d_2=(0,1,-1)\sqrt{2}$ respectively. Using the invariant term form the decomposition of products $E_2 \otimes E_2$ and $E_2 \otimes E_2 \otimes E_2$, the Ginzburg-Landau expansion of the free energy up to cubic term reads~\cite{SM_Sigrist1991}
	\begin{equation}
		\begin{aligned}
			f=\alpha (\eta_1^2 +\eta_2^2) +\beta (\eta_1,\eta_2) \cdot (\eta_1^2-\eta_2^2,2\eta_1\eta_2)^{T} 
			=\alpha (\eta_1^2 +\eta_2^2) +
			\beta(\eta_1^3-3\eta_1\eta_2^2) \, .
			% \beta(3\eta_1^2\eta_2-\eta_2^3) \, .
		\end{aligned}
	\end{equation}
	Letting $(\eta_1, \eta_2 ) = \eta_0(\cos \theta, \sin \theta)$, the cubic term in free energy is written as $\beta\eta_0^2\cos(3\theta)$ which is minimized by $\theta=2n\pi/3$ for $\beta<0$ and $\theta=(2n+1)\pi/3$ for $\beta>0$. 
	Then the system chooses one out of three equivalent gro
    und states $(2,-1,-1)/\sqrt{6}, (-1,2,-1)/\sqrt{6},
	(-1,-1,2)/\sqrt{6}$ that break lattice $C^\ann_3$ rotational symmetry and thus represent a three state Potts nematic order.
	
	\section{Enhancement of nCDW through SC channel fluctuations for weak $V_2$ }\label{supp:anchor:ncdw-enhance}\label{supp:sec:ncdw-enhance}
	\begin{figure}[t]
		\centering
		\includegraphics[width=0.7\columnwidth]{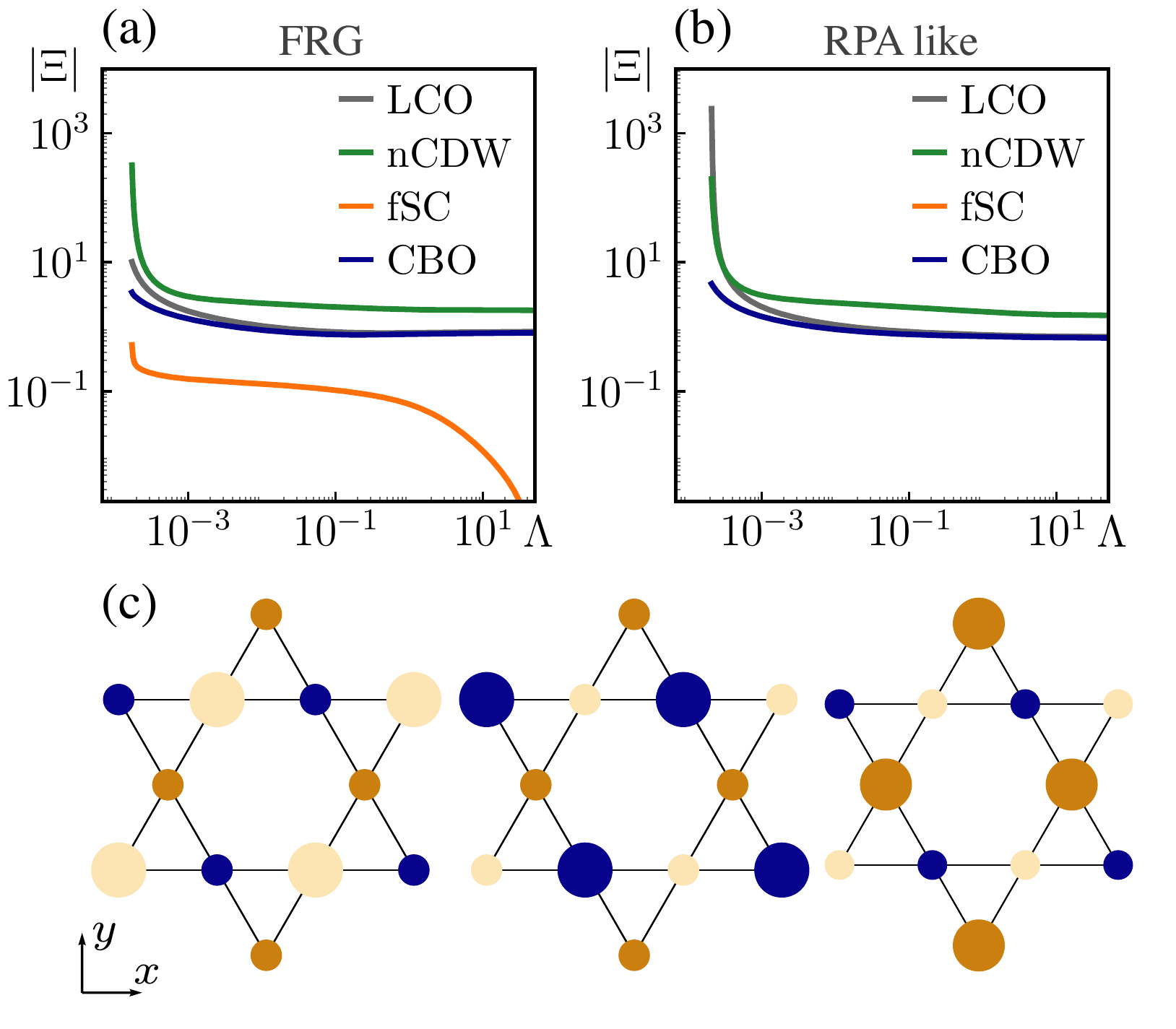}
		\caption{ Nematic CDW in the weak coupling regime. (a) Expectation value flow of nematic CDW, fSC, 2nn CBO and 2nn LCO with $V_1=0$ and $V_2=0.6t$. (b) Flow of (a) without cross-channel projections between particle-particle and particle-hole channels. Here, the 2nn LCO emerges as the leading instability, aligning with RPA results. This underscore the importance of FRG achieving accurate many-body ground states. (c) Real-space pattern for the three different nematic CDWs that minimizes the free energy, with the sizes of circles denoting charge density values.
			% (d)Expectation value flow of the leading charge instabilities without coupling to the pairing channel for $V_1=0$ and $V_2=0.5t$.
		}
		\label{fig:CDW}
	\end{figure}

	%Surprisingly, as shown in Fig.~\ref{fig:CDW}(a), lowering the interaction scale triggers a transition from the LCO to a reentrant two-fold CDW order at $\bm{q}=0$, corresponding to the two dimensional $E_{2g}$ irrep of $D_{6h}$.
	%The minimization of the free energy within the associated two dimensional eigenspace leads to a nematic order that breaks the six-fold rotational symmetry. The three possible configurations of this nematic order are depicted Fig.~\ref{fig:CDW}(c), where the sizes of circles represent the charge occupation.
	%This state is identical to classical solution where the repulsive interactions are significiant.  
	
	To understand the emergence of nCDW in the weak $V_2$ regime, we perform FRG
	calculations by treating pairing fluctuations and particle-hole excitations
	independently, \textit{i.e.} switching off the cross-channel projection between
	the particle-particle and particle-hole channels.
	The obtained flows, shown in Fig.~\ref{fig:CDW}(b), reveal a close competition between
	the LCO and nCDW, with the LCO marginally prevailing, thus recovering
	previously obtained RPA results~\cite{SM_RPA_paper}.
	Comparing these with the full FRG calculations in Fig.~\ref{fig:CDW}(a) poses
	an intriguing scenario: Although the superconducting fluctuations are
	weak, they become relevant at low cutoffs owing to the double logarithmic
	divergence in the particle-particle bubble. Their imprint on the particle-hole
	channels promotes the nCDW as the leading instability while
	simultaneously suppressing the LCO .
	This interplay is intimately related to the internal sublattice structure of
	fluctuations, which involves intra-sublattice channels in nCDW and $f$-wave
	superconductivity cases, as opposed to inter-sublattice channels for the LCO.

	%\subsection{Analytical analysis}

\begin{figure}[t]
		\centering
		\includegraphics[width=0.8\linewidth]{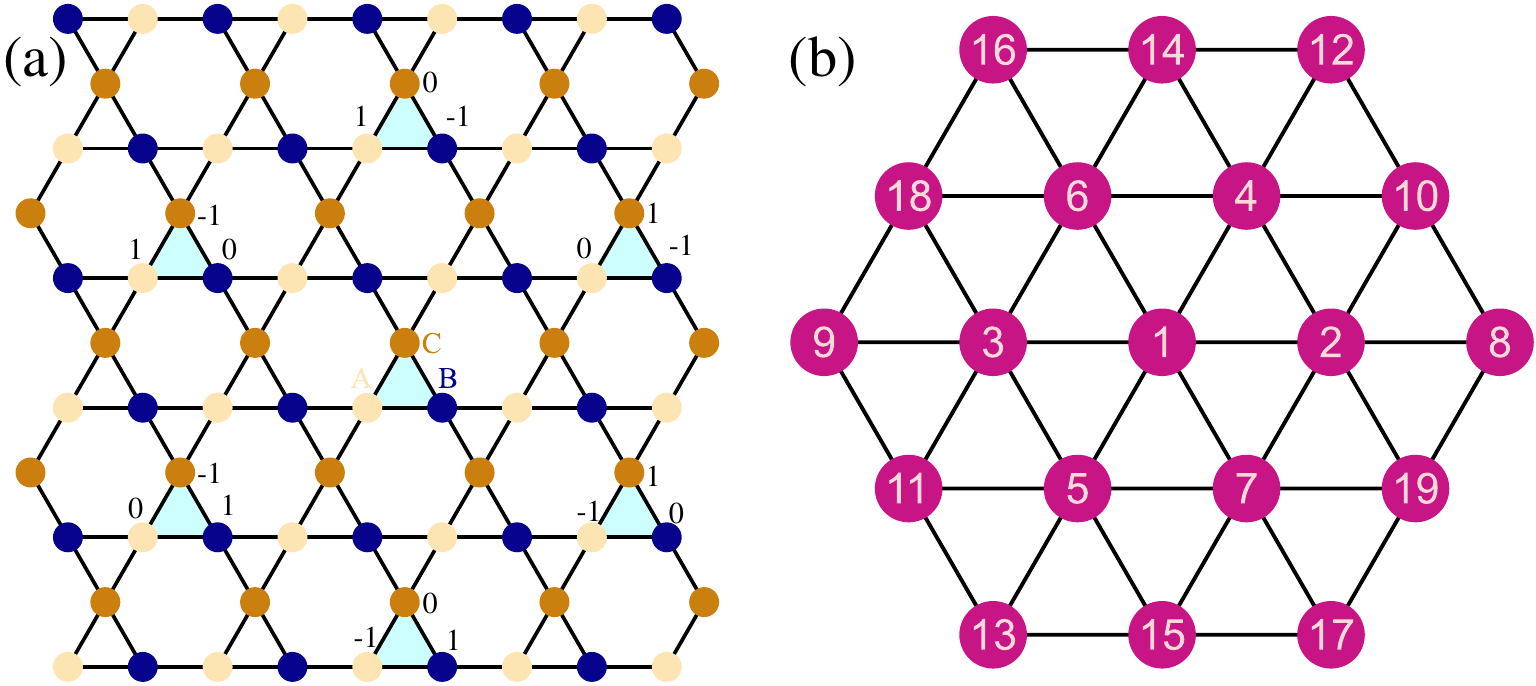}
		\caption{(a)Real space $B_2$ paring on 6NN bonds with $f_{x^3-3xy^2}$ symmetry. (b) Numbering of intra-sublattice formfactors}
		\label{fig:Fig6NNfSC}
	\end{figure}

	In the nCDW phase, the dominant fluctuation in the particle-particle channel is along intra-sublattice bonds with $f_{x^3-3xy^2}$ symmetry. 
	Here we take $6$NN $f$-wave pair fluctuation as an example and analyze its feedback to particle-hole channels. 
	We start by getting real space $6$NN $f$-wave formfactors which transform as $B_2$ representation of $C^\ann_{6v}$ using projection operators of irreducible representation decomposition (Fig.~\ref{fig:Fig6NNfSC}), 
	
	\begin{equation}
		\begin{aligned}
			\Phi_{AA11}&=-\Phi_{AA10}=0,\Phi_{AA14}=-\Phi_{AA15}=\Delta,\Phi_{AA19}=-\Phi_{AA18}=-\Delta\\\Phi_{BB11}&=-\Phi_{BB10}=\Delta,\Phi_{BB14}=-\Phi_{BB15}=-\Delta,\Phi_{CC19}=-\Phi_{CC18}=0\\\Phi_{CC11}&=-\Phi_{CC10}=-\Delta,\Phi_{CC14}=-\Phi_{CC15}=0,\Phi_{CC19}=-\Phi_{CC18}=\Delta \, .
		\end{aligned}
	\end{equation}
	The $\mathrm{P}$ vertix forming 6NN $B_2$ $\mathbf{Q}=\mathbf{0}$ SC fluctuations is given by $\tilde{P}_{o_1o_2\mathbf{R}_m,o_3o_4\mathbf{R}_n}(\mathbf{R})=-\Phi_{o_1o_2m}\Phi_{o_3o_4n}^*$. Using the $P$ to $D$ channel projection
	\begin{equation}
		\begin{aligned}
			V_{o_1o_3m,o_4o_2n}^{\mathrm{D} \leftarrow \mathrm{P}}(\mathbf{q})&=\sum_l \tilde{P}_{o_1o_2 \mathbf{R}_l ,o_3o_4 \mathbf{R}_l-\mathbf{R}_m-\mathbf{R}_n}\left(-\mathbf{R}_m\right) e^{i\left(\mathbf{R}_n-\mathbf{R}_l\right) \cdot \mathbf{q}}
			=-\sum_{l}\Phi_{o_1o_2m}\Phi_{o_3o_4n}^* e^{i\left(\mathbf{R}_n-\mathbf{R}_l\right) \cdot \mathbf{q}}
		\end{aligned}
	\end{equation}
	the effective interaction for the on-site CDW can be constructed as
	\begin{equation}
		\begin{aligned}
			V^{\mathrm{CDW}}_{AA1,AA1}&=\sum_l\Phi_{AA,l}\Phi_{AA,l}^*=4\Delta^2,V^{\mathrm{nCDW}}_{AA1,BB1}=\sum_l\Phi_{AB,l}\Phi_{BA,l}^*=0 \, .
		\end{aligned}
	\end{equation}
	In the basis $\big(c^\cre_{A}(\mathbf{R})c^\ann_{A}(\mathbf{R}) ,
	c^\cre_{B}(\mathbf{R})c^\ann_{B}(\mathbf{R}),
	c^\cre_{C}(\mathbf{R})c^\ann_{C}(\mathbf{R}) \big)$, the paring matrix for nCDW is written as
	\begin{equation}
		\begin{aligned}
			M^{\mathrm{CDW}}=4\Delta^2\begin{pmatrix}1&0&0\\0&1&0\\0&0&1\end{pmatrix} \, .
		\end{aligned}
	\end{equation}
	By diagonalizing $M^{\mathrm{CDW}}$, the effective interaction for nCDW is identified as two fold degenerate eigenvalue $\lambda=4\Delta^2$ with eigenvectors $d_1=(2,-1,-1)/\sqrt{6}$ and $d_2=(0,1,-1)\sqrt{2}$.
	
	Similarly, we can project the P vertices onto the 1NN bond order channel. Non-zero matrix elements are \textit{e.g.}
	\begin{equation}
		\begin{aligned}
			V_{AB1,BA1}=\sum_l\Phi_{AA,l}\Phi_{BB,l}^*e^{i(-\mathbf{R}_l+\mathbf{R}_1)\cdot\mathbf{M}_c}=-2\Delta^2,
			V_{AB2,BA3}=\sum_l\Phi_{AA,l}\Phi_{BB,l}^*e^{i(-\mathbf{R}_l+\mathbf{R}_3)\cdot\mathbf{M}_c}=-2\Delta^2 \, .
		\end{aligned}
	\end{equation}
	Then the paring matrix for for 1NN $A-B$ bond order is written as
	\begin{equation}
		\begin{aligned}
			M^{\mathrm{1NNAB}}=\Delta^2\begin{pmatrix}0&-2&0&0\\-2&0&0&0\\0&0&0&-2\\0&0&-2&0\end{pmatrix}
		\end{aligned}
	\end{equation}
	with the basis $\big( c^\cre_{A}(\mathbf{R})c^\ann_{B}(\mathbf{R}) ,
	c^\cre_{B}(\mathbf{R})c^\ann_{A}(\mathbf{R}),
	c^\cre_{A}(\mathbf{R}+\mathbf{R}_2)c^\ann_{B}(\mathbf{R}),
	c^\cre_{B}(\mathbf{R}+\mathbf{R}_3)c^\ann_{A}(\mathbf{R}) \big)$.
	Diagonalizing $M^{\mathrm{1NNAB}}$, the effective interaction for 1NN LCO along the AB bond direction exhibits a two fold degenerate eigenvalue $\lambda=2\Delta^2$. The two eigenvectors $(1,-1,1,-1)/2$ and $(1,-1,-1,1)/2$ correspond to symmetrized and anti-symmetrized LCO respectively.
	
	For the 2NN bond order channel, non-zero matrix elements are given by
	\begin{equation}
		\begin{aligned}
			V_{AB4,BA5}=\sum_l\Phi_{AA,l}\Phi_{BB,l}^*e^{i(-\mathbf{R}_l+\mathbf{R}_5)\cdot\mathbf{M}_c}=2\Delta^2,
			V_{AB7,BA6}=\sum_l\Phi_{AA,l}\Phi_{BB,l}^*e^{i(-\mathbf{R}_l+\mathbf{R}_6)\cdot\mathbf{M}_c}=2\Delta^2
		\end{aligned}
	\end{equation}
	and the paring matrix for for 2NN $A-B$ bond order is written as
	\begin{equation}
		\begin{aligned}
			M^{\mathrm{2NNAB}}=\Delta^2\begin{pmatrix}0&2&0&0\\2&0&0&0\\0&0&0&2\\0&0&2&0\end{pmatrix}
		\end{aligned}
	\end{equation}
	in the basis $\big( c^\cre_{A}(\mathbf{R}+\mathbf{R}_4))c^\ann_{B}(\mathbf{R}) ,
	c^\cre_{B}(\mathbf{R}+\mathbf{R}_5))c^\ann_{A}(\mathbf{R}),
	c^\cre_{A}(\mathbf{R}+\mathbf{R}_7)c^\ann_{B}(\mathbf{R}),
	c^\cre_{B}(\mathbf{R}+\mathbf{R}_6)c^\ann_{A}(\mathbf{R}) \big)$.
	Diagonalizing $M^{\mathrm{2NNAB}}$, the effective interaction for 2NN LCO along AB bond direction is identified as two fold degenerate eigenvalue $\lambda=2\Delta^2$. The two eigenvectors $(1,1,1,1)/2$ and $(1,1,-1,-1)/2$ correspond to symmetrized and anti-symmetrized LCO respectively.
	
	Then from 6NN $B_2$ f-wave SC fluctuation, the nCDW enhancement is twice of LCO. This will affect the competition between nCDW and LCO and finally elevates the nCDW to exceed LCO in the small $V_2$ regime.

\begin{figure}[t]
	\centering
	\includegraphics[width=0.8\linewidth]{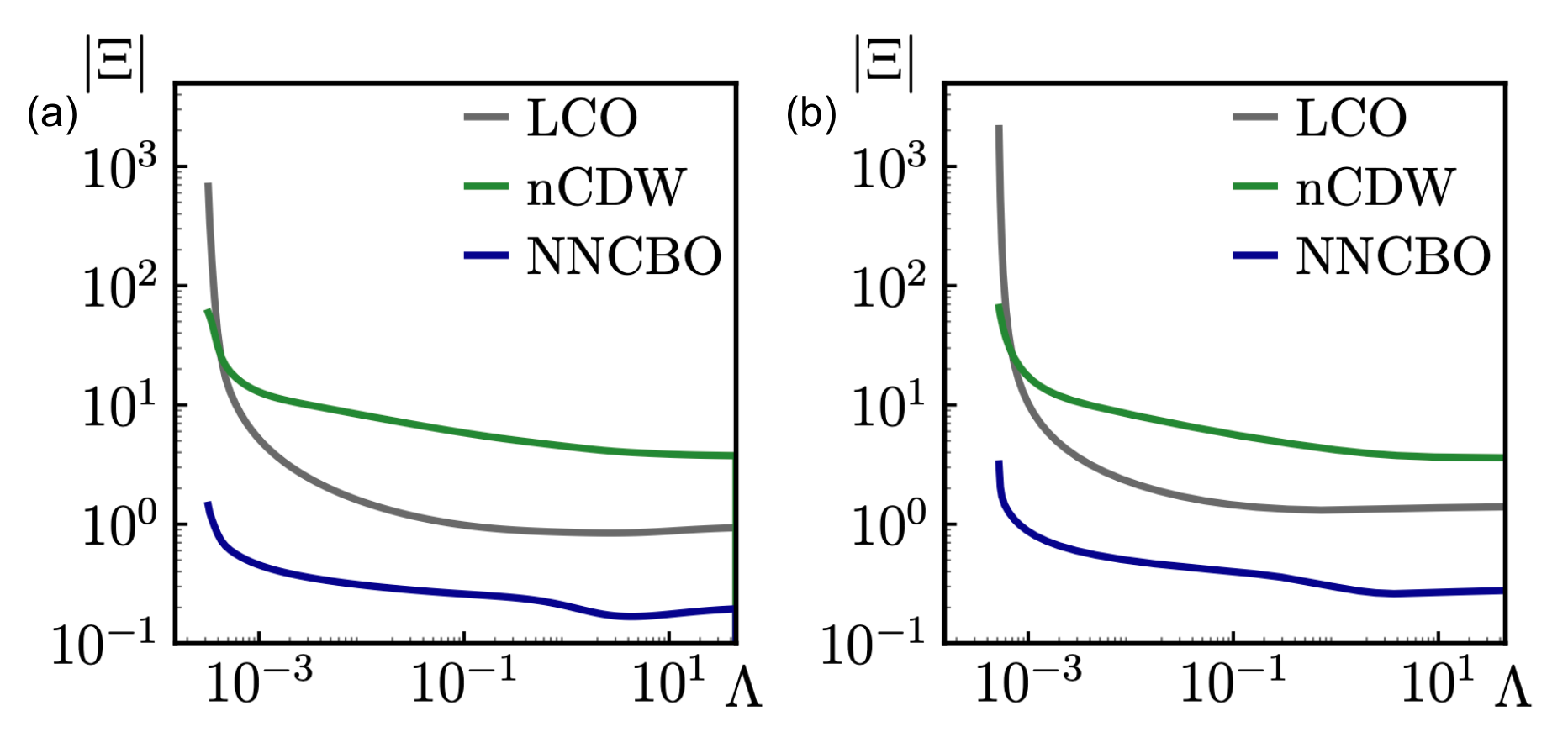}
	\caption{Representative FRG flows of the expectation values of LCO, nCDW and 1nn CBO for the spinful case with a Zeeman splitting of 6t (a) and the spinless case (b). In the spinful case, the onsite interaction is $U=t$. The nonlocal interactions in both cases are $V_1=0.2t, V_2=t$.}
	\label{fig:spinfulFRG}
\end{figure}

	\section{LCO in spinful model with Zeeman splitting}\label{supp:anchor:spinful}\label{supp:sec:spinful}
%	To demonstrate that our spinless model is relevant for FeGe, we further study spinful model with ferromagnetic spin splinting,
%	\begin{equation}
%		\begin{split}
%			H=&-t\sum_{\langle i, j\rangle \sigma}  (c_{i\sigma}^{\dagger} c_{j\sigma} + \text{h.c.})  -\mu\sum_{i}n_{i} + B\sum_{i}(n_{i\uparrow}-n_{i\downarrow}) \\  
%			+&U\sum_{i}n_{i\uparrow}n_{i\downarrow}+V_1\sum_{\langle i, j\rangle } n_i n_j + V_2 \sum_{\langle\langle i, j\rangle \rangle } n_i n_j,
%		\end{split}
%%		\label{eqn:hamiltonian}
%	\end{equation} 
%	where $n_{i}=n_{i\uparrow}+n_{i\downarrow}$, $B$ is energy splitting of spin up and down band. Fig. \ref{fig:spinfulFRG}(a) shows the FRG results with large spin splitting $B=-6t$ and Fermi level at p-type Van Hove filling of spin up band. With moderate $V_2$, the leading instability of system is LCO as shown in the FRG flow. As a comparison, we also show the result for the effectiove spinless model with same nolocal interaction strength in  Fig. \ref{fig:spinfulFRG}(b). The corresponding FRG flow is similar to the spinful case, which shows that effective spinless model is equivalent to spinful model with with large spin splitting at low energy.

	To demonstrate the relevance of our spinless model to FeGe, we further study a spinful model with ferromagnetic spin splitting as provided by the stacked ferromagnetic order below the antiferromagnetic transition temperature of $T_{\text{N}}\sim 400$K in FeGe~\cite{SM_TengXK2022},
%	\begin{equation}
%		\begin{split}
%			H&=-t\sum_{\langle i, j\rangle \sigma}  (c_{i\sigma}^{\dagger} c_{j\sigma} + \text{h.c.})  -\mu\sum_{i}n_{i} +\frac{B}{2}\sum_{i}(n_{i\uparrow}-n_{i\downarrow}) \\  
%			&\quad+U\sum_{i}n_{i\uparrow}n_{i\downarrow}+V_1\sum_{\langle i, j\rangle } n_i n_j + V_2 \sum_{\langle\langle i, j\rangle \rangle } n_i n_j,
%		\end{split}
%		%		\label{eqn:hamiltonian}
%	\end{equation} 
	\begin{equation}
			H=-t\sum_{\langle i, j\rangle \sigma}  (c_{i\sigma}^{\dagger} c_{j\sigma} + \text{h.c.})  -\mu\sum_{i}n_{i} +B\sum_{i}(n_{i\uparrow}-n_{i\downarrow}) 
			+U\sum_{i}n_{i\uparrow}n_{i\downarrow}+V_1\sum_{\langle i, j\rangle } n_i n_j + V_2 \sum_{\langle\langle i, j\rangle \rangle } n_i n_j,
	\end{equation} 
	where $n_{i}=n_{i\uparrow}+n_{i\downarrow}$ and $B$ is Zeeman splitting of spin up and down bands. Fig. \ref{fig:spinfulFRG}(a) shows the FRG result for a large spin splitting $B=-6t$, with the Fermi level set at the p-type Van Hove filling of the spin-up band. With moderate $V_2$, the leading instability of the system is still a LCO in the spin majority sector, as indicated by the FRG flow.
	
	For comparison, we also present the result of the effective spinless model with the same nonlocal interaction strengths in Fig. \ref{fig:spinfulFRG}(b). The corresponding FRG flow closely resembles that of the spinful case, demonstrating that the effective spinless model captures the low-energy physics of the spinful model in the presence of large spin splitting. Furthermore, the LCO phase remains robust even for smaller spin splittings down to $B=-2.5t$, provided that the Fermi level is located at the p-type Van Hove filling of the spin-split band. This scale is consistent with band splitting in realistic kagome magnets like FeGe, strongly reinforcing its direct relevance to the physics observed in this material family. 
	
%	 These findings indicate that our spinless model is directly relevant to magnetic materials such as FeGe.
%	Our results extend to moderate spin splittings below B=-2.5t observed  in ARPES measurements of the spin polarized bandstructure of  FeGe close to the charge ordering temperature~\cite{}.

%	\begin{figure}[t]
%		\centering
%		\includegraphics[width=0.8\linewidth]{frg_V2.pdf}
%		\caption{FRG expectation value flow of nematic CDW, 1nn CBO and LCO for (a)spinful model with $U=1t, V_1=0.2t, V_2=1t, B=-6t$ and (b) spinless model with $V_1=0.2t, V_2=1t$.}
%		\label{fig:spinfulFRG}
%	\end{figure}
%	\begin{figure}[t]
%		\centering
%		\includegraphics[width=0.8\linewidth]{frg_V2.pdf}
%		\caption{Representative FRG flows of the expectation values of LCO, nCDW and 1nn CBO for the spinful case with a Zeeman splitting of 6t (a) and the spinless case (b). In the spinful case, the onsite interaction is $U=t$. The nonlocal interactions in both cases are $V_1=0.2t, V_2=t$.}
%		\label{fig:spinfulFRG}
%	\end{figure}
	
	% \nocite{*}
	%\bibliographystyle{plain}

\phantomsection\label{supp:sec:refs}
\input{sm_refs.bbl}

%\vspace{1em}
%\noindent\textit{Equal contribution:} These authors contributed equally to this work.\par
%\noindent\textit{Corresponding authors:} \href{mailto:xxwu@itp.ac.cn}{xxwu@itp.ac.cn}; \href{mailto:jphu@iphy.ac.cn}{jphu@iphy.ac.cn}

\end{document}

%% file: sm_refs.bbl
%merlin.mbs apsrev4-1.bst 2010-07-25 4.21a (PWD, AO, DPC) hacked
%Control: key (0)
%Control: author (72) initials jnrlst
%Control: editor formatted (1) identically to author
%Control: production of article title (-1) disabled
%Control: page (0) single
%Control: year (1) truncated
%Control: production of eprint (0) enabled
%

%% file: main.bbl
\begin{thebibliography}{69}%
\makeatletter
\providecommand \@ifxundefined [1]{%
 \@ifx{#1\undefined}
}%
\providecommand \@ifnum [1]{%
 \ifnum #1\expandafter \@firstoftwo
 \else \expandafter \@secondoftwo
 \fi
}%
\providecommand \@ifx [1]{%
 \ifx #1\expandafter \@firstoftwo
 \else \expandafter \@secondoftwo
 \fi
}%
\providecommand \natexlab [1]{#1}%
\providecommand \enquote  [1]{``#1''}%
\providecommand \bibnamefont  [1]{#1}%
\providecommand \bibfnamefont [1]{#1}%
\providecommand \citenamefont [1]{#1}%
\providecommand \href@noop [0]{\@secondoftwo}%
\providecommand \href [0]{\begingroup \@sanitize@url \@href}%
\providecommand \@href[1]{\@@startlink{#1}\@@href}%
\providecommand \@@href[1]{\endgroup#1\@@endlink}%
\providecommand \@sanitize@url [0]{\catcode `\\12\catcode `\$12\catcode
  `\&12\catcode `\#12\catcode `\^12\catcode `\_12\catcode `\%12\relax}%
\providecommand \@@startlink[1]{}%
\providecommand \@@endlink[0]{}%
\providecommand \url  [0]{\begingroup\@sanitize@url \@url }%
\providecommand \@url [1]{\endgroup\@href {#1}{\urlprefix }}%
\providecommand \urlprefix  [0]{URL }%
\providecommand \Eprint [0]{\href }%
\providecommand \doibase [0]{http://dx.doi.org/}%
\providecommand \selectlanguage [0]{\@gobble}%
\providecommand \bibinfo  [0]{\@secondoftwo}%
\providecommand \bibfield  [0]{\@secondoftwo}%
\providecommand \translation [1]{[#1]}%
\providecommand \BibitemOpen [0]{}%
\providecommand \bibitemStop [0]{}%
\providecommand \bibitemNoStop [0]{.\EOS\space}%
\providecommand \EOS [0]{\spacefactor3000\relax}%
\providecommand \BibitemShut  [1]{\csname bibitem#1\endcsname}%
\let\auto@bib@innerbib\@empty
%</preamble>
\bibitem [{\citenamefont {Norman}(2016)}]{RevModPhys.88.041002}%
  \BibitemOpen
  \bibfield  {author} {\bibinfo {author} {\bibfnamefont {M.~R.}\ \bibnamefont
  {Norman}},\ }\href {\doibase 10.1103/RevModPhys.88.041002} {\bibfield
  {journal} {\bibinfo  {journal} {Rev. Mod. Phys.}\ }\textbf {\bibinfo {volume}
  {88}},\ \bibinfo {pages} {041002} (\bibinfo {year} {2016})}\BibitemShut
  {NoStop}%
\bibitem [{\citenamefont {Mielke}(1991)}]{AMielke_1991}%
  \BibitemOpen
  \bibfield  {author} {\bibinfo {author} {\bibfnamefont {A.}~\bibnamefont
  {Mielke}},\ }\href {\doibase 10.1088/0305-4470/24/2/005} {\bibfield
  {journal} {\bibinfo  {journal} {Journal of Physics A: Mathematical and
  General}\ }\textbf {\bibinfo {volume} {24}},\ \bibinfo {pages} {L73}
  (\bibinfo {year} {1991})}\BibitemShut {NoStop}%
\bibitem [{\citenamefont {Pollmann}\ \emph {et~al.}(2008)\citenamefont
  {Pollmann}, \citenamefont {Fulde},\ and\ \citenamefont
  {Shtengel}}]{PhysRevLett.100.136404}%
  \BibitemOpen
  \bibfield  {author} {\bibinfo {author} {\bibfnamefont {F.}~\bibnamefont
  {Pollmann}}, \bibinfo {author} {\bibfnamefont {P.}~\bibnamefont {Fulde}}, \
  and\ \bibinfo {author} {\bibfnamefont {K.}~\bibnamefont {Shtengel}},\ }\href
  {\doibase 10.1103/PhysRevLett.100.136404} {\bibfield  {journal} {\bibinfo
  {journal} {Phys. Rev. Lett.}\ }\textbf {\bibinfo {volume} {100}},\ \bibinfo
  {pages} {136404} (\bibinfo {year} {2008})}\BibitemShut {NoStop}%
\bibitem [{\citenamefont {Ye}\ \emph {et~al.}(2018)\citenamefont {Ye},
  \citenamefont {Kang}, \citenamefont {Liu}, \citenamefont {von Cube},
  \citenamefont {Wicker}, \citenamefont {Suzuki}, \citenamefont {Jozwiak},
  \citenamefont {Bostwick}, \citenamefont {Rotenberg}, \citenamefont {Bell},
  \citenamefont {Fu}, \citenamefont {Comin},\ and\ \citenamefont
  {Checkelsky}}]{ye2018massive}%
  \BibitemOpen
  \bibfield  {author} {\bibinfo {author} {\bibfnamefont {L.}~\bibnamefont
  {Ye}}, \bibinfo {author} {\bibfnamefont {M.}~\bibnamefont {Kang}}, \bibinfo
  {author} {\bibfnamefont {J.}~\bibnamefont {Liu}}, \bibinfo {author}
  {\bibfnamefont {F.}~\bibnamefont {von Cube}}, \bibinfo {author}
  {\bibfnamefont {C.~R.}\ \bibnamefont {Wicker}}, \bibinfo {author}
  {\bibfnamefont {T.}~\bibnamefont {Suzuki}}, \bibinfo {author} {\bibfnamefont
  {C.}~\bibnamefont {Jozwiak}}, \bibinfo {author} {\bibfnamefont
  {A.}~\bibnamefont {Bostwick}}, \bibinfo {author} {\bibfnamefont
  {E.}~\bibnamefont {Rotenberg}}, \bibinfo {author} {\bibfnamefont {D.~C.}\
  \bibnamefont {Bell}}, \bibinfo {author} {\bibfnamefont {L.}~\bibnamefont
  {Fu}}, \bibinfo {author} {\bibfnamefont {R.}~\bibnamefont {Comin}}, \ and\
  \bibinfo {author} {\bibfnamefont {J.~G.}\ \bibnamefont {Checkelsky}},\ }\href
  {\doibase 10.1038/nature25987} {\bibfield  {journal} {\bibinfo  {journal}
  {Nature}\ }\textbf {\bibinfo {volume} {555}},\ \bibinfo {pages} {638}
  (\bibinfo {year} {2018})}\BibitemShut {NoStop}%
\bibitem [{\citenamefont {Yin}\ \emph {et~al.}(2020)\citenamefont {Yin},
  \citenamefont {Ma}, \citenamefont {Cochran}, \citenamefont {Xu},
  \citenamefont {Zhang}, \citenamefont {Tien}, \citenamefont {Shumiya},
  \citenamefont {Cheng}, \citenamefont {Jiang}, \citenamefont {Lian},
  \citenamefont {Song}, \citenamefont {Chang}, \citenamefont {Belopolski},
  \citenamefont {Multer}, \citenamefont {Litskevich}, \citenamefont {Cheng},
  \citenamefont {Yang}, \citenamefont {Swidler}, \citenamefont {Zhou},
  \citenamefont {Lin}, \citenamefont {Neupert}, \citenamefont {Wang},
  \citenamefont {Yao}, \citenamefont {Chang}, \citenamefont {Jia},\ and\
  \citenamefont {Zahid~Hasan}}]{yin2020quantum}%
  \BibitemOpen
  \bibfield  {author} {\bibinfo {author} {\bibfnamefont {J.-X.}\ \bibnamefont
  {Yin}}, \bibinfo {author} {\bibfnamefont {W.}~\bibnamefont {Ma}}, \bibinfo
  {author} {\bibfnamefont {T.~A.}\ \bibnamefont {Cochran}}, \bibinfo {author}
  {\bibfnamefont {X.}~\bibnamefont {Xu}}, \bibinfo {author} {\bibfnamefont
  {S.~S.}\ \bibnamefont {Zhang}}, \bibinfo {author} {\bibfnamefont {H.-J.}\
  \bibnamefont {Tien}}, \bibinfo {author} {\bibfnamefont {N.}~\bibnamefont
  {Shumiya}}, \bibinfo {author} {\bibfnamefont {G.}~\bibnamefont {Cheng}},
  \bibinfo {author} {\bibfnamefont {K.}~\bibnamefont {Jiang}}, \bibinfo
  {author} {\bibfnamefont {B.}~\bibnamefont {Lian}}, \bibinfo {author}
  {\bibfnamefont {Z.}~\bibnamefont {Song}}, \bibinfo {author} {\bibfnamefont
  {G.}~\bibnamefont {Chang}}, \bibinfo {author} {\bibfnamefont
  {I.}~\bibnamefont {Belopolski}}, \bibinfo {author} {\bibfnamefont
  {D.}~\bibnamefont {Multer}}, \bibinfo {author} {\bibfnamefont
  {M.}~\bibnamefont {Litskevich}}, \bibinfo {author} {\bibfnamefont {Z.-J.}\
  \bibnamefont {Cheng}}, \bibinfo {author} {\bibfnamefont {X.~P.}\ \bibnamefont
  {Yang}}, \bibinfo {author} {\bibfnamefont {B.}~\bibnamefont {Swidler}},
  \bibinfo {author} {\bibfnamefont {H.}~\bibnamefont {Zhou}}, \bibinfo {author}
  {\bibfnamefont {H.}~\bibnamefont {Lin}}, \bibinfo {author} {\bibfnamefont
  {T.}~\bibnamefont {Neupert}}, \bibinfo {author} {\bibfnamefont
  {Z.}~\bibnamefont {Wang}}, \bibinfo {author} {\bibfnamefont {N.}~\bibnamefont
  {Yao}}, \bibinfo {author} {\bibfnamefont {T.-R.}\ \bibnamefont {Chang}},
  \bibinfo {author} {\bibfnamefont {S.}~\bibnamefont {Jia}}, \ and\ \bibinfo
  {author} {\bibfnamefont {M.}~\bibnamefont {Zahid~Hasan}},\ }\href {\doibase
  10.1038/s41586-020-2482-7} {\bibfield  {journal} {\bibinfo  {journal}
  {Nature}\ }\textbf {\bibinfo {volume} {583}},\ \bibinfo {pages} {533}
  (\bibinfo {year} {2020})}\BibitemShut {NoStop}%
\bibitem [{\citenamefont {Kang}\ \emph {et~al.}(2020)\citenamefont {Kang},
  \citenamefont {Ye}, \citenamefont {Fang}, \citenamefont {You}, \citenamefont
  {Levitan}, \citenamefont {Han}, \citenamefont {Facio}, \citenamefont
  {Jozwiak}, \citenamefont {Bostwick}, \citenamefont {Rotenberg}, \citenamefont
  {Chan}, \citenamefont {McDonald}, \citenamefont {Graf}, \citenamefont
  {Kaznatcheev}, \citenamefont {Vescovo}, \citenamefont {Bell}, \citenamefont
  {Kaxiras}, \citenamefont {van~den Brink}, \citenamefont {Richter},
  \citenamefont {Prasad~Ghimire}, \citenamefont {Checkelsky},\ and\
  \citenamefont {Comin}}]{kang2020dirac}%
  \BibitemOpen
  \bibfield  {author} {\bibinfo {author} {\bibfnamefont {M.}~\bibnamefont
  {Kang}}, \bibinfo {author} {\bibfnamefont {L.}~\bibnamefont {Ye}}, \bibinfo
  {author} {\bibfnamefont {S.}~\bibnamefont {Fang}}, \bibinfo {author}
  {\bibfnamefont {J.-S.}\ \bibnamefont {You}}, \bibinfo {author} {\bibfnamefont
  {A.}~\bibnamefont {Levitan}}, \bibinfo {author} {\bibfnamefont
  {M.}~\bibnamefont {Han}}, \bibinfo {author} {\bibfnamefont {J.~I.}\
  \bibnamefont {Facio}}, \bibinfo {author} {\bibfnamefont {C.}~\bibnamefont
  {Jozwiak}}, \bibinfo {author} {\bibfnamefont {A.}~\bibnamefont {Bostwick}},
  \bibinfo {author} {\bibfnamefont {E.}~\bibnamefont {Rotenberg}}, \bibinfo
  {author} {\bibfnamefont {M.~K.}\ \bibnamefont {Chan}}, \bibinfo {author}
  {\bibfnamefont {R.~D.}\ \bibnamefont {McDonald}}, \bibinfo {author}
  {\bibfnamefont {D.}~\bibnamefont {Graf}}, \bibinfo {author} {\bibfnamefont
  {K.}~\bibnamefont {Kaznatcheev}}, \bibinfo {author} {\bibfnamefont
  {E.}~\bibnamefont {Vescovo}}, \bibinfo {author} {\bibfnamefont {D.~C.}\
  \bibnamefont {Bell}}, \bibinfo {author} {\bibfnamefont {E.}~\bibnamefont
  {Kaxiras}}, \bibinfo {author} {\bibfnamefont {J.}~\bibnamefont {van~den
  Brink}}, \bibinfo {author} {\bibfnamefont {M.}~\bibnamefont {Richter}},
  \bibinfo {author} {\bibfnamefont {M.}~\bibnamefont {Prasad~Ghimire}},
  \bibinfo {author} {\bibfnamefont {J.~G.}\ \bibnamefont {Checkelsky}}, \ and\
  \bibinfo {author} {\bibfnamefont {R.}~\bibnamefont {Comin}},\ }\href
  {\doibase 10.1038/s41563-019-0531-0} {\bibfield  {journal} {\bibinfo
  {journal} {Nature Materials}\ }\textbf {\bibinfo {volume} {19}},\ \bibinfo
  {pages} {163} (\bibinfo {year} {2020})}\BibitemShut {NoStop}%
\bibitem [{\citenamefont {Ortiz}\ \emph {et~al.}(2019)\citenamefont {Ortiz},
  \citenamefont {Gomes}, \citenamefont {Morey}, \citenamefont {Winiarski},
  \citenamefont {Bordelon}, \citenamefont {Mangum}, \citenamefont {Oswald},
  \citenamefont {Rodriguez-Rivera}, \citenamefont {Neilson}, \citenamefont
  {Wilson}, \citenamefont {Ertekin}, \citenamefont {McQueen},\ and\
  \citenamefont {Toberer}}]{AV3Sb5_Ortiz_first_paper}%
  \BibitemOpen
  \bibfield  {author} {\bibinfo {author} {\bibfnamefont {B.~R.}\ \bibnamefont
  {Ortiz}}, \bibinfo {author} {\bibfnamefont {L.~C.}\ \bibnamefont {Gomes}},
  \bibinfo {author} {\bibfnamefont {J.~R.}\ \bibnamefont {Morey}}, \bibinfo
  {author} {\bibfnamefont {M.}~\bibnamefont {Winiarski}}, \bibinfo {author}
  {\bibfnamefont {M.}~\bibnamefont {Bordelon}}, \bibinfo {author}
  {\bibfnamefont {J.~S.}\ \bibnamefont {Mangum}}, \bibinfo {author}
  {\bibfnamefont {I.~W.~H.}\ \bibnamefont {Oswald}}, \bibinfo {author}
  {\bibfnamefont {J.~A.}\ \bibnamefont {Rodriguez-Rivera}}, \bibinfo {author}
  {\bibfnamefont {J.~R.}\ \bibnamefont {Neilson}}, \bibinfo {author}
  {\bibfnamefont {S.~D.}\ \bibnamefont {Wilson}}, \bibinfo {author}
  {\bibfnamefont {E.}~\bibnamefont {Ertekin}}, \bibinfo {author} {\bibfnamefont
  {T.~M.}\ \bibnamefont {McQueen}}, \ and\ \bibinfo {author} {\bibfnamefont
  {E.~S.}\ \bibnamefont {Toberer}},\ }\href {\doibase
  10.1103/PhysRevMaterials.3.094407} {\bibfield  {journal} {\bibinfo  {journal}
  {Phys. Rev. Materials}\ }\textbf {\bibinfo {volume} {3}},\ \bibinfo {pages}
  {094407} (\bibinfo {year} {2019})}\BibitemShut {NoStop}%
\bibitem [{\citenamefont {Neupert}\ \emph {et~al.}(2022)\citenamefont
  {Neupert}, \citenamefont {Denner}, \citenamefont {Yin}, \citenamefont
  {Thomale},\ and\ \citenamefont {Hasan}}]{AV3Sb5_nature_review}%
  \BibitemOpen
  \bibfield  {author} {\bibinfo {author} {\bibfnamefont {T.}~\bibnamefont
  {Neupert}}, \bibinfo {author} {\bibfnamefont {M.~M.}\ \bibnamefont {Denner}},
  \bibinfo {author} {\bibfnamefont {J.-X.}\ \bibnamefont {Yin}}, \bibinfo
  {author} {\bibfnamefont {R.}~\bibnamefont {Thomale}}, \ and\ \bibinfo
  {author} {\bibfnamefont {M.~Z.}\ \bibnamefont {Hasan}},\ }\href
  {https://doi.org/10.1038/s41567-021-01404-y} {\bibfield  {journal} {\bibinfo
  {journal} {Nature Physics}\ }\textbf {\bibinfo {volume} {18}},\ \bibinfo
  {pages} {137} (\bibinfo {year} {2022})}\BibitemShut {NoStop}%
\bibitem [{\citenamefont {Jiang}\ \emph {et~al.}(2022)\citenamefont {Jiang},
  \citenamefont {Wu}, \citenamefont {Yin}, \citenamefont {Wang}, \citenamefont
  {Hasan}, \citenamefont {Wilson}, \citenamefont {Chen},\ and\ \citenamefont
  {Hu}}]{jiangping_hu_review}%
  \BibitemOpen
  \bibfield  {author} {\bibinfo {author} {\bibfnamefont {K.}~\bibnamefont
  {Jiang}}, \bibinfo {author} {\bibfnamefont {T.}~\bibnamefont {Wu}}, \bibinfo
  {author} {\bibfnamefont {J.-X.}\ \bibnamefont {Yin}}, \bibinfo {author}
  {\bibfnamefont {Z.}~\bibnamefont {Wang}}, \bibinfo {author} {\bibfnamefont
  {M.~Z.}\ \bibnamefont {Hasan}}, \bibinfo {author} {\bibfnamefont {S.~D.}\
  \bibnamefont {Wilson}}, \bibinfo {author} {\bibfnamefont {X.}~\bibnamefont
  {Chen}}, \ and\ \bibinfo {author} {\bibfnamefont {J.}~\bibnamefont {Hu}},\
  }\href {\doibase 10.1093/nsr/nwac199} {\bibfield  {journal} {\bibinfo
  {journal} {National Science Review}\ }\textbf {\bibinfo {volume} {10}},\
  \bibinfo {pages} {nwac199} (\bibinfo {year} {2022})}\BibitemShut {NoStop}%
\bibitem [{\citenamefont {Yin}\ \emph {et~al.}(2022)\citenamefont {Yin},
  \citenamefont {Jiang}, \citenamefont {Teng}, \citenamefont {Hossain},
  \citenamefont {Mardanya}, \citenamefont {Chang}, \citenamefont {Ye},
  \citenamefont {Xu}, \citenamefont {Denner}, \citenamefont {Neupert},
  \citenamefont {Lienhard}, \citenamefont {Deng}, \citenamefont {Setty},
  \citenamefont {Si}, \citenamefont {Chang}, \citenamefont {Guguchia},
  \citenamefont {Gao}, \citenamefont {Shumiya}, \citenamefont {Zhang},
  \citenamefont {Cochran}, \citenamefont {Multer}, \citenamefont {Yi},
  \citenamefont {Dai},\ and\ \citenamefont {Hasan}}]{PhysRevLett.129.166401}%
  \BibitemOpen
  \bibfield  {author} {\bibinfo {author} {\bibfnamefont {J.-X.}\ \bibnamefont
  {Yin}}, \bibinfo {author} {\bibfnamefont {Y.-X.}\ \bibnamefont {Jiang}},
  \bibinfo {author} {\bibfnamefont {X.}~\bibnamefont {Teng}}, \bibinfo {author}
  {\bibfnamefont {M.~S.}\ \bibnamefont {Hossain}}, \bibinfo {author}
  {\bibfnamefont {S.}~\bibnamefont {Mardanya}}, \bibinfo {author}
  {\bibfnamefont {T.-R.}\ \bibnamefont {Chang}}, \bibinfo {author}
  {\bibfnamefont {Z.}~\bibnamefont {Ye}}, \bibinfo {author} {\bibfnamefont
  {G.}~\bibnamefont {Xu}}, \bibinfo {author} {\bibfnamefont {M.~M.}\
  \bibnamefont {Denner}}, \bibinfo {author} {\bibfnamefont {T.}~\bibnamefont
  {Neupert}}, \bibinfo {author} {\bibfnamefont {B.}~\bibnamefont {Lienhard}},
  \bibinfo {author} {\bibfnamefont {H.-B.}\ \bibnamefont {Deng}}, \bibinfo
  {author} {\bibfnamefont {C.}~\bibnamefont {Setty}}, \bibinfo {author}
  {\bibfnamefont {Q.}~\bibnamefont {Si}}, \bibinfo {author} {\bibfnamefont
  {G.}~\bibnamefont {Chang}}, \bibinfo {author} {\bibfnamefont
  {Z.}~\bibnamefont {Guguchia}}, \bibinfo {author} {\bibfnamefont
  {B.}~\bibnamefont {Gao}}, \bibinfo {author} {\bibfnamefont {N.}~\bibnamefont
  {Shumiya}}, \bibinfo {author} {\bibfnamefont {Q.}~\bibnamefont {Zhang}},
  \bibinfo {author} {\bibfnamefont {T.~A.}\ \bibnamefont {Cochran}}, \bibinfo
  {author} {\bibfnamefont {D.}~\bibnamefont {Multer}}, \bibinfo {author}
  {\bibfnamefont {M.}~\bibnamefont {Yi}}, \bibinfo {author} {\bibfnamefont
  {P.}~\bibnamefont {Dai}}, \ and\ \bibinfo {author} {\bibfnamefont {M.~Z.}\
  \bibnamefont {Hasan}},\ }\href {\doibase 10.1103/PhysRevLett.129.166401}
  {\bibfield  {journal} {\bibinfo  {journal} {Phys. Rev. Lett.}\ }\textbf
  {\bibinfo {volume} {129}},\ \bibinfo {pages} {166401} (\bibinfo {year}
  {2022})}\BibitemShut {NoStop}%
\bibitem [{\citenamefont {Teng}\ \emph {et~al.}(2022)\citenamefont {Teng},
  \citenamefont {Chen}, \citenamefont {Ye}, \citenamefont {Rosenberg},
  \citenamefont {Liu}, \citenamefont {Yin}, \citenamefont {Jiang},
  \citenamefont {Oh}, \citenamefont {Hasan}, \citenamefont {Neubauer},
  \citenamefont {Gao}, \citenamefont {Xie}, \citenamefont {Hashimoto},
  \citenamefont {Lu}, \citenamefont {Jozwiak}, \citenamefont {Bostwick},
  \citenamefont {Rotenberg}, \citenamefont {Birgeneau}, \citenamefont {Chu},
  \citenamefont {Yi},\ and\ \citenamefont {Dai}}]{TengXK2022}%
  \BibitemOpen
  \bibfield  {author} {\bibinfo {author} {\bibfnamefont {X.}~\bibnamefont
  {Teng}}, \bibinfo {author} {\bibfnamefont {L.}~\bibnamefont {Chen}}, \bibinfo
  {author} {\bibfnamefont {F.}~\bibnamefont {Ye}}, \bibinfo {author}
  {\bibfnamefont {E.}~\bibnamefont {Rosenberg}}, \bibinfo {author}
  {\bibfnamefont {Z.}~\bibnamefont {Liu}}, \bibinfo {author} {\bibfnamefont
  {J.-X.}\ \bibnamefont {Yin}}, \bibinfo {author} {\bibfnamefont {Y.-X.}\
  \bibnamefont {Jiang}}, \bibinfo {author} {\bibfnamefont {J.~S.}\ \bibnamefont
  {Oh}}, \bibinfo {author} {\bibfnamefont {M.~Z.}\ \bibnamefont {Hasan}},
  \bibinfo {author} {\bibfnamefont {K.~J.}\ \bibnamefont {Neubauer}}, \bibinfo
  {author} {\bibfnamefont {B.}~\bibnamefont {Gao}}, \bibinfo {author}
  {\bibfnamefont {Y.}~\bibnamefont {Xie}}, \bibinfo {author} {\bibfnamefont
  {M.}~\bibnamefont {Hashimoto}}, \bibinfo {author} {\bibfnamefont
  {D.}~\bibnamefont {Lu}}, \bibinfo {author} {\bibfnamefont {C.}~\bibnamefont
  {Jozwiak}}, \bibinfo {author} {\bibfnamefont {A.}~\bibnamefont {Bostwick}},
  \bibinfo {author} {\bibfnamefont {E.}~\bibnamefont {Rotenberg}}, \bibinfo
  {author} {\bibfnamefont {R.~J.}\ \bibnamefont {Birgeneau}}, \bibinfo {author}
  {\bibfnamefont {J.-H.}\ \bibnamefont {Chu}}, \bibinfo {author} {\bibfnamefont
  {M.}~\bibnamefont {Yi}}, \ and\ \bibinfo {author} {\bibfnamefont
  {P.}~\bibnamefont {Dai}},\ }\href {\doibase 10.1038/s41586-022-05034-z}
  {\bibfield  {journal} {\bibinfo  {journal} {Nature}\ }\textbf {\bibinfo
  {volume} {609}},\ \bibinfo {pages} {490} (\bibinfo {year}
  {2022})}\BibitemShut {NoStop}%
\bibitem [{\citenamefont {Teng}\ \emph {et~al.}(2023)\citenamefont {Teng},
  \citenamefont {Oh}, \citenamefont {Tan}, \citenamefont {Chen}, \citenamefont
  {Huang}, \citenamefont {Gao}, \citenamefont {Yin}, \citenamefont {Chu},
  \citenamefont {Hashimoto}, \citenamefont {Lu}, \citenamefont {Jozwiak},
  \citenamefont {Bostwick}, \citenamefont {Rotenberg}, \citenamefont
  {Granroth}, \citenamefont {Yan}, \citenamefont {Birgeneau}, \citenamefont
  {Dai},\ and\ \citenamefont {Yi}}]{TengXK2023}%
  \BibitemOpen
  \bibfield  {author} {\bibinfo {author} {\bibfnamefont {X.}~\bibnamefont
  {Teng}}, \bibinfo {author} {\bibfnamefont {J.~S.}\ \bibnamefont {Oh}},
  \bibinfo {author} {\bibfnamefont {H.}~\bibnamefont {Tan}}, \bibinfo {author}
  {\bibfnamefont {L.}~\bibnamefont {Chen}}, \bibinfo {author} {\bibfnamefont
  {J.}~\bibnamefont {Huang}}, \bibinfo {author} {\bibfnamefont
  {B.}~\bibnamefont {Gao}}, \bibinfo {author} {\bibfnamefont {J.-X.}\
  \bibnamefont {Yin}}, \bibinfo {author} {\bibfnamefont {J.-H.}\ \bibnamefont
  {Chu}}, \bibinfo {author} {\bibfnamefont {M.}~\bibnamefont {Hashimoto}},
  \bibinfo {author} {\bibfnamefont {D.}~\bibnamefont {Lu}}, \bibinfo {author}
  {\bibfnamefont {C.}~\bibnamefont {Jozwiak}}, \bibinfo {author} {\bibfnamefont
  {A.}~\bibnamefont {Bostwick}}, \bibinfo {author} {\bibfnamefont
  {E.}~\bibnamefont {Rotenberg}}, \bibinfo {author} {\bibfnamefont {G.~E.}\
  \bibnamefont {Granroth}}, \bibinfo {author} {\bibfnamefont {B.}~\bibnamefont
  {Yan}}, \bibinfo {author} {\bibfnamefont {R.~J.}\ \bibnamefont {Birgeneau}},
  \bibinfo {author} {\bibfnamefont {P.}~\bibnamefont {Dai}}, \ and\ \bibinfo
  {author} {\bibfnamefont {M.}~\bibnamefont {Yi}},\ }\href {\doibase
  10.1038/s41567-023-01985-w} {\bibfield  {journal} {\bibinfo  {journal}
  {Nature Physics}\ }\textbf {\bibinfo {volume} {19}},\ \bibinfo {pages} {814}
  (\bibinfo {year} {2023})}\BibitemShut {NoStop}%
\bibitem [{\citenamefont {Han}\ \emph {et~al.}(2024)\citenamefont {Han},
  \citenamefont {Li}, \citenamefont {Tang}, \citenamefont {Wang}, \citenamefont
  {Zhang}, \citenamefont {Diao}, \citenamefont {Zhao}, \citenamefont {Sun},
  \citenamefont {Tian}, \citenamefont {Breese}, \citenamefont {Cai},
  \citenamefont {Milosevic}, \citenamefont {Qi}, \citenamefont {Wee},\ and\
  \citenamefont {Yin}}]{han2024orbital}%
  \BibitemOpen
  \bibfield  {author} {\bibinfo {author} {\bibfnamefont {S.}~\bibnamefont
  {Han}}, \bibinfo {author} {\bibfnamefont {L.}~\bibnamefont {Li}}, \bibinfo
  {author} {\bibfnamefont {C.~S.}\ \bibnamefont {Tang}}, \bibinfo {author}
  {\bibfnamefont {Q.}~\bibnamefont {Wang}}, \bibinfo {author} {\bibfnamefont
  {L.}~\bibnamefont {Zhang}}, \bibinfo {author} {\bibfnamefont
  {C.}~\bibnamefont {Diao}}, \bibinfo {author} {\bibfnamefont {M.}~\bibnamefont
  {Zhao}}, \bibinfo {author} {\bibfnamefont {S.}~\bibnamefont {Sun}}, \bibinfo
  {author} {\bibfnamefont {L.}~\bibnamefont {Tian}}, \bibinfo {author}
  {\bibfnamefont {M.~B.~H.}\ \bibnamefont {Breese}}, \bibinfo {author}
  {\bibfnamefont {C.}~\bibnamefont {Cai}}, \bibinfo {author} {\bibfnamefont
  {M.~V.}\ \bibnamefont {Milosevic}}, \bibinfo {author} {\bibfnamefont
  {Y.}~\bibnamefont {Qi}}, \bibinfo {author} {\bibfnamefont {A.~T.~S.}\
  \bibnamefont {Wee}}, \ and\ \bibinfo {author} {\bibfnamefont
  {X.}~\bibnamefont {Yin}},\ }\href@noop {} {\enquote {\bibinfo {title}
  {Orbital origin of magnetic moment enhancement induced by charge density wave
  in kagome fege},}\ } (\bibinfo {year} {2024}),\ \Eprint
  {http://arxiv.org/abs/2407.01076} {arXiv:2407.01076 [cond-mat.str-el]}
  \BibitemShut {NoStop}%
\bibitem [{\citenamefont {Mielke}\ \emph {et~al.}(2022)\citenamefont {Mielke},
  \citenamefont {Das}, \citenamefont {Yin}, \citenamefont {Liu}, \citenamefont
  {Gupta}, \citenamefont {Jiang}, \citenamefont {Medarde}, \citenamefont {Wu},
  \citenamefont {Lei}, \citenamefont {Chang}, \citenamefont {Dai},
  \citenamefont {Si}, \citenamefont {Miao}, \citenamefont {Thomale},
  \citenamefont {Neupert}, \citenamefont {Shi}, \citenamefont {Khasanov},
  \citenamefont {Hasan}, \citenamefont {Luetkens},\ and\ \citenamefont
  {Guguchia}}]{Mielke2022}%
  \BibitemOpen
  \bibfield  {author} {\bibinfo {author} {\bibfnamefont {C.}~\bibnamefont
  {Mielke}}, \bibinfo {author} {\bibfnamefont {D.}~\bibnamefont {Das}},
  \bibinfo {author} {\bibfnamefont {J.~X.}\ \bibnamefont {Yin}}, \bibinfo
  {author} {\bibfnamefont {H.}~\bibnamefont {Liu}}, \bibinfo {author}
  {\bibfnamefont {R.}~\bibnamefont {Gupta}}, \bibinfo {author} {\bibfnamefont
  {Y.~X.}\ \bibnamefont {Jiang}}, \bibinfo {author} {\bibfnamefont
  {M.}~\bibnamefont {Medarde}}, \bibinfo {author} {\bibfnamefont
  {X.}~\bibnamefont {Wu}}, \bibinfo {author} {\bibfnamefont {H.~C.}\
  \bibnamefont {Lei}}, \bibinfo {author} {\bibfnamefont {J.}~\bibnamefont
  {Chang}}, \bibinfo {author} {\bibfnamefont {P.}~\bibnamefont {Dai}}, \bibinfo
  {author} {\bibfnamefont {Q.}~\bibnamefont {Si}}, \bibinfo {author}
  {\bibfnamefont {H.}~\bibnamefont {Miao}}, \bibinfo {author} {\bibfnamefont
  {R.}~\bibnamefont {Thomale}}, \bibinfo {author} {\bibfnamefont
  {T.}~\bibnamefont {Neupert}}, \bibinfo {author} {\bibfnamefont
  {Y.}~\bibnamefont {Shi}}, \bibinfo {author} {\bibfnamefont {R.}~\bibnamefont
  {Khasanov}}, \bibinfo {author} {\bibfnamefont {M.~Z.}\ \bibnamefont {Hasan}},
  \bibinfo {author} {\bibfnamefont {H.}~\bibnamefont {Luetkens}}, \ and\
  \bibinfo {author} {\bibfnamefont {Z.}~\bibnamefont {Guguchia}},\ }\href
  {\doibase 10.1038/s41586-021-04327-z} {\bibfield  {journal} {\bibinfo
  {journal} {Nature}\ }\textbf {\bibinfo {volume} {602}},\ \bibinfo {pages}
  {245} (\bibinfo {year} {2022})}\BibitemShut {NoStop}%
\bibitem [{\citenamefont {{Yu}}\ \emph {et~al.}(2021)\citenamefont {{Yu}},
  \citenamefont {{Wang}}, \citenamefont {{Zhang}}, \citenamefont {{Sander}},
  \citenamefont {{Ni}}, \citenamefont {{Lu}}, \citenamefont {{Ma}},
  \citenamefont {{Wang}}, \citenamefont {{Zhao}}, \citenamefont {{Chen}},
  \citenamefont {{Jiang}}, \citenamefont {{Zhang}}, \citenamefont {{Yang}},
  \citenamefont {{Zhou}}, \citenamefont {{Dong}}, \citenamefont {{Johnson}},
  \citenamefont {{Graf}}, \citenamefont {{Hu}}, \citenamefont {{Gao}},\ and\
  \citenamefont {{Zhao}}}]{YuL2021}%
  \BibitemOpen
  \bibfield  {author} {\bibinfo {author} {\bibfnamefont {L.}~\bibnamefont
  {{Yu}}}, \bibinfo {author} {\bibfnamefont {C.}~\bibnamefont {{Wang}}},
  \bibinfo {author} {\bibfnamefont {Y.}~\bibnamefont {{Zhang}}}, \bibinfo
  {author} {\bibfnamefont {M.}~\bibnamefont {{Sander}}}, \bibinfo {author}
  {\bibfnamefont {S.}~\bibnamefont {{Ni}}}, \bibinfo {author} {\bibfnamefont
  {Z.}~\bibnamefont {{Lu}}}, \bibinfo {author} {\bibfnamefont {S.}~\bibnamefont
  {{Ma}}}, \bibinfo {author} {\bibfnamefont {Z.}~\bibnamefont {{Wang}}},
  \bibinfo {author} {\bibfnamefont {Z.}~\bibnamefont {{Zhao}}}, \bibinfo
  {author} {\bibfnamefont {H.}~\bibnamefont {{Chen}}}, \bibinfo {author}
  {\bibfnamefont {K.}~\bibnamefont {{Jiang}}}, \bibinfo {author} {\bibfnamefont
  {Y.}~\bibnamefont {{Zhang}}}, \bibinfo {author} {\bibfnamefont
  {H.}~\bibnamefont {{Yang}}}, \bibinfo {author} {\bibfnamefont
  {F.}~\bibnamefont {{Zhou}}}, \bibinfo {author} {\bibfnamefont
  {X.}~\bibnamefont {{Dong}}}, \bibinfo {author} {\bibfnamefont {S.~L.}\
  \bibnamefont {{Johnson}}}, \bibinfo {author} {\bibfnamefont {M.~J.}\
  \bibnamefont {{Graf}}}, \bibinfo {author} {\bibfnamefont {J.}~\bibnamefont
  {{Hu}}}, \bibinfo {author} {\bibfnamefont {H.-J.}\ \bibnamefont {{Gao}}}, \
  and\ \bibinfo {author} {\bibfnamefont {Z.}~\bibnamefont {{Zhao}}},\ }\href
  {\doibase 10.48550/arXiv.2107.10714} {\bibfield  {journal} {\bibinfo
  {journal} {arXiv e-prints}\ ,\ \bibinfo {eid} {arXiv:2107.10714}} (\bibinfo
  {year} {2021})},\ \Eprint {http://arxiv.org/abs/2107.10714} {arXiv:2107.10714
  [cond-mat.supr-con]} \BibitemShut {NoStop}%
\bibitem [{\citenamefont {Wu}\ \emph {et~al.}(2022)\citenamefont {Wu},
  \citenamefont {Wang}, \citenamefont {Liu}, \citenamefont {Li}, \citenamefont
  {Xu}, \citenamefont {Yin}, \citenamefont {Gong}, \citenamefont {Tu},
  \citenamefont {Lei}, \citenamefont {Dong},\ and\ \citenamefont
  {Wang}}]{2021arXiv211011306W}%
  \BibitemOpen
  \bibfield  {author} {\bibinfo {author} {\bibfnamefont {Q.}~\bibnamefont
  {Wu}}, \bibinfo {author} {\bibfnamefont {Z.~X.}\ \bibnamefont {Wang}},
  \bibinfo {author} {\bibfnamefont {Q.~M.}\ \bibnamefont {Liu}}, \bibinfo
  {author} {\bibfnamefont {R.~S.}\ \bibnamefont {Li}}, \bibinfo {author}
  {\bibfnamefont {S.~X.}\ \bibnamefont {Xu}}, \bibinfo {author} {\bibfnamefont
  {Q.~W.}\ \bibnamefont {Yin}}, \bibinfo {author} {\bibfnamefont {C.~S.}\
  \bibnamefont {Gong}}, \bibinfo {author} {\bibfnamefont {Z.~J.}\ \bibnamefont
  {Tu}}, \bibinfo {author} {\bibfnamefont {H.~C.}\ \bibnamefont {Lei}},
  \bibinfo {author} {\bibfnamefont {T.}~\bibnamefont {Dong}}, \ and\ \bibinfo
  {author} {\bibfnamefont {N.~L.}\ \bibnamefont {Wang}},\ }\href {\doibase
  10.1103/PhysRevB.106.205109} {\bibfield  {journal} {\bibinfo  {journal}
  {Phys. Rev. B}\ }\textbf {\bibinfo {volume} {106}},\ \bibinfo {pages}
  {205109} (\bibinfo {year} {2022})}\BibitemShut {NoStop}%
\bibitem [{\citenamefont {Xu}\ \emph {et~al.}(2022)\citenamefont {Xu},
  \citenamefont {Ni}, \citenamefont {Liu}, \citenamefont {Ortiz}, \citenamefont
  {Deng}, \citenamefont {Wilson}, \citenamefont {Yan}, \citenamefont
  {Balents},\ and\ \citenamefont {Wu}}]{2022arXiv220410116X}%
  \BibitemOpen
  \bibfield  {author} {\bibinfo {author} {\bibfnamefont {Y.}~\bibnamefont
  {Xu}}, \bibinfo {author} {\bibfnamefont {Z.}~\bibnamefont {Ni}}, \bibinfo
  {author} {\bibfnamefont {Y.}~\bibnamefont {Liu}}, \bibinfo {author}
  {\bibfnamefont {B.~R.}\ \bibnamefont {Ortiz}}, \bibinfo {author}
  {\bibfnamefont {Q.}~\bibnamefont {Deng}}, \bibinfo {author} {\bibfnamefont
  {S.~D.}\ \bibnamefont {Wilson}}, \bibinfo {author} {\bibfnamefont
  {B.}~\bibnamefont {Yan}}, \bibinfo {author} {\bibfnamefont {L.}~\bibnamefont
  {Balents}}, \ and\ \bibinfo {author} {\bibfnamefont {L.}~\bibnamefont {Wu}},\
  }\href {\doibase 10.1038/s41567-022-01805-7} {\bibfield  {journal} {\bibinfo
  {journal} {Nature Physics}\ }\textbf {\bibinfo {volume} {18}},\ \bibinfo
  {pages} {1470} (\bibinfo {year} {2022})}\BibitemShut {NoStop}%
\bibitem [{\citenamefont {Guo}\ \emph {et~al.}(2022)\citenamefont {Guo},
  \citenamefont {Putzke}, \citenamefont {Konyzheva}, \citenamefont {Huang},
  \citenamefont {Gutierrez-Amigo}, \citenamefont {Errea}, \citenamefont {Chen},
  \citenamefont {Vergniory}, \citenamefont {Felser}, \citenamefont {Fischer},
  \citenamefont {Neupert},\ and\ \citenamefont {Moll}}]{GuoCY2022}%
  \BibitemOpen
  \bibfield  {author} {\bibinfo {author} {\bibfnamefont {C.}~\bibnamefont
  {Guo}}, \bibinfo {author} {\bibfnamefont {C.}~\bibnamefont {Putzke}},
  \bibinfo {author} {\bibfnamefont {S.}~\bibnamefont {Konyzheva}}, \bibinfo
  {author} {\bibfnamefont {X.}~\bibnamefont {Huang}}, \bibinfo {author}
  {\bibfnamefont {M.}~\bibnamefont {Gutierrez-Amigo}}, \bibinfo {author}
  {\bibfnamefont {I.}~\bibnamefont {Errea}}, \bibinfo {author} {\bibfnamefont
  {D.}~\bibnamefont {Chen}}, \bibinfo {author} {\bibfnamefont {M.~G.}\
  \bibnamefont {Vergniory}}, \bibinfo {author} {\bibfnamefont {C.}~\bibnamefont
  {Felser}}, \bibinfo {author} {\bibfnamefont {M.~H.}\ \bibnamefont {Fischer}},
  \bibinfo {author} {\bibfnamefont {T.}~\bibnamefont {Neupert}}, \ and\
  \bibinfo {author} {\bibfnamefont {P.~J.~W.}\ \bibnamefont {Moll}},\ }\href
  {\doibase 10.1038/s41586-022-05127-9} {\bibfield  {journal} {\bibinfo
  {journal} {Nature}\ }\textbf {\bibinfo {volume} {611}},\ \bibinfo {pages}
  {461} (\bibinfo {year} {2022})}\BibitemShut {NoStop}%
\bibitem [{\citenamefont {Xing}\ \emph {et~al.}(2024)\citenamefont {Xing},
  \citenamefont {Bae}, \citenamefont {Ritz}, \citenamefont {Yang},
  \citenamefont {Birol}, \citenamefont {Capa~Salinas}, \citenamefont {Ortiz},
  \citenamefont {Wilson}, \citenamefont {Wang}, \citenamefont {Fernandes},\
  and\ \citenamefont {Madhavan}}]{xing2024optical}%
  \BibitemOpen
  \bibfield  {author} {\bibinfo {author} {\bibfnamefont {Y.}~\bibnamefont
  {Xing}}, \bibinfo {author} {\bibfnamefont {S.}~\bibnamefont {Bae}}, \bibinfo
  {author} {\bibfnamefont {E.}~\bibnamefont {Ritz}}, \bibinfo {author}
  {\bibfnamefont {F.}~\bibnamefont {Yang}}, \bibinfo {author} {\bibfnamefont
  {T.}~\bibnamefont {Birol}}, \bibinfo {author} {\bibfnamefont {A.~N.}\
  \bibnamefont {Capa~Salinas}}, \bibinfo {author} {\bibfnamefont {B.~R.}\
  \bibnamefont {Ortiz}}, \bibinfo {author} {\bibfnamefont {S.~D.}\ \bibnamefont
  {Wilson}}, \bibinfo {author} {\bibfnamefont {Z.}~\bibnamefont {Wang}},
  \bibinfo {author} {\bibfnamefont {R.~M.}\ \bibnamefont {Fernandes}}, \ and\
  \bibinfo {author} {\bibfnamefont {V.}~\bibnamefont {Madhavan}},\ }\href
  {\doibase 10.1038/s41586-024-07519-5} {\bibfield  {journal} {\bibinfo
  {journal} {Nature}\ } (\bibinfo {year} {2024}),\
  10.1038/s41586-024-07519-5}\BibitemShut {NoStop}%
\bibitem [{\citenamefont {Chen}\ \emph
  {et~al.}(2021{\natexlab{a}})\citenamefont {Chen}, \citenamefont {Wang},
  \citenamefont {Yin}, \citenamefont {Gu}, \citenamefont {Jiang}, \citenamefont
  {Tu}, \citenamefont {Gong}, \citenamefont {Uwatoko}, \citenamefont {Sun},
  \citenamefont {Lei}, \citenamefont {Hu},\ and\ \citenamefont
  {Cheng}}]{PhysRevLett.126.247001}%
  \BibitemOpen
  \bibfield  {author} {\bibinfo {author} {\bibfnamefont {K.~Y.}\ \bibnamefont
  {Chen}}, \bibinfo {author} {\bibfnamefont {N.~N.}\ \bibnamefont {Wang}},
  \bibinfo {author} {\bibfnamefont {Q.~W.}\ \bibnamefont {Yin}}, \bibinfo
  {author} {\bibfnamefont {Y.~H.}\ \bibnamefont {Gu}}, \bibinfo {author}
  {\bibfnamefont {K.}~\bibnamefont {Jiang}}, \bibinfo {author} {\bibfnamefont
  {Z.~J.}\ \bibnamefont {Tu}}, \bibinfo {author} {\bibfnamefont {C.~S.}\
  \bibnamefont {Gong}}, \bibinfo {author} {\bibfnamefont {Y.}~\bibnamefont
  {Uwatoko}}, \bibinfo {author} {\bibfnamefont {J.~P.}\ \bibnamefont {Sun}},
  \bibinfo {author} {\bibfnamefont {H.~C.}\ \bibnamefont {Lei}}, \bibinfo
  {author} {\bibfnamefont {J.~P.}\ \bibnamefont {Hu}}, \ and\ \bibinfo {author}
  {\bibfnamefont {J.-G.}\ \bibnamefont {Cheng}},\ }\href {\doibase
  10.1103/PhysRevLett.126.247001} {\bibfield  {journal} {\bibinfo  {journal}
  {Phys. Rev. Lett.}\ }\textbf {\bibinfo {volume} {126}},\ \bibinfo {pages}
  {247001} (\bibinfo {year} {2021}{\natexlab{a}})}\BibitemShut {NoStop}%
\bibitem [{\citenamefont {Zhang}\ \emph {et~al.}(2021)\citenamefont {Zhang},
  \citenamefont {Chen}, \citenamefont {Zhou}, \citenamefont {Yuan},
  \citenamefont {Wang}, \citenamefont {Wang}, \citenamefont {Yang},
  \citenamefont {An}, \citenamefont {Zhang}, \citenamefont {Zhu}, \citenamefont
  {Zhou}, \citenamefont {Chen}, \citenamefont {Zhou},\ and\ \citenamefont
  {Yang}}]{PhysRevB.103.224513}%
  \BibitemOpen
  \bibfield  {author} {\bibinfo {author} {\bibfnamefont {Z.}~\bibnamefont
  {Zhang}}, \bibinfo {author} {\bibfnamefont {Z.}~\bibnamefont {Chen}},
  \bibinfo {author} {\bibfnamefont {Y.}~\bibnamefont {Zhou}}, \bibinfo {author}
  {\bibfnamefont {Y.}~\bibnamefont {Yuan}}, \bibinfo {author} {\bibfnamefont
  {S.}~\bibnamefont {Wang}}, \bibinfo {author} {\bibfnamefont {J.}~\bibnamefont
  {Wang}}, \bibinfo {author} {\bibfnamefont {H.}~\bibnamefont {Yang}}, \bibinfo
  {author} {\bibfnamefont {C.}~\bibnamefont {An}}, \bibinfo {author}
  {\bibfnamefont {L.}~\bibnamefont {Zhang}}, \bibinfo {author} {\bibfnamefont
  {X.}~\bibnamefont {Zhu}}, \bibinfo {author} {\bibfnamefont {Y.}~\bibnamefont
  {Zhou}}, \bibinfo {author} {\bibfnamefont {X.}~\bibnamefont {Chen}}, \bibinfo
  {author} {\bibfnamefont {J.}~\bibnamefont {Zhou}}, \ and\ \bibinfo {author}
  {\bibfnamefont {Z.}~\bibnamefont {Yang}},\ }\href {\doibase
  10.1103/PhysRevB.103.224513} {\bibfield  {journal} {\bibinfo  {journal}
  {Phys. Rev. B}\ }\textbf {\bibinfo {volume} {103}},\ \bibinfo {pages}
  {224513} (\bibinfo {year} {2021})}\BibitemShut {NoStop}%
\bibitem [{\citenamefont {Chen}\ \emph
  {et~al.}(2021{\natexlab{b}})\citenamefont {Chen}, \citenamefont {Zhan},
  \citenamefont {Wang}, \citenamefont {Deng}, \citenamefont {Liu},
  \citenamefont {Chen}, \citenamefont {Guo},\ and\ \citenamefont
  {Chen}}]{Chen_2021}%
  \BibitemOpen
  \bibfield  {author} {\bibinfo {author} {\bibfnamefont {X.}~\bibnamefont
  {Chen}}, \bibinfo {author} {\bibfnamefont {X.}~\bibnamefont {Zhan}}, \bibinfo
  {author} {\bibfnamefont {X.}~\bibnamefont {Wang}}, \bibinfo {author}
  {\bibfnamefont {J.}~\bibnamefont {Deng}}, \bibinfo {author} {\bibfnamefont
  {X.-B.}\ \bibnamefont {Liu}}, \bibinfo {author} {\bibfnamefont
  {X.}~\bibnamefont {Chen}}, \bibinfo {author} {\bibfnamefont {J.-G.}\
  \bibnamefont {Guo}}, \ and\ \bibinfo {author} {\bibfnamefont
  {X.}~\bibnamefont {Chen}},\ }\href {\doibase 10.1088/0256-307x/38/5/057402}
  {\bibfield  {journal} {\bibinfo  {journal} {Chinese Physics Letters}\
  }\textbf {\bibinfo {volume} {38}},\ \bibinfo {pages} {057402} (\bibinfo
  {year} {2021}{\natexlab{b}})}\BibitemShut {NoStop}%
\bibitem [{\citenamefont {Song}\ \emph {et~al.}(2021)\citenamefont {Song},
  \citenamefont {Ying}, \citenamefont {Chen}, \citenamefont {Han},
  \citenamefont {Wu}, \citenamefont {Schnyder}, \citenamefont {Huang},
  \citenamefont {Guo},\ and\ \citenamefont {Chen}}]{PhysRevLett.127.237001}%
  \BibitemOpen
  \bibfield  {author} {\bibinfo {author} {\bibfnamefont {Y.}~\bibnamefont
  {Song}}, \bibinfo {author} {\bibfnamefont {T.}~\bibnamefont {Ying}}, \bibinfo
  {author} {\bibfnamefont {X.}~\bibnamefont {Chen}}, \bibinfo {author}
  {\bibfnamefont {X.}~\bibnamefont {Han}}, \bibinfo {author} {\bibfnamefont
  {X.}~\bibnamefont {Wu}}, \bibinfo {author} {\bibfnamefont {A.~P.}\
  \bibnamefont {Schnyder}}, \bibinfo {author} {\bibfnamefont {Y.}~\bibnamefont
  {Huang}}, \bibinfo {author} {\bibfnamefont {J.-g.}\ \bibnamefont {Guo}}, \
  and\ \bibinfo {author} {\bibfnamefont {X.}~\bibnamefont {Chen}},\ }\href
  {\doibase 10.1103/PhysRevLett.127.237001} {\bibfield  {journal} {\bibinfo
  {journal} {Phys. Rev. Lett.}\ }\textbf {\bibinfo {volume} {127}},\ \bibinfo
  {pages} {237001} (\bibinfo {year} {2021})}\BibitemShut {NoStop}%
\bibitem [{\citenamefont {Oey}\ \emph {et~al.}(2022)\citenamefont {Oey},
  \citenamefont {Ortiz}, \citenamefont {Kaboudvand}, \citenamefont
  {Frassineti}, \citenamefont {Garcia}, \citenamefont {Cong}, \citenamefont
  {Sanna}, \citenamefont {Mitrovi\ifmmode~\acute{c}\else \'{c}\fi{}},
  \citenamefont {Seshadri},\ and\ \citenamefont
  {Wilson}}]{PhysRevMaterials.6.L041801}%
  \BibitemOpen
  \bibfield  {author} {\bibinfo {author} {\bibfnamefont {Y.~M.}\ \bibnamefont
  {Oey}}, \bibinfo {author} {\bibfnamefont {B.~R.}\ \bibnamefont {Ortiz}},
  \bibinfo {author} {\bibfnamefont {F.}~\bibnamefont {Kaboudvand}}, \bibinfo
  {author} {\bibfnamefont {J.}~\bibnamefont {Frassineti}}, \bibinfo {author}
  {\bibfnamefont {E.}~\bibnamefont {Garcia}}, \bibinfo {author} {\bibfnamefont
  {R.}~\bibnamefont {Cong}}, \bibinfo {author} {\bibfnamefont {S.}~\bibnamefont
  {Sanna}}, \bibinfo {author} {\bibfnamefont {V.~F.}\ \bibnamefont
  {Mitrovi\ifmmode~\acute{c}\else \'{c}\fi{}}}, \bibinfo {author}
  {\bibfnamefont {R.}~\bibnamefont {Seshadri}}, \ and\ \bibinfo {author}
  {\bibfnamefont {S.~D.}\ \bibnamefont {Wilson}},\ }\href {\doibase
  10.1103/PhysRevMaterials.6.L041801} {\bibfield  {journal} {\bibinfo
  {journal} {Phys. Rev. Materials}\ }\textbf {\bibinfo {volume} {6}},\ \bibinfo
  {pages} {L041801} (\bibinfo {year} {2022})}\BibitemShut {NoStop}%
\bibitem [{\citenamefont {Liu}\ \emph {et~al.}(2023)\citenamefont {Liu},
  \citenamefont {Wang}, \citenamefont {Cai}, \citenamefont {Hao}, \citenamefont
  {Ma}, \citenamefont {Wang}, \citenamefont {Liu}, \citenamefont {Chen},
  \citenamefont {Zhou}, \citenamefont {Wang}, \citenamefont {Wang},
  \citenamefont {He}, \citenamefont {Liu}, \citenamefont {Cui}, \citenamefont
  {Huang}, \citenamefont {Wang}, \citenamefont {Chen},\ and\ \citenamefont
  {Mei}}]{2021arXiv211012651L}%
  \BibitemOpen
  \bibfield  {author} {\bibinfo {author} {\bibfnamefont {Y.}~\bibnamefont
  {Liu}}, \bibinfo {author} {\bibfnamefont {Y.}~\bibnamefont {Wang}}, \bibinfo
  {author} {\bibfnamefont {Y.}~\bibnamefont {Cai}}, \bibinfo {author}
  {\bibfnamefont {Z.}~\bibnamefont {Hao}}, \bibinfo {author} {\bibfnamefont
  {X.-M.}\ \bibnamefont {Ma}}, \bibinfo {author} {\bibfnamefont
  {L.}~\bibnamefont {Wang}}, \bibinfo {author} {\bibfnamefont {C.}~\bibnamefont
  {Liu}}, \bibinfo {author} {\bibfnamefont {J.}~\bibnamefont {Chen}}, \bibinfo
  {author} {\bibfnamefont {L.}~\bibnamefont {Zhou}}, \bibinfo {author}
  {\bibfnamefont {J.}~\bibnamefont {Wang}}, \bibinfo {author} {\bibfnamefont
  {S.}~\bibnamefont {Wang}}, \bibinfo {author} {\bibfnamefont {H.}~\bibnamefont
  {He}}, \bibinfo {author} {\bibfnamefont {Y.}~\bibnamefont {Liu}}, \bibinfo
  {author} {\bibfnamefont {S.}~\bibnamefont {Cui}}, \bibinfo {author}
  {\bibfnamefont {B.}~\bibnamefont {Huang}}, \bibinfo {author} {\bibfnamefont
  {J.}~\bibnamefont {Wang}}, \bibinfo {author} {\bibfnamefont {C.}~\bibnamefont
  {Chen}}, \ and\ \bibinfo {author} {\bibfnamefont {J.-W.}\ \bibnamefont
  {Mei}},\ }\href {\doibase 10.1103/PhysRevMaterials.7.064801} {\bibfield
  {journal} {\bibinfo  {journal} {Phys. Rev. Mater.}\ }\textbf {\bibinfo
  {volume} {7}},\ \bibinfo {pages} {064801} (\bibinfo {year}
  {2023})}\BibitemShut {NoStop}%
\bibitem [{\citenamefont {Feng}\ \emph {et~al.}(2021)\citenamefont {Feng},
  \citenamefont {Jiang}, \citenamefont {Wang},\ and\ \citenamefont
  {Hu}}]{feng2021chiral}%
  \BibitemOpen
  \bibfield  {author} {\bibinfo {author} {\bibfnamefont {X.}~\bibnamefont
  {Feng}}, \bibinfo {author} {\bibfnamefont {K.}~\bibnamefont {Jiang}},
  \bibinfo {author} {\bibfnamefont {Z.}~\bibnamefont {Wang}}, \ and\ \bibinfo
  {author} {\bibfnamefont {J.}~\bibnamefont {Hu}},\ }\href@noop {} {\bibfield
  {journal} {\bibinfo  {journal} {Science bulletin}\ }\textbf {\bibinfo
  {volume} {66}},\ \bibinfo {pages} {1384} (\bibinfo {year}
  {2021})}\BibitemShut {NoStop}%
\bibitem [{\citenamefont {Denner}\ \emph {et~al.}(2021)\citenamefont {Denner},
  \citenamefont {Thomale},\ and\ \citenamefont
  {Neupert}}]{PhysRevLett.127.217601}%
  \BibitemOpen
  \bibfield  {author} {\bibinfo {author} {\bibfnamefont {M.~M.}\ \bibnamefont
  {Denner}}, \bibinfo {author} {\bibfnamefont {R.}~\bibnamefont {Thomale}}, \
  and\ \bibinfo {author} {\bibfnamefont {T.}~\bibnamefont {Neupert}},\ }\href
  {\doibase 10.1103/PhysRevLett.127.217601} {\bibfield  {journal} {\bibinfo
  {journal} {Phys. Rev. Lett.}\ }\textbf {\bibinfo {volume} {127}},\ \bibinfo
  {pages} {217601} (\bibinfo {year} {2021})}\BibitemShut {NoStop}%
\bibitem [{\citenamefont {Lin}\ and\ \citenamefont
  {Nandkishore}(2021)}]{PhysRevB.104.045122}%
  \BibitemOpen
  \bibfield  {author} {\bibinfo {author} {\bibfnamefont {Y.-P.}\ \bibnamefont
  {Lin}}\ and\ \bibinfo {author} {\bibfnamefont {R.~M.}\ \bibnamefont
  {Nandkishore}},\ }\href {\doibase 10.1103/PhysRevB.104.045122} {\bibfield
  {journal} {\bibinfo  {journal} {Phys. Rev. B}\ }\textbf {\bibinfo {volume}
  {104}},\ \bibinfo {pages} {045122} (\bibinfo {year} {2021})}\BibitemShut
  {NoStop}%
\bibitem [{\citenamefont {Park}\ \emph {et~al.}(2021)\citenamefont {Park},
  \citenamefont {Ye},\ and\ \citenamefont {Balents}}]{PhysRevB.104.035142}%
  \BibitemOpen
  \bibfield  {author} {\bibinfo {author} {\bibfnamefont {T.}~\bibnamefont
  {Park}}, \bibinfo {author} {\bibfnamefont {M.}~\bibnamefont {Ye}}, \ and\
  \bibinfo {author} {\bibfnamefont {L.}~\bibnamefont {Balents}},\ }\href
  {\doibase 10.1103/PhysRevB.104.035142} {\bibfield  {journal} {\bibinfo
  {journal} {Phys. Rev. B}\ }\textbf {\bibinfo {volume} {104}},\ \bibinfo
  {pages} {035142} (\bibinfo {year} {2021})}\BibitemShut {NoStop}%
\bibitem [{\citenamefont {Varma}(1997)}]{PhysRevB.55.14554}%
  \BibitemOpen
  \bibfield  {author} {\bibinfo {author} {\bibfnamefont {C.~M.}\ \bibnamefont
  {Varma}},\ }\href {\doibase 10.1103/PhysRevB.55.14554} {\bibfield  {journal}
  {\bibinfo  {journal} {Phys. Rev. B}\ }\textbf {\bibinfo {volume} {55}},\
  \bibinfo {pages} {14554} (\bibinfo {year} {1997})}\BibitemShut {NoStop}%
\bibitem [{\citenamefont {Chakravarty}\ \emph {et~al.}(2001)\citenamefont
  {Chakravarty}, \citenamefont {Laughlin}, \citenamefont {Morr},\ and\
  \citenamefont {Nayak}}]{PhysRevB.63.094503}%
  \BibitemOpen
  \bibfield  {author} {\bibinfo {author} {\bibfnamefont {S.}~\bibnamefont
  {Chakravarty}}, \bibinfo {author} {\bibfnamefont {R.~B.}\ \bibnamefont
  {Laughlin}}, \bibinfo {author} {\bibfnamefont {D.~K.}\ \bibnamefont {Morr}},
  \ and\ \bibinfo {author} {\bibfnamefont {C.}~\bibnamefont {Nayak}},\ }\href
  {\doibase 10.1103/PhysRevB.63.094503} {\bibfield  {journal} {\bibinfo
  {journal} {Phys. Rev. B}\ }\textbf {\bibinfo {volume} {63}},\ \bibinfo
  {pages} {094503} (\bibinfo {year} {2001})}\BibitemShut {NoStop}%
\bibitem [{\citenamefont {Varma}(1999)}]{PhysRevLett.83.3538}%
  \BibitemOpen
  \bibfield  {author} {\bibinfo {author} {\bibfnamefont {C.~M.}\ \bibnamefont
  {Varma}},\ }\href {\doibase 10.1103/PhysRevLett.83.3538} {\bibfield
  {journal} {\bibinfo  {journal} {Phys. Rev. Lett.}\ }\textbf {\bibinfo
  {volume} {83}},\ \bibinfo {pages} {3538} (\bibinfo {year}
  {1999})}\BibitemShut {NoStop}%
\bibitem [{\citenamefont {Haldane}(1988)}]{haldane1988model}%
  \BibitemOpen
  \bibfield  {author} {\bibinfo {author} {\bibfnamefont {F.~D.~M.}\
  \bibnamefont {Haldane}},\ }\href {\doibase 10.1103/PhysRevLett.61.2015}
  {\bibfield  {journal} {\bibinfo  {journal} {Phys. Rev. Lett.}\ }\textbf
  {\bibinfo {volume} {61}},\ \bibinfo {pages} {2015} (\bibinfo {year}
  {1988})}\BibitemShut {NoStop}%
\bibitem [{\citenamefont {Raghu}\ \emph {et~al.}(2008)\citenamefont {Raghu},
  \citenamefont {Qi}, \citenamefont {Honerkamp},\ and\ \citenamefont
  {Zhang}}]{PhysRevLett.100.156401}%
  \BibitemOpen
  \bibfield  {author} {\bibinfo {author} {\bibfnamefont {S.}~\bibnamefont
  {Raghu}}, \bibinfo {author} {\bibfnamefont {X.-L.}\ \bibnamefont {Qi}},
  \bibinfo {author} {\bibfnamefont {C.}~\bibnamefont {Honerkamp}}, \ and\
  \bibinfo {author} {\bibfnamefont {S.-C.}\ \bibnamefont {Zhang}},\ }\href
  {\doibase 10.1103/PhysRevLett.100.156401} {\bibfield  {journal} {\bibinfo
  {journal} {Phys. Rev. Lett.}\ }\textbf {\bibinfo {volume} {100}},\ \bibinfo
  {pages} {156401} (\bibinfo {year} {2008})}\BibitemShut {NoStop}%
\bibitem [{\citenamefont {Macridin}\ \emph {et~al.}(2004)\citenamefont
  {Macridin}, \citenamefont {Jarrell},\ and\ \citenamefont
  {Maier}}]{PhysRevB.70.113105}%
  \BibitemOpen
  \bibfield  {author} {\bibinfo {author} {\bibfnamefont {A.}~\bibnamefont
  {Macridin}}, \bibinfo {author} {\bibfnamefont {M.}~\bibnamefont {Jarrell}}, \
  and\ \bibinfo {author} {\bibfnamefont {T.}~\bibnamefont {Maier}},\ }\href
  {\doibase 10.1103/PhysRevB.70.113105} {\bibfield  {journal} {\bibinfo
  {journal} {Phys. Rev. B}\ }\textbf {\bibinfo {volume} {70}},\ \bibinfo
  {pages} {113105} (\bibinfo {year} {2004})}\BibitemShut {NoStop}%
\bibitem [{\citenamefont {Assaad}(2005)}]{PhysRevB.71.075103}%
  \BibitemOpen
  \bibfield  {author} {\bibinfo {author} {\bibfnamefont {F.~F.}\ \bibnamefont
  {Assaad}},\ }\href {\doibase 10.1103/PhysRevB.71.075103} {\bibfield
  {journal} {\bibinfo  {journal} {Phys. Rev. B}\ }\textbf {\bibinfo {volume}
  {71}},\ \bibinfo {pages} {075103} (\bibinfo {year} {2005})}\BibitemShut
  {NoStop}%
\bibitem [{\citenamefont {Weber}\ \emph {et~al.}(2009)\citenamefont {Weber},
  \citenamefont {L\"auchli}, \citenamefont {Mila},\ and\ \citenamefont
  {Giamarchi}}]{PhysRevLett.102.017005}%
  \BibitemOpen
  \bibfield  {author} {\bibinfo {author} {\bibfnamefont {C.}~\bibnamefont
  {Weber}}, \bibinfo {author} {\bibfnamefont {A.}~\bibnamefont {L\"auchli}},
  \bibinfo {author} {\bibfnamefont {F.}~\bibnamefont {Mila}}, \ and\ \bibinfo
  {author} {\bibfnamefont {T.}~\bibnamefont {Giamarchi}},\ }\href {\doibase
  10.1103/PhysRevLett.102.017005} {\bibfield  {journal} {\bibinfo  {journal}
  {Phys. Rev. Lett.}\ }\textbf {\bibinfo {volume} {102}},\ \bibinfo {pages}
  {017005} (\bibinfo {year} {2009})}\BibitemShut {NoStop}%
\bibitem [{\citenamefont {Greiter}\ and\ \citenamefont
  {Thomale}(2007)}]{PhysRevLett.99.027005}%
  \BibitemOpen
  \bibfield  {author} {\bibinfo {author} {\bibfnamefont {M.}~\bibnamefont
  {Greiter}}\ and\ \bibinfo {author} {\bibfnamefont {R.}~\bibnamefont
  {Thomale}},\ }\href {\doibase 10.1103/PhysRevLett.99.027005} {\bibfield
  {journal} {\bibinfo  {journal} {Phys. Rev. Lett.}\ }\textbf {\bibinfo
  {volume} {99}},\ \bibinfo {pages} {027005} (\bibinfo {year}
  {2007})}\BibitemShut {NoStop}%
\bibitem [{\citenamefont {Weber}\ \emph {et~al.}(2014)\citenamefont {Weber},
  \citenamefont {Giamarchi},\ and\ \citenamefont
  {Varma}}]{PhysRevLett.112.117001}%
  \BibitemOpen
  \bibfield  {author} {\bibinfo {author} {\bibfnamefont {C.}~\bibnamefont
  {Weber}}, \bibinfo {author} {\bibfnamefont {T.}~\bibnamefont {Giamarchi}}, \
  and\ \bibinfo {author} {\bibfnamefont {C.~M.}\ \bibnamefont {Varma}},\ }\href
  {\doibase 10.1103/PhysRevLett.112.117001} {\bibfield  {journal} {\bibinfo
  {journal} {Phys. Rev. Lett.}\ }\textbf {\bibinfo {volume} {112}},\ \bibinfo
  {pages} {117001} (\bibinfo {year} {2014})}\BibitemShut {NoStop}%
\bibitem [{\citenamefont {Garc\'{\i}a-Mart\'{\i}nez}\ \emph
  {et~al.}(2013)\citenamefont {Garc\'{\i}a-Mart\'{\i}nez}, \citenamefont
  {Grushin}, \citenamefont {Neupert}, \citenamefont {Valenzuela},\ and\
  \citenamefont {Castro}}]{PhysRevB.88.245123}%
  \BibitemOpen
  \bibfield  {author} {\bibinfo {author} {\bibfnamefont {N.~A.}\ \bibnamefont
  {Garc\'{\i}a-Mart\'{\i}nez}}, \bibinfo {author} {\bibfnamefont {A.~G.}\
  \bibnamefont {Grushin}}, \bibinfo {author} {\bibfnamefont {T.}~\bibnamefont
  {Neupert}}, \bibinfo {author} {\bibfnamefont {B.}~\bibnamefont {Valenzuela}},
  \ and\ \bibinfo {author} {\bibfnamefont {E.~V.}\ \bibnamefont {Castro}},\
  }\href {\doibase 10.1103/PhysRevB.88.245123} {\bibfield  {journal} {\bibinfo
  {journal} {Phys. Rev. B}\ }\textbf {\bibinfo {volume} {88}},\ \bibinfo
  {pages} {245123} (\bibinfo {year} {2013})}\BibitemShut {NoStop}%
\bibitem [{\citenamefont {Daghofer}\ and\ \citenamefont
  {Hohenadler}(2014)}]{PhysRevB.89.035103}%
  \BibitemOpen
  \bibfield  {author} {\bibinfo {author} {\bibfnamefont {M.}~\bibnamefont
  {Daghofer}}\ and\ \bibinfo {author} {\bibfnamefont {M.}~\bibnamefont
  {Hohenadler}},\ }\href {\doibase 10.1103/PhysRevB.89.035103} {\bibfield
  {journal} {\bibinfo  {journal} {Phys. Rev. B}\ }\textbf {\bibinfo {volume}
  {89}},\ \bibinfo {pages} {035103} (\bibinfo {year} {2014})}\BibitemShut
  {NoStop}%
\bibitem [{\citenamefont {Motruk}\ \emph {et~al.}(2015)\citenamefont {Motruk},
  \citenamefont {Grushin}, \citenamefont {de~Juan},\ and\ \citenamefont
  {Pollmann}}]{PhysRevB.92.085147}%
  \BibitemOpen
  \bibfield  {author} {\bibinfo {author} {\bibfnamefont {J.}~\bibnamefont
  {Motruk}}, \bibinfo {author} {\bibfnamefont {A.~G.}\ \bibnamefont {Grushin}},
  \bibinfo {author} {\bibfnamefont {F.}~\bibnamefont {de~Juan}}, \ and\
  \bibinfo {author} {\bibfnamefont {F.}~\bibnamefont {Pollmann}},\ }\href
  {\doibase 10.1103/PhysRevB.92.085147} {\bibfield  {journal} {\bibinfo
  {journal} {Phys. Rev. B}\ }\textbf {\bibinfo {volume} {92}},\ \bibinfo
  {pages} {085147} (\bibinfo {year} {2015})}\BibitemShut {NoStop}%
\bibitem [{\citenamefont {Capponi}\ and\ \citenamefont
  {L\"auchli}(2015)}]{PhysRevB.92.085146}%
  \BibitemOpen
  \bibfield  {author} {\bibinfo {author} {\bibfnamefont {S.}~\bibnamefont
  {Capponi}}\ and\ \bibinfo {author} {\bibfnamefont {A.~M.}\ \bibnamefont
  {L\"auchli}},\ }\href {\doibase 10.1103/PhysRevB.92.085146} {\bibfield
  {journal} {\bibinfo  {journal} {Phys. Rev. B}\ }\textbf {\bibinfo {volume}
  {92}},\ \bibinfo {pages} {085146} (\bibinfo {year} {2015})}\BibitemShut
  {NoStop}%
\bibitem [{\citenamefont {Scherer}\ \emph {et~al.}(2015)\citenamefont
  {Scherer}, \citenamefont {Scherer},\ and\ \citenamefont
  {Honerkamp}}]{PhysRevB.92.155137}%
  \BibitemOpen
  \bibfield  {author} {\bibinfo {author} {\bibfnamefont {D.~D.}\ \bibnamefont
  {Scherer}}, \bibinfo {author} {\bibfnamefont {M.~M.}\ \bibnamefont
  {Scherer}}, \ and\ \bibinfo {author} {\bibfnamefont {C.}~\bibnamefont
  {Honerkamp}},\ }\href {\doibase 10.1103/PhysRevB.92.155137} {\bibfield
  {journal} {\bibinfo  {journal} {Phys. Rev. B}\ }\textbf {\bibinfo {volume}
  {92}},\ \bibinfo {pages} {155137} (\bibinfo {year} {2015})}\BibitemShut
  {NoStop}%
\bibitem [{\citenamefont {Kiesel}\ and\ \citenamefont
  {Thomale}(2012)}]{Kiesel2012}%
  \BibitemOpen
  \bibfield  {author} {\bibinfo {author} {\bibfnamefont {M.~L.}\ \bibnamefont
  {Kiesel}}\ and\ \bibinfo {author} {\bibfnamefont {R.}~\bibnamefont
  {Thomale}},\ }\href {\doibase 10.1103/PhysRevB.86.121105} {\bibfield
  {journal} {\bibinfo  {journal} {Phys. Rev. B}\ }\textbf {\bibinfo {volume}
  {86}},\ \bibinfo {pages} {121105} (\bibinfo {year} {2012})}\BibitemShut
  {NoStop}%
\bibitem [{\citenamefont {Wu}\ \emph {et~al.}(2021)\citenamefont {Wu},
  \citenamefont {Schwemmer}, \citenamefont {M\"uller}, \citenamefont
  {Consiglio}, \citenamefont {Sangiovanni}, \citenamefont {Di~Sante},
  \citenamefont {Iqbal}, \citenamefont {Hanke}, \citenamefont {Schnyder},
  \citenamefont {Denner}, \citenamefont {Fischer}, \citenamefont {Neupert},\
  and\ \citenamefont {Thomale}}]{Wu2021}%
  \BibitemOpen
  \bibfield  {author} {\bibinfo {author} {\bibfnamefont {X.}~\bibnamefont
  {Wu}}, \bibinfo {author} {\bibfnamefont {T.}~\bibnamefont {Schwemmer}},
  \bibinfo {author} {\bibfnamefont {T.}~\bibnamefont {M\"uller}}, \bibinfo
  {author} {\bibfnamefont {A.}~\bibnamefont {Consiglio}}, \bibinfo {author}
  {\bibfnamefont {G.}~\bibnamefont {Sangiovanni}}, \bibinfo {author}
  {\bibfnamefont {D.}~\bibnamefont {Di~Sante}}, \bibinfo {author}
  {\bibfnamefont {Y.}~\bibnamefont {Iqbal}}, \bibinfo {author} {\bibfnamefont
  {W.}~\bibnamefont {Hanke}}, \bibinfo {author} {\bibfnamefont {A.~P.}\
  \bibnamefont {Schnyder}}, \bibinfo {author} {\bibfnamefont {M.~M.}\
  \bibnamefont {Denner}}, \bibinfo {author} {\bibfnamefont {M.~H.}\
  \bibnamefont {Fischer}}, \bibinfo {author} {\bibfnamefont {T.}~\bibnamefont
  {Neupert}}, \ and\ \bibinfo {author} {\bibfnamefont {R.}~\bibnamefont
  {Thomale}},\ }\href {\doibase 10.1103/PhysRevLett.127.177001} {\bibfield
  {journal} {\bibinfo  {journal} {Phys. Rev. Lett.}\ }\textbf {\bibinfo
  {volume} {127}},\ \bibinfo {pages} {177001} (\bibinfo {year}
  {2021})}\BibitemShut {NoStop}%
\bibitem [{\citenamefont {Kiesel}\ \emph {et~al.}(2013)\citenamefont {Kiesel},
  \citenamefont {Platt},\ and\ \citenamefont {Thomale}}]{Kiesel2013}%
  \BibitemOpen
  \bibfield  {author} {\bibinfo {author} {\bibfnamefont {M.~L.}\ \bibnamefont
  {Kiesel}}, \bibinfo {author} {\bibfnamefont {C.}~\bibnamefont {Platt}}, \
  and\ \bibinfo {author} {\bibfnamefont {R.}~\bibnamefont {Thomale}},\ }\href
  {\doibase 10.1103/PhysRevLett.110.126405} {\bibfield  {journal} {\bibinfo
  {journal} {Phys. Rev. Lett.}\ }\textbf {\bibinfo {volume} {110}},\ \bibinfo
  {pages} {126405} (\bibinfo {year} {2013})}\BibitemShut {NoStop}%
\bibitem [{\citenamefont {Yu}\ and\ \citenamefont {Li}(2012)}]{SYu2012}%
  \BibitemOpen
  \bibfield  {author} {\bibinfo {author} {\bibfnamefont {S.-L.}\ \bibnamefont
  {Yu}}\ and\ \bibinfo {author} {\bibfnamefont {J.-X.}\ \bibnamefont {Li}},\
  }\href {\doibase 10.1103/PhysRevB.85.144402} {\bibfield  {journal} {\bibinfo
  {journal} {Phys. Rev. B}\ }\textbf {\bibinfo {volume} {85}},\ \bibinfo
  {pages} {144402} (\bibinfo {year} {2012})}\BibitemShut {NoStop}%
\bibitem [{\citenamefont {Wang}\ \emph {et~al.}(2013)\citenamefont {Wang},
  \citenamefont {Li}, \citenamefont {Xiang},\ and\ \citenamefont
  {Wang}}]{Wang2013}%
  \BibitemOpen
  \bibfield  {author} {\bibinfo {author} {\bibfnamefont {W.-S.}\ \bibnamefont
  {Wang}}, \bibinfo {author} {\bibfnamefont {Z.-Z.}\ \bibnamefont {Li}},
  \bibinfo {author} {\bibfnamefont {Y.-Y.}\ \bibnamefont {Xiang}}, \ and\
  \bibinfo {author} {\bibfnamefont {Q.-H.}\ \bibnamefont {Wang}},\ }\href
  {\doibase 10.1103/PhysRevB.87.115135} {\bibfield  {journal} {\bibinfo
  {journal} {Phys. Rev. B}\ }\textbf {\bibinfo {volume} {87}},\ \bibinfo
  {pages} {115135} (\bibinfo {year} {2013})}\BibitemShut {NoStop}%
\bibitem [{\citenamefont {Ferrari}\ \emph {et~al.}(2022)\citenamefont
  {Ferrari}, \citenamefont {Becca},\ and\ \citenamefont
  {Valent\'{\i}}}]{FerrariPRB2022}%
  \BibitemOpen
  \bibfield  {author} {\bibinfo {author} {\bibfnamefont {F.}~\bibnamefont
  {Ferrari}}, \bibinfo {author} {\bibfnamefont {F.}~\bibnamefont {Becca}}, \
  and\ \bibinfo {author} {\bibfnamefont {R.}~\bibnamefont {Valent\'{\i}}},\
  }\href {\doibase 10.1103/PhysRevB.106.L081107} {\bibfield  {journal}
  {\bibinfo  {journal} {Phys. Rev. B}\ }\textbf {\bibinfo {volume} {106}},\
  \bibinfo {pages} {L081107} (\bibinfo {year} {2022})}\BibitemShut {NoStop}%
\bibitem [{\citenamefont {Dong}\ \emph {et~al.}(2023)\citenamefont {Dong},
  \citenamefont {Wang},\ and\ \citenamefont {Zhou}}]{Dong2023}%
  \BibitemOpen
  \bibfield  {author} {\bibinfo {author} {\bibfnamefont {J.-W.}\ \bibnamefont
  {Dong}}, \bibinfo {author} {\bibfnamefont {Z.}~\bibnamefont {Wang}}, \ and\
  \bibinfo {author} {\bibfnamefont {S.}~\bibnamefont {Zhou}},\ }\href {\doibase
  10.1103/PhysRevB.107.045127} {\bibfield  {journal} {\bibinfo  {journal}
  {Phys. Rev. B}\ }\textbf {\bibinfo {volume} {107}},\ \bibinfo {pages}
  {045127} (\bibinfo {year} {2023})}\BibitemShut {NoStop}%
\bibitem [{\citenamefont {Fu}\ \emph {et~al.}(2025)\citenamefont {Fu},
  \citenamefont {Zhan}, \citenamefont {Dürrnagel}, \citenamefont {Hohmann},
  \citenamefont {Thomale}, \citenamefont {Hu}, \citenamefont {Wang},
  \citenamefont {Zhou},\ and\ \citenamefont {Wu}}]{RPA_paper}%
  \BibitemOpen
  \bibfield  {author} {\bibinfo {author} {\bibfnamefont {R.}~\bibnamefont
  {Fu}}, \bibinfo {author} {\bibfnamefont {J.}~\bibnamefont {Zhan}}, \bibinfo
  {author} {\bibfnamefont {M.}~\bibnamefont {Dürrnagel}}, \bibinfo {author}
  {\bibfnamefont {H.}~\bibnamefont {Hohmann}}, \bibinfo {author} {\bibfnamefont
  {R.}~\bibnamefont {Thomale}}, \bibinfo {author} {\bibfnamefont
  {J.}~\bibnamefont {Hu}}, \bibinfo {author} {\bibfnamefont {Z.}~\bibnamefont
  {Wang}}, \bibinfo {author} {\bibfnamefont {S.}~\bibnamefont {Zhou}}, \ and\
  \bibinfo {author} {\bibfnamefont {X.}~\bibnamefont {Wu}},\ }\href {\doibase
  10.1093/nsr/nwaf414} {\bibfield  {journal} {\bibinfo  {journal} {National
  Science Review}\ }\textbf {\bibinfo {volume} {12}},\ \bibinfo {pages}
  {nwaf414} (\bibinfo {year} {2025})}\BibitemShut {NoStop}%
\bibitem [{\citenamefont {Wen}\ \emph {et~al.}(2010)\citenamefont {Wen},
  \citenamefont {R\"uegg}, \citenamefont {Wang},\ and\ \citenamefont
  {Fiete}}]{PhysRevB.82.075125}%
  \BibitemOpen
  \bibfield  {author} {\bibinfo {author} {\bibfnamefont {J.}~\bibnamefont
  {Wen}}, \bibinfo {author} {\bibfnamefont {A.}~\bibnamefont {R\"uegg}},
  \bibinfo {author} {\bibfnamefont {C.-C.~J.}\ \bibnamefont {Wang}}, \ and\
  \bibinfo {author} {\bibfnamefont {G.~A.}\ \bibnamefont {Fiete}},\ }\href
  {\doibase 10.1103/PhysRevB.82.075125} {\bibfield  {journal} {\bibinfo
  {journal} {Phys. Rev. B}\ }\textbf {\bibinfo {volume} {82}},\ \bibinfo
  {pages} {075125} (\bibinfo {year} {2010})}\BibitemShut {NoStop}%
\bibitem [{SM()}]{SM}%
  \BibitemOpen
  \href@noop {} {}\bibinfo {note} {See Supplemental Material for details of the
  FRG methods employed, the Ginzburg--Landau analysis, and the physical
  properties of LCO, as well as its realizations in a spinful model with Zeeman
  splitting.}\BibitemShut {Stop}%
\bibitem [{\citenamefont {Metzner}\ \emph {et~al.}(2012)\citenamefont
  {Metzner}, \citenamefont {Salmhofer}, \citenamefont {Honerkamp},
  \citenamefont {Meden},\ and\ \citenamefont {Sch\"onhammer}}]{Metzner2012}%
  \BibitemOpen
  \bibfield  {author} {\bibinfo {author} {\bibfnamefont {W.}~\bibnamefont
  {Metzner}}, \bibinfo {author} {\bibfnamefont {M.}~\bibnamefont {Salmhofer}},
  \bibinfo {author} {\bibfnamefont {C.}~\bibnamefont {Honerkamp}}, \bibinfo
  {author} {\bibfnamefont {V.}~\bibnamefont {Meden}}, \ and\ \bibinfo {author}
  {\bibfnamefont {K.}~\bibnamefont {Sch\"onhammer}},\ }\href {\doibase
  10.1103/RevModPhys.84.299} {\bibfield  {journal} {\bibinfo  {journal} {Rev.
  Mod. Phys.}\ }\textbf {\bibinfo {volume} {84}},\ \bibinfo {pages} {299}
  (\bibinfo {year} {2012})}\BibitemShut {NoStop}%
\bibitem [{\citenamefont {Platt}\ \emph {et~al.}(2013)\citenamefont {Platt},
  \citenamefont {Hanke},\ and\ \citenamefont {Thomale}}]{Platt2013}%
  \BibitemOpen
  \bibfield  {author} {\bibinfo {author} {\bibfnamefont {C.}~\bibnamefont
  {Platt}}, \bibinfo {author} {\bibfnamefont {W.}~\bibnamefont {Hanke}}, \ and\
  \bibinfo {author} {\bibfnamefont {R.}~\bibnamefont {Thomale}},\ }\href
  {\doibase 10.1080/00018732.2013.862020} {\bibfield  {journal} {\bibinfo
  {journal} {Advances in Physics}\ }\textbf {\bibinfo {volume} {62}},\ \bibinfo
  {pages} {453} (\bibinfo {year} {2013})}\BibitemShut {NoStop}%
\bibitem [{\citenamefont {Beyer}\ \emph {et~al.}(2022)\citenamefont {Beyer},
  \citenamefont {Hauck},\ and\ \citenamefont {Klebl}}]{Beyer2022}%
  \BibitemOpen
  \bibfield  {author} {\bibinfo {author} {\bibfnamefont {J.}~\bibnamefont
  {Beyer}}, \bibinfo {author} {\bibfnamefont {J.~B.}\ \bibnamefont {Hauck}}, \
  and\ \bibinfo {author} {\bibfnamefont {L.}~\bibnamefont {Klebl}},\ }\href
  {\doibase 10.1140/epjb/s10051-022-00323-y} {\bibfield  {journal} {\bibinfo
  {journal} {The European Physical Journal B}\ }\textbf {\bibinfo {volume}
  {95}},\ \bibinfo {pages} {65} (\bibinfo {year} {2022})}\BibitemShut {NoStop}%
\bibitem [{\citenamefont {O}\ \emph {et~al.}(2021)\citenamefont {O},
  \citenamefont {Kim}, \citenamefont {Pak}, \citenamefont {Jong}, \citenamefont
  {Ri},\ and\ \citenamefont {Pak}}]{O2021}%
  \BibitemOpen
  \bibfield  {author} {\bibinfo {author} {\bibfnamefont {S.-J.}\ \bibnamefont
  {O}}, \bibinfo {author} {\bibfnamefont {Y.-H.}\ \bibnamefont {Kim}}, \bibinfo
  {author} {\bibfnamefont {O.-G.}\ \bibnamefont {Pak}}, \bibinfo {author}
  {\bibfnamefont {K.-H.}\ \bibnamefont {Jong}}, \bibinfo {author}
  {\bibfnamefont {C.-W.}\ \bibnamefont {Ri}}, \ and\ \bibinfo {author}
  {\bibfnamefont {H.-C.}\ \bibnamefont {Pak}},\ }\href {\doibase
  10.1103/PhysRevB.103.235150} {\bibfield  {journal} {\bibinfo  {journal}
  {Phys. Rev. B}\ }\textbf {\bibinfo {volume} {103}},\ \bibinfo {pages}
  {235150} (\bibinfo {year} {2021})}\BibitemShut {NoStop}%
\bibitem [{\citenamefont {Gneist}\ \emph {et~al.}(2022)\citenamefont {Gneist},
  \citenamefont {Kiese}, \citenamefont {Henkel}, \citenamefont {Thomale},
  \citenamefont {Classen},\ and\ \citenamefont {Scherer}}]{Gneist2022}%
  \BibitemOpen
  \bibfield  {author} {\bibinfo {author} {\bibfnamefont {N.}~\bibnamefont
  {Gneist}}, \bibinfo {author} {\bibfnamefont {D.}~\bibnamefont {Kiese}},
  \bibinfo {author} {\bibfnamefont {R.}~\bibnamefont {Henkel}}, \bibinfo
  {author} {\bibfnamefont {R.}~\bibnamefont {Thomale}}, \bibinfo {author}
  {\bibfnamefont {L.}~\bibnamefont {Classen}}, \ and\ \bibinfo {author}
  {\bibfnamefont {M.~M.}\ \bibnamefont {Scherer}},\ }\href
  {https://link.springer.com/article/10.1140/epjb/s10051-022-00395-w}
  {\bibfield  {journal} {\bibinfo  {journal} {The European Physical Journal B}\
  }\textbf {\bibinfo {volume} {95}},\ \bibinfo {pages} {157} (\bibinfo {year}
  {2022})}\BibitemShut {NoStop}%
\bibitem [{\citenamefont {Profe}\ \emph {et~al.}(2024)\citenamefont {Profe},
  \citenamefont {Klebl}, \citenamefont {Grandi}, \citenamefont {Hohmann},
  \citenamefont {D\"urrnagel}, \citenamefont {Schwemmer}, \citenamefont
  {Thomale},\ and\ \citenamefont {Kennes}}]{Profe2024}%
  \BibitemOpen
  \bibfield  {author} {\bibinfo {author} {\bibfnamefont {J.~B.}\ \bibnamefont
  {Profe}}, \bibinfo {author} {\bibfnamefont {L.}~\bibnamefont {Klebl}},
  \bibinfo {author} {\bibfnamefont {F.}~\bibnamefont {Grandi}}, \bibinfo
  {author} {\bibfnamefont {H.}~\bibnamefont {Hohmann}}, \bibinfo {author}
  {\bibfnamefont {M.}~\bibnamefont {D\"urrnagel}}, \bibinfo {author}
  {\bibfnamefont {T.}~\bibnamefont {Schwemmer}}, \bibinfo {author}
  {\bibfnamefont {R.}~\bibnamefont {Thomale}}, \ and\ \bibinfo {author}
  {\bibfnamefont {D.~M.}\ \bibnamefont {Kennes}},\ }\href {\doibase
  10.1103/PhysRevResearch.6.043078} {\bibfield  {journal} {\bibinfo  {journal}
  {Phys. Rev. Res.}\ }\textbf {\bibinfo {volume} {6}},\ \bibinfo {pages}
  {043078} (\bibinfo {year} {2024})}\BibitemShut {NoStop}%
\bibitem [{\citenamefont {Tazai}\ \emph {et~al.}(2023)\citenamefont {Tazai},
  \citenamefont {Yamakawa},\ and\ \citenamefont {Kontani}}]{Tazai2023}%
  \BibitemOpen
  \bibfield  {author} {\bibinfo {author} {\bibfnamefont {R.}~\bibnamefont
  {Tazai}}, \bibinfo {author} {\bibfnamefont {Y.}~\bibnamefont {Yamakawa}}, \
  and\ \bibinfo {author} {\bibfnamefont {H.}~\bibnamefont {Kontani}},\ }\href
  {\doibase 10.1038/s41467-023-42952-6} {\bibfield  {journal} {\bibinfo
  {journal} {Nature Communications}\ }\textbf {\bibinfo {volume} {14}},\
  \bibinfo {pages} {7845} (\bibinfo {year} {2023})}\BibitemShut {NoStop}%
\bibitem [{\citenamefont {Schwemmer}\ \emph {et~al.}(2024)\citenamefont
  {Schwemmer}, \citenamefont {Hohmann}, \citenamefont {D\"urrnagel},
  \citenamefont {Potten}, \citenamefont {Beyer}, \citenamefont {Rachel},
  \citenamefont {Wu}, \citenamefont {Raghu}, \citenamefont {M\"uller},
  \citenamefont {Hanke},\ and\ \citenamefont {Thomale}}]{Schwemmer2023}%
  \BibitemOpen
  \bibfield  {author} {\bibinfo {author} {\bibfnamefont {T.}~\bibnamefont
  {Schwemmer}}, \bibinfo {author} {\bibfnamefont {H.}~\bibnamefont {Hohmann}},
  \bibinfo {author} {\bibfnamefont {M.}~\bibnamefont {D\"urrnagel}}, \bibinfo
  {author} {\bibfnamefont {J.}~\bibnamefont {Potten}}, \bibinfo {author}
  {\bibfnamefont {J.}~\bibnamefont {Beyer}}, \bibinfo {author} {\bibfnamefont
  {S.}~\bibnamefont {Rachel}}, \bibinfo {author} {\bibfnamefont {Y.-M.}\
  \bibnamefont {Wu}}, \bibinfo {author} {\bibfnamefont {S.}~\bibnamefont
  {Raghu}}, \bibinfo {author} {\bibfnamefont {T.}~\bibnamefont {M\"uller}},
  \bibinfo {author} {\bibfnamefont {W.}~\bibnamefont {Hanke}}, \ and\ \bibinfo
  {author} {\bibfnamefont {R.}~\bibnamefont {Thomale}},\ }\href {\doibase
  10.1103/PhysRevB.110.024501} {\bibfield  {journal} {\bibinfo  {journal}
  {Phys. Rev. B}\ }\textbf {\bibinfo {volume} {110}},\ \bibinfo {pages}
  {024501} (\bibinfo {year} {2024})}\BibitemShut {NoStop}%
\bibitem [{\citenamefont {Hu}\ \emph {et~al.}(2022)\citenamefont {Hu},
  \citenamefont {Wu}, \citenamefont {Ortiz}, \citenamefont {Ju}, \citenamefont
  {Han}, \citenamefont {Ma}, \citenamefont {Plumb}, \citenamefont {Radovic},
  \citenamefont {Thomale}, \citenamefont {Wilson}, \citenamefont {Schnyder},\
  and\ \citenamefont {Shi}}]{Hu2022}%
  \BibitemOpen
  \bibfield  {author} {\bibinfo {author} {\bibfnamefont {Y.}~\bibnamefont
  {Hu}}, \bibinfo {author} {\bibfnamefont {X.}~\bibnamefont {Wu}}, \bibinfo
  {author} {\bibfnamefont {B.~R.}\ \bibnamefont {Ortiz}}, \bibinfo {author}
  {\bibfnamefont {S.}~\bibnamefont {Ju}}, \bibinfo {author} {\bibfnamefont
  {X.}~\bibnamefont {Han}}, \bibinfo {author} {\bibfnamefont {J.}~\bibnamefont
  {Ma}}, \bibinfo {author} {\bibfnamefont {N.~C.}\ \bibnamefont {Plumb}},
  \bibinfo {author} {\bibfnamefont {M.}~\bibnamefont {Radovic}}, \bibinfo
  {author} {\bibfnamefont {R.}~\bibnamefont {Thomale}}, \bibinfo {author}
  {\bibfnamefont {S.~D.}\ \bibnamefont {Wilson}}, \bibinfo {author}
  {\bibfnamefont {A.~P.}\ \bibnamefont {Schnyder}}, \ and\ \bibinfo {author}
  {\bibfnamefont {M.}~\bibnamefont {Shi}},\ }\href {\doibase
  10.1038/s41467-022-29828-x} {\bibfield  {journal} {\bibinfo  {journal}
  {Nature Communications}\ }\textbf {\bibinfo {volume} {13}},\ \bibinfo {pages}
  {2220} (\bibinfo {year} {2022})}\BibitemShut {NoStop}%
\bibitem [{\citenamefont {Kang}\ \emph {et~al.}(2022)\citenamefont {Kang},
  \citenamefont {Fang}, \citenamefont {Kim}, \citenamefont {Ortiz},
  \citenamefont {Ryu}, \citenamefont {Kim}, \citenamefont {Yoo}, \citenamefont
  {Sangiovanni}, \citenamefont {Di~Sante}, \citenamefont {Park}, \citenamefont
  {Jozwiak}, \citenamefont {Bostwick}, \citenamefont {Rotenberg}, \citenamefont
  {Kaxiras}, \citenamefont {Wilson}, \citenamefont {Park},\ and\ \citenamefont
  {Comin}}]{Kang2022}%
  \BibitemOpen
  \bibfield  {author} {\bibinfo {author} {\bibfnamefont {M.}~\bibnamefont
  {Kang}}, \bibinfo {author} {\bibfnamefont {S.}~\bibnamefont {Fang}}, \bibinfo
  {author} {\bibfnamefont {J.-K.}\ \bibnamefont {Kim}}, \bibinfo {author}
  {\bibfnamefont {B.~R.}\ \bibnamefont {Ortiz}}, \bibinfo {author}
  {\bibfnamefont {S.~H.}\ \bibnamefont {Ryu}}, \bibinfo {author} {\bibfnamefont
  {J.}~\bibnamefont {Kim}}, \bibinfo {author} {\bibfnamefont {J.}~\bibnamefont
  {Yoo}}, \bibinfo {author} {\bibfnamefont {G.}~\bibnamefont {Sangiovanni}},
  \bibinfo {author} {\bibfnamefont {D.}~\bibnamefont {Di~Sante}}, \bibinfo
  {author} {\bibfnamefont {B.-G.}\ \bibnamefont {Park}}, \bibinfo {author}
  {\bibfnamefont {C.}~\bibnamefont {Jozwiak}}, \bibinfo {author} {\bibfnamefont
  {A.}~\bibnamefont {Bostwick}}, \bibinfo {author} {\bibfnamefont
  {E.}~\bibnamefont {Rotenberg}}, \bibinfo {author} {\bibfnamefont
  {E.}~\bibnamefont {Kaxiras}}, \bibinfo {author} {\bibfnamefont {S.~D.}\
  \bibnamefont {Wilson}}, \bibinfo {author} {\bibfnamefont {J.-H.}\
  \bibnamefont {Park}}, \ and\ \bibinfo {author} {\bibfnamefont
  {R.}~\bibnamefont {Comin}},\ }\href {\doibase 10.1038/s41567-021-01451-5}
  {\bibfield  {journal} {\bibinfo  {journal} {Nature Physics}\ }\textbf
  {\bibinfo {volume} {18}},\ \bibinfo {pages} {301} (\bibinfo {year}
  {2022})}\BibitemShut {NoStop}%
\bibitem [{\citenamefont {Jiang}\ \emph {et~al.}(2025)\citenamefont {Jiang},
  \citenamefont {Hu}, \citenamefont {C\ifmmode \u{a}\else
  \u{a}\fi{}lug\ifmmode~\u{a}\else \u{a}\fi{}ru}, \citenamefont {Felser},
  \citenamefont {Blanco-Canosa}, \citenamefont {Weng}, \citenamefont {Xu},\
  and\ \citenamefont {Bernevig}}]{PhysRevB.111.125163}%
  \BibitemOpen
  \bibfield  {author} {\bibinfo {author} {\bibfnamefont {Y.}~\bibnamefont
  {Jiang}}, \bibinfo {author} {\bibfnamefont {H.}~\bibnamefont {Hu}}, \bibinfo
  {author} {\bibfnamefont {D.}~\bibnamefont {C\ifmmode \u{a}\else
  \u{a}\fi{}lug\ifmmode~\u{a}\else \u{a}\fi{}ru}}, \bibinfo {author}
  {\bibfnamefont {C.}~\bibnamefont {Felser}}, \bibinfo {author} {\bibfnamefont
  {S.}~\bibnamefont {Blanco-Canosa}}, \bibinfo {author} {\bibfnamefont
  {H.}~\bibnamefont {Weng}}, \bibinfo {author} {\bibfnamefont {Y.}~\bibnamefont
  {Xu}}, \ and\ \bibinfo {author} {\bibfnamefont {B.~A.}\ \bibnamefont
  {Bernevig}},\ }\href {\doibase 10.1103/PhysRevB.111.125163} {\bibfield
  {journal} {\bibinfo  {journal} {Phys. Rev. B}\ }\textbf {\bibinfo {volume}
  {111}},\ \bibinfo {pages} {125163} (\bibinfo {year} {2025})}\BibitemShut
  {NoStop}%
\bibitem [{\citenamefont {Tan}\ \emph {et~al.}(2021)\citenamefont {Tan},
  \citenamefont {Liu}, \citenamefont {Wang},\ and\ \citenamefont
  {Yan}}]{Tan2021}%
  \BibitemOpen
  \bibfield  {author} {\bibinfo {author} {\bibfnamefont {H.}~\bibnamefont
  {Tan}}, \bibinfo {author} {\bibfnamefont {Y.}~\bibnamefont {Liu}}, \bibinfo
  {author} {\bibfnamefont {Z.}~\bibnamefont {Wang}}, \ and\ \bibinfo {author}
  {\bibfnamefont {B.}~\bibnamefont {Yan}},\ }\href {\doibase
  10.1103/PhysRevLett.127.046401} {\bibfield  {journal} {\bibinfo  {journal}
  {Phys. Rev. Lett.}\ }\textbf {\bibinfo {volume} {127}},\ \bibinfo {pages}
  {046401} (\bibinfo {year} {2021})}\BibitemShut {NoStop}%
\bibitem [{\citenamefont {Gutierrez-Amigo}\ \emph {et~al.}(2024)\citenamefont
  {Gutierrez-Amigo}, \citenamefont {Dangic}, \citenamefont {Guo}, \citenamefont
  {Felser}, \citenamefont {Moll}, \citenamefont {Vergniory},\ and\
  \citenamefont {Errea}}]{Gutierrez-Amigo2024}%
  \BibitemOpen
  \bibfield  {author} {\bibinfo {author} {\bibfnamefont {M.}~\bibnamefont
  {Gutierrez-Amigo}}, \bibinfo {author} {\bibfnamefont {D.}~\bibnamefont
  {Dangic}}, \bibinfo {author} {\bibfnamefont {C.}~\bibnamefont {Guo}},
  \bibinfo {author} {\bibfnamefont {C.}~\bibnamefont {Felser}}, \bibinfo
  {author} {\bibfnamefont {P.~J.~W.}\ \bibnamefont {Moll}}, \bibinfo {author}
  {\bibfnamefont {M.~G.}\ \bibnamefont {Vergniory}}, \ and\ \bibinfo {author}
  {\bibfnamefont {I.}~\bibnamefont {Errea}},\ }\href {\doibase
  10.1038/s43246-024-00676-0} {\bibfield  {journal} {\bibinfo  {journal}
  {Communications Materials}\ }\textbf {\bibinfo {volume} {5}},\ \bibinfo
  {pages} {234} (\bibinfo {year} {2024})}\BibitemShut {NoStop}%
\bibitem [{\citenamefont {Li}\ \emph {et~al.}(2024)\citenamefont {Li},
  \citenamefont {Kim},\ and\ \citenamefont {Kee}}]{PhysRevLett.132.146501}%
  \BibitemOpen
  \bibfield  {author} {\bibinfo {author} {\bibfnamefont {H.}~\bibnamefont
  {Li}}, \bibinfo {author} {\bibfnamefont {Y.~B.}\ \bibnamefont {Kim}}, \ and\
  \bibinfo {author} {\bibfnamefont {H.-Y.}\ \bibnamefont {Kee}},\ }\href
  {\doibase 10.1103/PhysRevLett.132.146501} {\bibfield  {journal} {\bibinfo
  {journal} {Phys. Rev. Lett.}\ }\textbf {\bibinfo {volume} {132}},\ \bibinfo
  {pages} {146501} (\bibinfo {year} {2024})}\BibitemShut {NoStop}%
\bibitem [{\citenamefont {Zhou}\ and\ \citenamefont {Wang}(2022)}]{Zhou2022}%
  \BibitemOpen
  \bibfield  {author} {\bibinfo {author} {\bibfnamefont {S.}~\bibnamefont
  {Zhou}}\ and\ \bibinfo {author} {\bibfnamefont {Z.}~\bibnamefont {Wang}},\
  }\href {\doibase 10.1038/s41467-022-34832-2} {\bibfield  {journal} {\bibinfo
  {journal} {Nature Communications}\ }\textbf {\bibinfo {volume} {13}},\
  \bibinfo {pages} {7288} (\bibinfo {year} {2022})}\BibitemShut {NoStop}%
\end{thebibliography}

\begin{thebibliography}{15}%
\makeatletter
\providecommand \@ifxundefined [1]{%
 \@ifx{#1\undefined}
}%
\providecommand \@ifnum [1]{%
 \ifnum #1\expandafter \@firstoftwo
 \else \expandafter \@secondoftwo
 \fi
}%
\providecommand \@ifx [1]{%
 \ifx #1\expandafter \@firstoftwo
 \else \expandafter \@secondoftwo
 \fi
}%
\providecommand \natexlab [1]{#1}%
\providecommand \enquote  [1]{``#1''}%
\providecommand \bibnamefont  [1]{#1}%
\providecommand \bibfnamefont [1]{#1}%
\providecommand \citenamefont [1]{#1}%
\providecommand \href@noop [0]{\@secondoftwo}%
\providecommand \href [0]{\begingroup \@sanitize@url \@href}%
\providecommand \@href[1]{\@@startlink{#1}\@@href}%
\providecommand \@@href[1]{\endgroup#1\@@endlink}%
\providecommand \@sanitize@url [0]{\catcode `\\12\catcode `\$12\catcode
  `\&12\catcode `\#12\catcode `\^12\catcode `\_12\catcode `\%12\relax}%
\providecommand \@@startlink[1]{}%
\providecommand \@@endlink[0]{}%
\providecommand \url  [0]{\begingroup\@sanitize@url \@url }%
\providecommand \@url [1]{\endgroup\@href {#1}{\urlprefix }}%
\providecommand \urlprefix  [0]{URL }%
\providecommand \Eprint [0]{\href }%
\providecommand \doibase [0]{http://dx.doi.org/}%
\providecommand \selectlanguage [0]{\@gobble}%
\providecommand \bibinfo  [0]{\@secondoftwo}%
\providecommand \bibfield  [0]{\@secondoftwo}%
\providecommand \translation [1]{[#1]}%
\providecommand \BibitemOpen [0]{}%
\providecommand \bibitemStop [0]{}%
\providecommand \bibitemNoStop [0]{.\EOS\space}%
\providecommand \EOS [0]{\spacefactor3000\relax}%
\providecommand \BibitemShut  [1]{\csname bibitem#1\endcsname}%
\let\auto@bib@innerbib\@empty
%</preamble>
\bibitem [{\citenamefont {Metzner}\ \emph {et~al.}(2012)\citenamefont
  {Metzner}, \citenamefont {Salmhofer}, \citenamefont {Honerkamp},
  \citenamefont {Meden},\ and\ \citenamefont {Sch\"onhammer}}]{SM_Metzner2012}%
  \BibitemOpen
  \bibfield  {author} {\bibinfo {author} {\bibfnamefont {W.}~\bibnamefont
  {Metzner}}, \bibinfo {author} {\bibfnamefont {M.}~\bibnamefont {Salmhofer}},
  \bibinfo {author} {\bibfnamefont {C.}~\bibnamefont {Honerkamp}}, \bibinfo
  {author} {\bibfnamefont {V.}~\bibnamefont {Meden}}, \ and\ \bibinfo {author}
  {\bibfnamefont {K.}~\bibnamefont {Sch\"onhammer}},\ }\href {\doibase
  10.1103/RevModPhys.84.299} {\bibfield  {journal} {\bibinfo  {journal} {Rev.
  Mod. Phys.}\ }\textbf {\bibinfo {volume} {84}},\ \bibinfo {pages} {299}
  (\bibinfo {year} {2012})}\BibitemShut {NoStop}%
\bibitem [{\citenamefont {Platt}\ \emph {et~al.}(2013)\citenamefont {Platt},
  \citenamefont {Hanke},\ and\ \citenamefont {Thomale}}]{SM_Platt2013}%
  \BibitemOpen
  \bibfield  {author} {\bibinfo {author} {\bibfnamefont {C.}~\bibnamefont
  {Platt}}, \bibinfo {author} {\bibfnamefont {W.}~\bibnamefont {Hanke}}, \ and\
  \bibinfo {author} {\bibfnamefont {R.}~\bibnamefont {Thomale}},\ }\href@noop
  {} {\bibfield  {journal} {\bibinfo  {journal} {Advances in Physics}\ }\textbf
  {\bibinfo {volume} {62}},\ \bibinfo {pages} {453} (\bibinfo {year}
  {2013})}\BibitemShut {NoStop}%
\bibitem [{\citenamefont {Beyer}\ \emph {et~al.}(2022)\citenamefont {Beyer},
  \citenamefont {Hauck},\ and\ \citenamefont {Klebl}}]{SM_Beyer2022}%
  \BibitemOpen
  \bibfield  {author} {\bibinfo {author} {\bibfnamefont {J.}~\bibnamefont
  {Beyer}}, \bibinfo {author} {\bibfnamefont {J.~B.}\ \bibnamefont {Hauck}}, \
  and\ \bibinfo {author} {\bibfnamefont {L.}~\bibnamefont {Klebl}},\ }\href
  {\doibase 10.1140/epjb/s10051-022-00323-y} {\bibfield  {journal} {\bibinfo
  {journal} {The European Physical Journal B}\ }\textbf {\bibinfo {volume}
  {95}},\ \bibinfo {pages} {65} (\bibinfo {year} {2022})}\BibitemShut {NoStop}%
\bibitem [{\citenamefont {Wang}\ \emph {et~al.}(2012)\citenamefont {Wang},
  \citenamefont {Xiang}, \citenamefont {Wang}, \citenamefont {Wang},
  \citenamefont {Yang},\ and\ \citenamefont {Lee}}]{SM_QHWangGraphene}%
  \BibitemOpen
  \bibfield  {author} {\bibinfo {author} {\bibfnamefont {W.-S.}\ \bibnamefont
  {Wang}}, \bibinfo {author} {\bibfnamefont {Y.-Y.}\ \bibnamefont {Xiang}},
  \bibinfo {author} {\bibfnamefont {Q.-H.}\ \bibnamefont {Wang}}, \bibinfo
  {author} {\bibfnamefont {F.}~\bibnamefont {Wang}}, \bibinfo {author}
  {\bibfnamefont {F.}~\bibnamefont {Yang}}, \ and\ \bibinfo {author}
  {\bibfnamefont {D.-H.}\ \bibnamefont {Lee}},\ }\href {\doibase
  10.1103/PhysRevB.85.035414} {\bibfield  {journal} {\bibinfo  {journal} {Phys.
  Rev. B}\ }\textbf {\bibinfo {volume} {85}},\ \bibinfo {pages} {035414}
  (\bibinfo {year} {2012})}\BibitemShut {NoStop}%
\bibitem [{\citenamefont {Lichtenstein}\ \emph {et~al.}(2017)\citenamefont
  {Lichtenstein}, \citenamefont {S{\'a}nchez de~la Pe{\~{n}}a}, \citenamefont
  {Rohe}, \citenamefont {Di~Napoli}, \citenamefont {Honerkamp},\ and\
  \citenamefont {Maier}}]{SM_Lichtenstein2017}%
  \BibitemOpen
  \bibfield  {author} {\bibinfo {author} {\bibfnamefont {J.}~\bibnamefont
  {Lichtenstein}}, \bibinfo {author} {\bibfnamefont {D.}~\bibnamefont
  {S{\'a}nchez de~la Pe{\~{n}}a}}, \bibinfo {author} {\bibfnamefont
  {D.}~\bibnamefont {Rohe}}, \bibinfo {author} {\bibfnamefont {E.}~\bibnamefont
  {Di~Napoli}}, \bibinfo {author} {\bibfnamefont {C.}~\bibnamefont
  {Honerkamp}}, \ and\ \bibinfo {author} {\bibfnamefont {S.~A.}\ \bibnamefont
  {Maier}},\ }\href
  {https://www.sciencedirect.com/science/article/pii/S0010465516303927}
  {\bibfield  {journal} {\bibinfo  {journal} {Computer Physics Communications}\
  }\textbf {\bibinfo {volume} {213}},\ \bibinfo {pages} {100} (\bibinfo {year}
  {2017})}\BibitemShut {NoStop}%
\bibitem [{\citenamefont {Husemann}\ and\ \citenamefont
  {Salmhofer}(2009)}]{SM_PhysRevB.79.195125}%
  \BibitemOpen
  \bibfield  {author} {\bibinfo {author} {\bibfnamefont {C.}~\bibnamefont
  {Husemann}}\ and\ \bibinfo {author} {\bibfnamefont {M.}~\bibnamefont
  {Salmhofer}},\ }\href {\doibase 10.1103/PhysRevB.79.195125} {\bibfield
  {journal} {\bibinfo  {journal} {Phys. Rev. B}\ }\textbf {\bibinfo {volume}
  {79}},\ \bibinfo {pages} {195125} (\bibinfo {year} {2009})}\BibitemShut
  {NoStop}%
\bibitem [{\citenamefont {Honerkamp}\ and\ \citenamefont
  {Salmhofer}(2001)}]{SM_Honerkamp2001}%
  \BibitemOpen
  \bibfield  {author} {\bibinfo {author} {\bibfnamefont {C.}~\bibnamefont
  {Honerkamp}}\ and\ \bibinfo {author} {\bibfnamefont {M.}~\bibnamefont
  {Salmhofer}},\ }\href {\doibase 10.1103/PhysRevB.64.184516} {\bibfield
  {journal} {\bibinfo  {journal} {Phys. Rev. B}\ }\textbf {\bibinfo {volume}
  {64}},\ \bibinfo {pages} {184516} (\bibinfo {year} {2001})}\BibitemShut
  {NoStop}%
\bibitem [{\citenamefont {Denner}\ \emph {et~al.}(2021)\citenamefont {Denner},
  \citenamefont {Thomale},\ and\ \citenamefont {Neupert}}]{SM_Denner2021}%
  \BibitemOpen
  \bibfield  {author} {\bibinfo {author} {\bibfnamefont {M.~M.}\ \bibnamefont
  {Denner}}, \bibinfo {author} {\bibfnamefont {R.}~\bibnamefont {Thomale}}, \
  and\ \bibinfo {author} {\bibfnamefont {T.}~\bibnamefont {Neupert}},\ }\href
  {\doibase 10.1103/PhysRevLett.127.217601} {\bibfield  {journal} {\bibinfo
  {journal} {Phys. Rev. Lett.}\ }\textbf {\bibinfo {volume} {127}},\ \bibinfo
  {pages} {217601} (\bibinfo {year} {2021})}\BibitemShut {NoStop}%
\bibitem [{\citenamefont {Fu}\ \emph {et~al.}(2025)\citenamefont {Fu},
  \citenamefont {Zhan}, \citenamefont {Dürrnagel}, \citenamefont {Hohmann},
  \citenamefont {Thomale}, \citenamefont {Hu}, \citenamefont {Wang},
  \citenamefont {Zhou},\ and\ \citenamefont {Wu}}]{SM_RPA_paper}%
  \BibitemOpen
  \bibfield  {author} {\bibinfo {author} {\bibfnamefont {R.}~\bibnamefont
  {Fu}}, \bibinfo {author} {\bibfnamefont {J.}~\bibnamefont {Zhan}}, \bibinfo
  {author} {\bibfnamefont {M.}~\bibnamefont {Dürrnagel}}, \bibinfo {author}
  {\bibfnamefont {H.}~\bibnamefont {Hohmann}}, \bibinfo {author} {\bibfnamefont
  {R.}~\bibnamefont {Thomale}}, \bibinfo {author} {\bibfnamefont
  {J.}~\bibnamefont {Hu}}, \bibinfo {author} {\bibfnamefont {Z.}~\bibnamefont
  {Wang}}, \bibinfo {author} {\bibfnamefont {S.}~\bibnamefont {Zhou}}, \ and\
  \bibinfo {author} {\bibfnamefont {X.}~\bibnamefont {Wu}},\ }\href {\doibase
  10.1093/nsr/nwaf414} {\bibfield  {journal} {\bibinfo  {journal} {National
  Science Review}\ }\textbf {\bibinfo {volume} {12}},\ \bibinfo {pages}
  {nwaf414} (\bibinfo {year} {2025})}\BibitemShut {NoStop}%
\bibitem [{\citenamefont {Dong}\ \emph {et~al.}(2023)\citenamefont {Dong},
  \citenamefont {Wang},\ and\ \citenamefont {Zhou}}]{SM_Dong2023}%
  \BibitemOpen
  \bibfield  {author} {\bibinfo {author} {\bibfnamefont {J.-W.}\ \bibnamefont
  {Dong}}, \bibinfo {author} {\bibfnamefont {Z.}~\bibnamefont {Wang}}, \ and\
  \bibinfo {author} {\bibfnamefont {S.}~\bibnamefont {Zhou}},\ }\href {\doibase
  10.1103/PhysRevB.107.045127} {\bibfield  {journal} {\bibinfo  {journal}
  {Phys. Rev. B}\ }\textbf {\bibinfo {volume} {107}},\ \bibinfo {pages}
  {045127} (\bibinfo {year} {2023})}\BibitemShut {NoStop}%
\bibitem [{\citenamefont {Nandkishore}\ \emph {et~al.}(2012)\citenamefont
  {Nandkishore}, \citenamefont {Chern},\ and\ \citenamefont
  {Chubukov}}]{SM_Nandkishore2012}%
  \BibitemOpen
  \bibfield  {author} {\bibinfo {author} {\bibfnamefont {R.}~\bibnamefont
  {Nandkishore}}, \bibinfo {author} {\bibfnamefont {G.-W.}\ \bibnamefont
  {Chern}}, \ and\ \bibinfo {author} {\bibfnamefont {A.~V.}\ \bibnamefont
  {Chubukov}},\ }\href {\doibase 10.1103/PhysRevLett.108.227204} {\bibfield
  {journal} {\bibinfo  {journal} {Phys. Rev. Lett.}\ }\textbf {\bibinfo
  {volume} {108}},\ \bibinfo {pages} {227204} (\bibinfo {year}
  {2012})}\BibitemShut {NoStop}%
\bibitem [{\citenamefont {Xiao}\ \emph {et~al.}(2010)\citenamefont {Xiao},
  \citenamefont {Chang},\ and\ \citenamefont {Niu}}]{SM_RevModPhys.82.1959}%
  \BibitemOpen
  \bibfield  {author} {\bibinfo {author} {\bibfnamefont {D.}~\bibnamefont
  {Xiao}}, \bibinfo {author} {\bibfnamefont {M.-C.}\ \bibnamefont {Chang}}, \
  and\ \bibinfo {author} {\bibfnamefont {Q.}~\bibnamefont {Niu}},\ }\href
  {\doibase 10.1103/RevModPhys.82.1959} {\bibfield  {journal} {\bibinfo
  {journal} {Rev. Mod. Phys.}\ }\textbf {\bibinfo {volume} {82}},\ \bibinfo
  {pages} {1959} (\bibinfo {year} {2010})}\BibitemShut {NoStop}%
\bibitem [{\citenamefont {Zhou}\ and\ \citenamefont {Wang}(2022)}]{SM_Zhou2022}%
  \BibitemOpen
  \bibfield  {author} {\bibinfo {author} {\bibfnamefont {S.}~\bibnamefont
  {Zhou}}\ and\ \bibinfo {author} {\bibfnamefont {Z.}~\bibnamefont {Wang}},\
  }\href {\doibase 10.1038/s41467-022-34832-2} {\bibfield  {journal} {\bibinfo
  {journal} {Nature Communications}\ }\textbf {\bibinfo {volume} {13}},\
  \bibinfo {pages} {7288} (\bibinfo {year} {2022})}\BibitemShut {NoStop}%
\bibitem [{\citenamefont {Teng}\ \emph {et~al.}(2022)\citenamefont {Teng},
  \citenamefont {Chen}, \citenamefont {Ye}, \citenamefont {Rosenberg},
  \citenamefont {Liu}, \citenamefont {Yin}, \citenamefont {Jiang},
  \citenamefont {Oh}, \citenamefont {Hasan}, \citenamefont {Neubauer},
  \citenamefont {Gao}, \citenamefont {Xie}, \citenamefont {Hashimoto},
  \citenamefont {Lu}, \citenamefont {Jozwiak}, \citenamefont {Bostwick},
  \citenamefont {Rotenberg}, \citenamefont {Birgeneau}, \citenamefont {Chu},
  \citenamefont {Yi},\ and\ \citenamefont {Dai}}]{SM_TengXK2022}%
  \BibitemOpen
  \bibfield  {author} {\bibinfo {author} {\bibfnamefont {X.}~\bibnamefont
  {Teng}}, \bibinfo {author} {\bibfnamefont {L.}~\bibnamefont {Chen}}, \bibinfo
  {author} {\bibfnamefont {F.}~\bibnamefont {Ye}}, \bibinfo {author}
  {\bibfnamefont {E.}~\bibnamefont {Rosenberg}}, \bibinfo {author}
  {\bibfnamefont {Z.}~\bibnamefont {Liu}}, \bibinfo {author} {\bibfnamefont
  {J.-X.}\ \bibnamefont {Yin}}, \bibinfo {author} {\bibfnamefont {Y.-X.}\
  \bibnamefont {Jiang}}, \bibinfo {author} {\bibfnamefont {J.~S.}\ \bibnamefont
  {Oh}}, \bibinfo {author} {\bibfnamefont {M.~Z.}\ \bibnamefont {Hasan}},
  \bibinfo {author} {\bibfnamefont {K.~J.}\ \bibnamefont {Neubauer}}, \bibinfo
  {author} {\bibfnamefont {B.}~\bibnamefont {Gao}}, \bibinfo {author}
  {\bibfnamefont {Y.}~\bibnamefont {Xie}}, \bibinfo {author} {\bibfnamefont
  {M.}~\bibnamefont {Hashimoto}}, \bibinfo {author} {\bibfnamefont
  {D.}~\bibnamefont {Lu}}, \bibinfo {author} {\bibfnamefont {C.}~\bibnamefont
  {Jozwiak}}, \bibinfo {author} {\bibfnamefont {A.}~\bibnamefont {Bostwick}},
  \bibinfo {author} {\bibfnamefont {E.}~\bibnamefont {Rotenberg}}, \bibinfo
  {author} {\bibfnamefont {R.~J.}\ \bibnamefont {Birgeneau}}, \bibinfo {author}
  {\bibfnamefont {J.-H.}\ \bibnamefont {Chu}}, \bibinfo {author} {\bibfnamefont
  {M.}~\bibnamefont {Yi}}, \ and\ \bibinfo {author} {\bibfnamefont
  {P.}~\bibnamefont {Dai}},\ }\href {\doibase 10.1038/s41586-022-05034-z}
  {\bibfield  {journal} {\bibinfo  {journal} {Nature}\ }\textbf {\bibinfo
  {volume} {609}},\ \bibinfo {pages} {490} (\bibinfo {year}
  {2022})}\BibitemShut {NoStop}%
\bibitem [{\citenamefont {Sigrist}\ and\ \citenamefont
  {Ueda}(1991)}]{SM_Sigrist1991}%
  \BibitemOpen
  \bibfield  {author} {\bibinfo {author} {\bibfnamefont {M.}~\bibnamefont
  {Sigrist}}\ and\ \bibinfo {author} {\bibfnamefont {K.}~\bibnamefont {Ueda}},\
  }\href {\doibase 10.1103/RevModPhys.63.239} {\bibfield  {journal} {\bibinfo
  {journal} {Rev. Mod. Phys.}\ }\textbf {\bibinfo {volume} {63}},\ \bibinfo
  {pages} {239} (\bibinfo {year} {1991})}\BibitemShut {NoStop}%
\end{thebibliography}
